\documentclass[a4paper,fleqn,usenatbib]{mnras}

\usepackage{amsmath}	

\usepackage{amssymb}	
\usepackage{txfonts}
\usepackage{natbib}

\usepackage[T1]{fontenc}
\usepackage{ae,aecompl}

\usepackage{graphicx}	
\usepackage{subfig}

\usepackage[utf8]{inputenc}
\usepackage{color}

\title[Particle acceleration in MHD reconnection I]{Reconnection and particle acceleration in interacting flux ropes I. Magnetohydrodynamics and test particles in 2.5D}

\author[B. Ripperda et al.]{
B. Ripperda,$^{1}$\thanks{E-mail: bart.ripperda@kuleuven.be}
O. Porth,$^{2}$
C. Xia$^{1}$
and R. Keppens$^{1}$
\\
$^{1}$Centre for mathematical Plasma Astrophysics, Department of Mathematics, KU Leuven, Celestijnenlaan 200B, B-3001 Leuven, Belgium\\
$^{2}$Institut fur Theoretische Physik, Max-von-Laue-Str. 1, D-60438 Frankfurt, Germany\\
}

\date{Accepted XXX. Received YYY; in original form ZZZ}

\pubyear{2016}

\begin{document}
\label{firstpage}
\pagerange{\pageref{firstpage}--\pageref{lastpage}}
\maketitle
\begin{abstract}
Magnetic reconnection and non-thermal particle distributions associated with current-driven instabilities are investigated by means of resistive magnetohydrodynamics (MHD) simulations combined with relativistic test particle methods. We propose a system with two parallel, repelling current channels in an initially force-free equilibrium, as a simplified representation of flux ropes in a stellar magnetosphere. The current channels undergo a rotation and separation on Alfv\'enic timescales, forming secondary islands and (up to tearing unstable) current sheets in which non-thermal energy distributions are expected to develop. Using the recently developed particle module of our open-source grid-adaptive MPI-AMRVAC software, we simulate MHD evolution combined with test particle treatments in MHD snapshots. We explore under which plasma-$\beta$ conditions the fastest reconnection occurs in two-and-a-half dimensional (2.5D) scenarios and in these settings test particles are evolved. We quantify energy distributions, acceleration mechanisms, relativistic corrections to the particle equations of motion and effects of resistivity in magnetically dominated proton-electron plasmas. Due to large resistive electric fields and indefinite acceleration of particles in the infinitely long current channels, hard energy spectra are found in 2.5D configurations. Solutions to these numerical artifacts are proposed for both 2.5D setups and future 3D work. We discuss the magnetohydrodynamics of an additional kink instability in 3D setups and the expected effects on energy distributions. The obtained results hold as a proof-of-principle for test particle approaches in MHD simulations, relevant to explore less idealised scenarios like solar flares and more exotic astrophysical phenomena, like black hole flares, magnetar magnetospheres and pulsar wind nebulae. 
\end{abstract}


\section{Introduction}
Astrophysical plasmas are systems in which physical phenomena are coupled from macroscopic to microscopic scales in space and time. The complexity of modelling such a system comes from the coupling of relatively slow macroscopic processes and faster processes on the particle scale. One of the most important processes exhibiting these distinctive differences, which are tightly coupled on the different scales, is magnetic reconnection. On the fundamental particle level, the energy released by magnetic reconnection goes into the kinetic energy of individual particles, however investigating how instabilities lead to large scale dynamics in astrophysical systems is traditionally done with an MHD approach. The macroscopic approximation describes the time dependent evolution of magnetic fields, bulk velocity fields and density of the plasma. During reconnection this approximation breaks down. However, reconnection can be studied in the MHD context by a parametrisation approximating particle interactions on scales below the characteristic MHD length scales through viscosity and resistivity. That, however, does not give any information on particle acceleration within reconnection regions. Nor does it include the effect of accelerated particles on the global fields.
There are currently many efforts to overcome this issue and to bridge the gap between the two approaches. Fully kinetic particle-in-cell (PIC) codes (e.g. \citealt{Lapenta3}, \citealt{Guo} for relativistic plasmas and \citealt{Lapenta2}, \citealt{Li} for non-relativistic plasmas) treat both electrons and ions as particles and iteratively move these particles and update the electromagnetic fields accordingly. This is the most complete method, but as mentioned, it has to resolve the microscopic scales of the plasma, demanding extreme computational cost. Another approach is to treat a part of the plasma as a fluid and another (typically non-thermal) ensemble of particles with a PIC technique. The separation can be done based on the energy of the particles (e.g. \citealt{Sironi2}) or based on locating zones in which kinetic effects may play an important role (e.g. \citealt{Daldorff}, \citealt{Markidis2}, \citealt{Markidis3}, \citealt{Markidis}, \citealt{Vaidya}). The caveat in this approach is that there has to be a clear separation of MHD scale physics and kinetic scale physics, either based on energy or location, to choose where kinetic physics are incorporated or not. This can be problematic for plasmas with reconnection occurring over the whole domain and when a majority of particles is accelerated up to non-thermal energies, since then the whole domain has to be treated with the particle-in-cell method.

The aim of this research is twofold. Our first goal is to use high-resolution simulations both with a fixed grid and with adaptive mesh refinement (AMR) to study magnetohydrodynamic reconnection due to current-driven instabilities in Newtonian, initially force-free plasmas in 2.5D and 3D scenarios. We present an unstable configuration in which the poloidal magnetic field forms oppositely directed, repelling current channels. In this so-called tilt instability (closely related to the coalescence instability, see e.g. \citealt{Finn}; \citealt{longcope}) the current channels rotate and translate causing fast reconnection and development of nearly-singular current layers. A similar two-dimensional (2D) setup with an initially force-free equilibrium has been studied by~\cite{Richard}. In 3D scenarios the currents are liable to an additional kink instability that interacts with the tilt instability, which was studied in a non-force-free equilibrium setting in~\cite{Keppens}. By considering force-free equilibrium setups we extend the work of \citealt{Richard} and \citealt{Keppens} allowing us to explore magnetically dominated plasmas in the low plasma-$\beta$ regime at high resolution. This setup represents the top parts of adjacent flux ropes as seen in coronae of stars. The tilt instability may be accessible if the flux ropes develop anti-parallel currents. The kink instability redistributes the poloidal field and this is known to relate to violent plasmas eruptions in astrophysical systems. It is typically triggered by strong twist in the magnetic field, which in 3D scenarios is provided by the tilting and rotating of the current channels. Current channels undergoing a tilt and/or kink instability typically undergo a linear phase in which kinetic energy grows exponentially, after which the stored magnetic energy is converted into other forms of energy via reconnection. This idealised representation is proposed as a novel route to the formation of non-thermal energy distributions in solar flares.
The second goal is to give a methodological proof-of-principle for the test particle module of the MPI-AMRVAC code (\citealt{Keppensporth}). This module allows us to dynamically evolve particle populations during MHD evolutions and in MHD snapshots. We treat electrons and ions as test particles embedded in a thermal (MHD) plasma. The test particles are guided by the electromagnetic fields without giving feedback to these fields. Particle kinetic energy is assumed to dominate over magnetic energy, indicated by the so called $\sigma$-parameter, $\sigma \ll 1$, meaning that the fields in the plasma are non-relativistic but particles can reach relativistic energies. In Newtonian, low plasma-$\beta$ conditions the gyration of a particle is considered negligible and a guiding centre approximation (GCA) can be applied, simplifying the equations of motion and taking relativistic drift velocities and additional, purely relativistic drifts into account. Particle are traced with this method in the global fields obtained from snapshots of high resolution 2.5D MHD runs. This method provides information on particle dynamics, energy distributions and acceleration mechanisms in selected MHD snapshots showing fast reconnection due to interacting flux ropes. Based on these results more elaborate and realistic 3D particle simulations are proposed for future work, in which particles are traced simultaneously to the MHD evolution. This is relevant for solar flares, but also for more extreme settings encountered in strongly magnetised astrophysical plasmas. The guiding centre approach has been widely used to study test particle acceleration in solar corona conditions. In 2.5D simulations of reconnection initiated by shrinking, thin current sheets (\citealt{Zhou}, \citealt{Zhou2}) and in reconnection driven by the kink instability in 3D setups (\citealt{Rosdahl}, \citealt{Gordovskyy}, \citealt{Pinto}). In these solar corona simulations particles are not considered to reach high Lorentz factors, allowing to make a semi-relativistic approximation for particle motion where the well-known classical plasma drifts are limited by the speed of light. In \cite{Porth3} the fully relativistic equation of motions, including particle gyration, are solved to study transport of high-energy particles in pulsar wind nebulae. Particle acceleration due to idealised current driven instabilities in magnetically dominated, relativistic plasmas has been studied with both MHD and particle-in-cell approaches in~\cite{SironiPorth}.

The MHD equilibrium setup and the numerical methods employed are described in detail in Section~\ref{sec:MHDtheory} and the test particle approach is discussed in Section~\ref{sec:GCA}. In Section~\ref{sec:2Dresults} 2.5D MHD simulations are discussed, in Section~\ref{sec:3Dresults} we discuss 3D simulations and the effect of the kink instability on the tilt instability. In Section~\ref{sec:particles} the behaviour of test particles in 2.5D MHD snapshots is discussed.
\section{Numerical setup}
\subsection{MHD setup and model description}
\label{sec:MHDtheory}
We simulate two parallel, adjacent, repelling current channels in a square region $[-3L,3L] \times [-3L,3L]$ in Cartesian coordinates ($x,y$) with vector $z$ components orthogonal to the plane in both the 2.5D and 3D setups. A typical unit of length for the astrophysical systems under consideration is $L = 10 Mm$. This length scale will be used to scale the MHD results in accordance with the dimension-full particle simulations. In equilibrium the currents are described by the initial conditions for the flux function $\psi_0(x,y)$
\begin{equation}
\psi_0(x,y) = \begin{cases} \frac{2}{j_0^1 J_0 (j_0^1)}J_1(j_0^1 r) \cos(\theta) & \mbox{for } r < 1. \\ (r-\frac{1}{r}) \cos(\theta) & \mbox{for } r \geq 1. \end{cases}
\label{eq:fluxinitial}
\end{equation}
where we use polar coordinates in the ($x,y$) plane with $(r,\theta) = (\sqrt{x^2+y^2},\arctan(y/x))$, $J_1$ is the Bessel function of the first kind and $j_0^1 \approx 3.831706$ is the first root of $J_1$. From the flux function the magnetic field is found as $\mathbf{B} = \nabla \psi_0 \times \mathbf{\hat{z}} +B_z\mathbf{\hat{z}} $. An ideal MHD equilibrium can be established in two ways. One option is to find a pressure gradient balancing the Lorentz force
\begin{equation}
p(x,y) = \begin{cases} p_0 + \frac{(j_0^1)^2}{2}(\psi_0(x,y))^2 & \mbox{for } r < 1. \\ p_0 & \mbox{for } r \geq 1. \end{cases}
\label{eq:eqp}
\end{equation}
and an additional parameter $B_z = B_{z0}$, to be chosen freely to analyse different cases of plasma-$\beta$, indicating the strength of the magnetic field component in the $z$-direction. A second option is to employ a force-free magnetic field with spatially varying, vertical component $B_{z}(x,y)$ and a uniform plasma pressure $p_0$ such that the Lorentz force $\mathbf{J}\times\mathbf{B} = \nabla p = 0$, here 
\begin{equation}
B_{z}(x,y) = \begin{cases} (j_0^1)(\psi_0(x,y)) & \mbox{for} r < 1. \\ 0 & \mbox{for } r \geq 1. \end{cases}
\label{eq:eqB}
\end{equation}
The constant pressure can be chosen freely to analyse different cases of plasma-$\beta$. These equilibria result in a current distribution with two anti-parallel current channels, where $J_z = (\nabla \times \mathbf{B})_z = -\nabla^2\psi_0$. In one half of the unit circle $J_z > 0$ and in the other half $J_z < 0$. In both cases there are no currents in the region $r \geq 1$ initially and the pressure is constant there. Note that a force-free configuration is more realistic since magnetic forces are so dominant in most areas of the corona that all other forces, including gravity and plasma pressure gradients are negligible (\citealt{Schrijver};~\citealt{Wiegelmann}). The vertical magnetic field component is strong inside the current channels and zero outside the current channels, which is also more realistic physically than a setup with a constant and uniform vertical magnetic field component both in the current channels and in the background. Another advantage is that a force-free configuration allows us to explore the low plasma-$\beta$ regime.

The set of resistive, compressible 3D MHD equations is solved, 
\begin{align}
 \partial_t\rho + \nabla \cdot (\mathbf{u}\rho) &= 0,\\
 \partial_t(\rho\mathbf{u})+\nabla\cdot(\mathbf{u}\rho\mathbf{u}-\mathbf{B}\mathbf{B})+\nabla(p+\mathbf{B}^2/2) &= 0,\\
\partial_t(e)+\nabla\cdot(\mathbf{u}e-\mathbf{B}\mathbf{B}\cdot\mathbf{u}+\mathbf{u}(p+\mathbf{B}^2/2)) &= \nabla \cdot(\mathbf{B}\times \eta \mathbf{J}), \\
 \partial_t(\mathbf{B})+\nabla\cdot(\mathbf{u}\mathbf{B}-\mathbf{B}\mathbf{u}) &= -\nabla \times (\eta \mathbf{J}),
\label{eq:MHD}
\end{align}
where $\mathbf{u}$ is the bulk fluid velocity field, $e$ is the total energy density, $\mathbf{J} = \nabla \times \mathbf{B}$ the total current density and total pressure is $p = (\Gamma-1)(e-\rho\mathbf{u}^2/2-\mathbf{B}^2/2)$. 

The resistivity parameter is set to $\eta_0=10^{-4}$. We apply two different resistivity models. A uniform resistivity $\eta_0$ and an anomalous, current dependent resistivity (e.g.~\citealt{Otto})
\begin{equation}
\eta(j) = \eta_0 S(j-j_c)
\label{eq:anomalousres}
\end{equation}
with $S(\xi)$ a step function, which is unity for $\xi \geq 0$ and zero otherwise. The critical current density $j_c$ is set such that the resistivity is nonzero for values larger the equilibrium current density, such that there is no diffusion in the equilibrium, but that the threshold is much lower than the peak current reached. Different values of $j_c$ will be used to see the effect on the diffusion of the gradient of the magnetic field, and hence, the current.

We will quantify differences between the two different equilibrium setups, and we will vary both $B_{z0}$ and $p_0$ to model different plasma-$\beta$ conditions. In both cases we set the density $\rho$ to unity initially and the ratio of specific heats $\Gamma=5/3$. In the non-force-free case we fix $p_0=1/\Gamma$ and vary $B_{z0}$ over a range of $0$ to $5$ and in the force-free case, we vary $p_0$ over a range of $0.01/\Gamma$ to $5/\Gamma$. The normalisation used implies the sound speed outside the current channels as the unit of speed, the radius of the double current channel as the unit of length and the density to fix the unit of mass. We employ magnetic units where $\mu_0=1$.

Both equilibrium setups are unstable to ideal MHD instabilities with Alfv\'enic growth rates (\citealt{Richard}). To trigger these instabilities, the equilibrium is perturbed by an incompressible velocity field
\begin{align}
 u_x &= + \frac{\partial \phi_0}{\partial y} \left[ \times \sin(k_zz) \right], \nonumber \\
 u_y &= - \frac{\partial \phi_0}{\partial x} \left[ \times \sin(k_zz) \right], \nonumber \\
 u_z &= 0,
\label{eq:vpert}
\end{align}
where $\phi_0(x,y) = \xi \exp(-x^2-y^2)$ is the stream function with a perturbation amplitude $\xi = 0.0001$. In case of a 3D simulation, $k_z = 2\pi/L_z$ with $L_z=6$ the typical simulation box size and $z \in \left[-3L,3L\right]$, again with $L=10 \cdot 10^6 m$. In all 2.5D simulations, the dependence on the $z$ coordinate doesn't apply for the perturbation field. A linear stability analysis for 2D incompressible MHD, based on an energy principle, has been carried out by~\cite{Richard}, showing that the equilibrium is unstable to a tilt instability in the $(x,y)$-plane. It is shown by~\cite{Keppens} that two additional effects come into play in a non-force-free 3D setup, namely that the field lines can bend with respect to the vertical direction and that the current channels may be unstable to an ideal kink instability, depending on typical magnetic field strengths and system size. If the $z$-component of the magnetic field is strong enough, the magnetic tension may also stabilize or delay kink deformations and even prevent tilt development in the $(x,y)$ plane.

The tilt instability is an ideal MHD instability, from which it can be concluded that the resistivity has little effect on the (linear) onset phase of the instability (\citealt{Richard}). Once the instability develops and the physics becomes naturally nonlinear, it allows for fast reconnection of the field lines. To show the difference between numerical resistivity and the implemented physical resistivity, we conduct simulations with varying grid resolutions, which mainly becomes important in the chaotic reconnection regime.

We employ a zero-gradient boundary condition on all boundaries in the $(x,y)$-plane for the primitive variables $\rho$, $\mathbf{v}$ and $p$ in all setups. In 2.5D setups the $z$-direction is invariant and in 3D setups periodic boundary conditions are employed for the $z$-boundaries. In the ghost cells, the magnetic field fixes the analytic profiles from the equilibrium and subsequently employs a second order, central difference evaluation of $\nabla \cdot \mathbf{B}$, to correct the component normal to the boundary. This approach was shown to work well by~\cite{Keppens} for a non-force-free case and will therefore be used here in combination with a diffusive approach on the monopole error control. We use the same discretisation to quantify the divergence of the magnetic field and then add it as a diffusion part to the induction equation~(\ref{eq:MHD}) as $\nabla((\Delta x)^2\nabla \cdot \mathbf{B})$. For the integration we use a three-step Runge-Kutta-type scheme with a third-order limiter and a Harten-Lax-van Leer (HLL) flux prescription (\citealt{Cada}; \citealt{Keppensporth}).

Resolutions of $300^2$ up to $4800^2$ are used, where the higher resolutions are achieved by employing $4-5$ AMR grid levels. Runs with a uniform grid at the highest resolutions have confirmed that numerical instabilities due to negative pressure development at refinement boundaries are avoided, even for the lowest $\beta$ case. In the 3D setups, we use a base resolution of $150^3$ with $1-2$ AMR grid levels to achieve either $150^3$ or $300^3$ effective resolution respectively. In Table~\ref{tab:example_table} we list the most important parameters for the different cases (where \textit{ff} indicates a force-free equilibrium configuration and \textit{nff} a non-force-free equilibrium configuration and the addition of \textit{AR} indicates that an anomalous resistivity model is used with critical current density threshold $j_c$), quantifying typical plasma conditions and effective resolution used. We calculate mean initial (equilibrium) values of prevailing current density $\bar{J}$ and plasma beta $\bar{\beta}$ for both the non-force-free and force-free cases with different choices of $B_{z0}$ and $p_0$. The mean value for a scalar is computed as
\begin{equation}
\bar{f}^{\pm} \equiv \frac{\iint_{j_z(t=0)^{>}_{<}0}f dx dy}{\iint_{j_z(t=0)^{>}_{<}0} dx dy}
\label{eq:mean}
\end{equation}
over the current cross-section. The $\pm$ stands for the positive or negative current density in the $z$-direction, respectively. Initially, the area in the denominator is exactly half the unit circle ($\pi/2$), however, at later times we cannot depend purely on the sign of $J_z(t)$ due to dynamic changes in the current channels. As a solution, we advect a tracer which identifies the displaced location of the positive and negative current channels. This tracer also allows us to quantify energetics in either one of the current channels.
\begin{table}
	\centering
	\caption{The simulated cases and several characteristic parameters.}
	\label{tab:example_table}
	\begin{tabular}{lcccccr}
		\hline
		Run & $p_0$ & $\bar{B}_{z}$ & $\bar{\beta}$ & effective res. & $j_c$ & $\gamma_{tilt}$\\
		\hline
		A2dnff & $1.0/\Gamma$  & 0.0 & 12.7 & $2400^2$ & 0&  1.5000 \\
		B2dnff & $1.0/\Gamma$  & 0.1 & 5.8& $2400^2$& 0&  1.4969\\
		C2dnff & $1.0/\Gamma$  & 0.5& 2.6& $2400^2$& 0&  1.4893\\
		D2dnff & $1.0/\Gamma$  & 1.0& 1.4& $2400^2$& 0&  1.4842\\
		E2dnff & $1.0/\Gamma$  & 5.0& 0.12& $2400^2$& 0&  1.3565\\
		f2dff & $0.01/\Gamma$ & 1.14& 0.04& $300^2$& 0&  \\
		F2dff & $0.01/\Gamma$  & 1.14& 0.04& $2400^2$& 0&  1.6578\\
		FF2dff& $0.01/\Gamma$  & 1.14& 0.04& $4800^2$& 0&  \\
		F2dffAR & $0.01/\Gamma$  & 1.14& 0.04& $2400^2$& 12 &  1.8206 \\
		F2dffAR2 & $0.01/\Gamma$  & 1.14& 0.04& $2400^2$& 500 &  1.6404 \\
		G2dff & $0.05/\Gamma$  & 1.14& 0.18& $2400^2$& 0&  1.6510\\
		G2dffAR & $0.05/\Gamma$  & 1.14& 0.18& $2400^2$& 12 &  1.8012 \\
		H2dff & $0.1/\Gamma$  & 1.14& 0.36& $2400^2$& 0&  1.6492\\
		I2dff & $0.5/\Gamma$  & 1.14& 1.8& $2400^2$& 0& 1.5793\\
		J2dff & $1.0/\Gamma$  & 1.14& 4.0& $2400^2$& 0&  1.4968\\
		K2dff & $5.0/\Gamma$  & 1.14& 18.1& $2400^2$& 0 & 1.4018\\
		
		\hline
		Run & $p_0$ & $\bar{B}_{z}$ & $\bar{\beta}$ &effective res. & $j_c$ & $K_{cr}$ \\
		\hline
		
		A3dnff& $1.0/\Gamma$ & 0.0& 12.7&$300^3$& 0&  $\infty$ \\
		B3dnff& $1.0/\Gamma$ & 0.1& 5.8&$300^3$& 0&  $43.5$ \\
		C3dnff& $1.0/\Gamma$ & 0.5& 2.6&$300^3$& 0&  $8.7$ \\
		D3dnff& $1.0/\Gamma$ & 1.0& 1.4&$300^3$& 0& $4.35$ \\
		E3dnff& $1.0/\Gamma$ & 5.0& 0.12&$300^3$& 0& $0.87$  \\
		f3dff &  $0.01/\Gamma$ & 1.14& 0.04&$150^3$& 0& $3.83$ \\
		F3dff&  $0.01/\Gamma$ & 1.14& 0.04&$300^3$& 0&  $3.83$ \\
    F3dffAR & $0.01/\Gamma$  & 1.14& 0.04& $300^3$&  12 &  $3.83$\\
		G3dff &  $0.05/\Gamma$ & 1.14& 0.18&$300^3$& 0&  $3.83$ \\
   	G3dffAR &  $0.05/\Gamma$ & 1.14& 0.18&$300^3$&  12 & $3.83$  \\
		H3dff &  $0.1/\Gamma$ & 1.14& 0.36&$300^3$& 0& $3.83$ \\
		I3dff &  $0.5/\Gamma$ & 1.14& 1.8&$300^3$& 0& $3.83$ \\
		J3dff &  $1.0/\Gamma$ & 1.14& 4.0&$300^3$& 0& $3.83$  \\
		K3dff &  $5.0/\Gamma$ & 1.14& 18.1&$300^3$& 0& $3.83$  \\
		\hline
			\end{tabular}

\textbf{Note}: The leftmost column labels the various runs, \textit{ff} and \textit{nff} indicate force-\\free and non-force-free equilibrium configurations respectively. The right\\ column quantifies the tilt mode growth rate for 2.5D runs and the liability to a kink instability for $3D$ runs (see text for details).
\end{table}
\subsection{Relativistic test particle dynamics and the guiding centre approximation}
\label{sec:GCA}
The MHD approach as described in Section~\ref{sec:MHDtheory} provides the macroscopic time dependent evolution of the plasma, in terms of the magnetic field, density, pressure and bulk velocity field. The current distribution and the electric field can be calculated from these fields. For a plasma in local thermodynamic equilibrium and on typical length scales much larger than the mean free path of particles, the large scale evolution is well described by these macroscopic parameters. If reconnection takes place, localised, near-singular current sheets form and sub-MHD-scale physics becomes important. The motion of charged particles, allowed to reach high energies, in electromagnetic fields is described by a relativistic, tensorial equation of motion (e.g. ~\citealt{Landau}):
\begin{equation}
\frac{d^2x_i}{d\tau^2} = \frac{q}{mc}F_{ik}\frac{dx^k}{d\tau},
\label{eq:lorentztens}
\end{equation}
in which the Einstein summation convention is used and with $x_i = (ct, \mathbf{r})$ the particles position in spacetime, $c$ the speed of light in vacuum and $\tau$ the proper time. $F_{ik}$ is the electromagnetic field tensor consisting of the components of $\mathbf{B}$ and $\mathbf{E}$. The equations can be split by choosing a certain frame of reference. The first three components of equation~(\ref{eq:lorentztens}) are similar to the equation of motion $d\mathbf{p}/dt = q(\mathbf{E}+\mathbf{v}\times\mathbf{B}/c)$, in three-space, where $\mathbf{p}=m_0 \gamma \mathbf{v}$ is the relativistic momentum, with $m_0$ the rest mass of the particle and $\gamma = 1/\sqrt{1-v^2/c^2}$ the Lorentz factor, $q$ the particles charge, $\mathbf{E}$ and $\mathbf{B}$ the electromagnetic fields guiding the particle and $\mathbf{v}$ the particle velocity. These equations are solved in \cite{Porth3} for test particles in pulsar wind nebulae. The fourth component of (\ref{eq:lorentztens}) is the rate of change of energy of the particle.

To simplify the description of the (relativistic) motion of a charged particle in an electromagnetic field one can express the position of the particle in terms of variables which represent the gyration around the magnetic field lines and the motion of the point around which the particle gyrates, called the guiding centre (\citealt{Northropbook}, \citealt{Northrop}). The gyration can be separated from the motion of the guiding centre under the assumption that the electromagnetic field varies only over an interval of spacetime much larger than the gyroradius, and that the particle undergoes many gyration cycles before the field has varied significantly in space, respectively.  More specifically, this guiding centre approximation is based on an expansion in powers of $\Omega^{-1}d/dt$, with $\Omega = qB/m_0$ the Larmor frequency and $B$ the magnitude of the magnetic field. The gyration at frequency $\Omega$ is averaged out, for a particle with gyroradius $R_L = \gamma m_0 v_{\perp}/Bq$ and $v_{\perp}$ the velocity component perpendicular to the magnetic field. For the low plasma-$\beta$ conditions used, the gyroradius can be compared to the typical size over which the fields change. If this size is very large compared to the gyroradius, the gyration of the test particles can be neglected and the guiding centre approximation gives accurate results, comparable to the solution given by the full equation of motion~(\ref{eq:lorentztens}). Typical parameters for low plasma-$\beta$ plasma in the solar corona are, for magnetic field magnitude $B=0.03 T$, temperature $T = 10^6 K$, plasma-$\beta = 0.0004$, number density $n = 10^{16}$ $m^{-3}$, thermal speed $v_{th,e} = 5.5 \times 10^7 m s^{-1}$ for electrons and $v_{th,p} = 1.3 \times 10^6 m s^{-1}$ for protons, both giving a thermal Lorentz factor of $\gamma \approx 1$ (\citealt{Goedbloed}). In solar coronal plasmas the typical gyroradius of the particles ($R_L = 10^{-3} m$ for electrons and $R_L = 4.4 \times 10^{-2} m$ for protons) is much smaller than the length scales over which MHD fields evolve and typical timescales of the particle dynamics are much smaller than dynamic timescales of the MHD, making the test particle approach valid.

From the expansion of (\ref{eq:lorentztens}) we obtain the relativistic guiding centre equations in a covariant form. The drift velocities are allowed to approach the speed of light, removing restrictions on the electric field strength. We assume that the variations of the field in which the particle moves, in space and time, are sufficiently slow such that it changes only during intervals of proper time which are long compared to a gyration period. This is the case for non-relativistic plasmas with $\sigma \equiv B^2/4\pi\rho_0 c^2 \ll 1$ and hence non-relativistic Alfv\'en velocities, with $\rho_0 c^2$ the rest energy density of the plasma. In the case of slowly varying global fields the temporal variations of the fields are much smaller than the variations due to particle motion and hence, temporal derivatives in the guiding centre equations can be neglected. This reduces computing time further without compromising accuracy. Using these approximations and neglecting the higher order terms of the expansion, allows to write an equation which solely depends on the guiding centre position of the particle. In the solar corona this assumption is accurate, however in flares emerging from black holes or neutron stars, the global flow may be strongly relativistic and this assumption breaks down. The complete solution for the guiding centre still depends on the particular electromagnetic field and its spatial derivatives as obtained from MHD. After choosing a specific hypersurface to split the spacetime components, applying the approximations to the guiding centre equations gives three equations, corresponding to the spatial components of (\ref{eq:lorentztens}), describing the (change in) guiding centre position $\mathbf{R}$, parallel relativistic momentum $p_{\|} = m_0\gamma v_{\|}$ and relativistic magnetic moment $\mu_r = m_0 \gamma^{2} v^{2}_{\perp}/2B$ in three-space (\citealt{Vandervoort})
\begin{align*}
\frac{d\mathbf{R}}{dt} = \frac{\left(\gamma v_{\|}\right)}{\gamma}\mathbf{\hat{b}}+\frac{\mathbf{\hat{b}}}{B\left(1-\frac{E_{\perp}^{2}}{B^2}\right)} \times \Biggl\{ -\left(1-\frac{E_{\perp}^{2}}{B^2}\right)c\mathbf{E} + \Biggr. \nonumber
\end{align*}
\begin{align*}
\frac{cm_0\gamma}{q}\left(v_{\|}^{2}\left(\mathbf{\hat{b}}\cdot\nabla\right)\mathbf{\hat{b}}+v_{\|}\left(\mathbf{u_E}\cdot\nabla\right)\mathbf{\hat{b}} + v_{\|}\left(\mathbf{\hat{b}}\cdot\nabla\right)\mathbf{u_E} + \left(\mathbf{u_E}\cdot \nabla\right)\mathbf{u_E}\right) + \nonumber
\end{align*}
\begin{align}
\Biggl. \frac{\mu_r c}{\gamma q}\nabla\left[B\left(1-\frac{E_{\perp}^{2}}{B^2}\right)^{1/2}\right]  + \frac{v_{\|}E_{\|}}{c}\mathbf{u_E}  \Biggr\},
\label{eq:gcastatic1}
\end{align}
\begin{align}
\frac{d \left(m_0 \gamma v_{\|}\right)}{dt} =  m_0\gamma\mathbf{u_E}\cdot \left(v_{\|}^{2}\left(\mathbf{\hat{b}}\cdot\nabla\right)\mathbf{\hat{b}}+v_{\|}\left(\mathbf{u_E}\cdot\nabla\right)\mathbf{\hat{b}}\right) +\nonumber \\ 
qE_{\|} -\frac{\mu_r}{\gamma}\mathbf{\hat{b}}\cdot\nabla\left[B\left(1-\frac{E^{2}_{\perp}}{B^2}\right)^{1/2}\right],
\label{eq:gcastatic2}
\end{align}
\begin{equation}
\frac{d \left(m_0 \gamma^{*2} v^{*2}_{\perp}/2B^*\right)}{dt} = \frac{d \mu_{r}^{*}}{dt} = 0.
\label{eq:gcastatic3}
\end{equation}
Here, $\mathbf{\hat{b}}$ is the unit vector in the direction of the magnetic field, $E$, which can be split as $E = \sqrt{E^{2}_{\perp} +E^{2}_{\|}}$, the amplitude of the electric field vector and $v_{\|}$ the component of the particle velocity vector parallel to $\mathbf{\hat{b}}$. In resistive plasmas the electric field typically has a large component parallel to the magnetic field, $E_{\|}$, called a resistive electric field. These parallel electric fields typically produce high energy particles with a small pitch angle $\alpha = \arctan(v_{\perp}/v_{\|})$, the angle between the velocity vector of a particle and the unit vector $\mathbf{\hat{b}}$. Furthermore, due to several drift terms in the equation of motion, particles are allowed to move in a direction not parallel to the magnetic field lines guiding them (hence, possibly obtaining a larger pitch angle). The drift velocity, perpendicular to $\mathbf{B}$ is written as $\mathbf{u_E} = c\mathbf{E}\times\mathbf{\mathbf{\hat{b}}}/B$ and $v^{*}_{\perp}$ is the perpendicular velocity the particle has, in the frame of reference moving at $\mathbf{u_E}$. The magnetic field in that frame is given by $B^* = B(1-E^{2}_{\perp}/B^2)^{1/2}$ up to first order. The relativistic magnetic moment $\mu_{r}^{*}$ is proportional to the flux through the gyration circle, again in the frame of reference moving at $\mathbf{u_E}$, and this is constant. The Lorentz factor is not constant, but oscillates at the gyrofrequency. This oscillation is averaged out as well, to give $\gamma = \gamma^{*}(1-E^{2}_{\perp}/B^2)^{-1/2}$. Because of the appearance of the denominator $1-E^{2}_{\perp}/B^2$ in the guiding centre equations of motion, it is evident that the perpendicular electric field $E_{\perp}$ has to be smaller than $B$ for the equations not to reach an unphysical singularity. The fourth equation following from this analysis describes the average rate of increase of the particles energy, which does not give more information up to first order, after applying the assumption of slowly varying fields and is therefore not evolved.

If the particle velocity is restricted to $v^2 \ll c^2$ and the magnitude of its drift velocity $u_E^2 \ll c^2$ then $\gamma \rightarrow 1$ and $1/\sqrt{(1-E_{\perp}^2/B^2)} = 1/\sqrt{1-u_{E}^2/c^2} \rightarrow 1$. Then also the relativistic magnetic moment, a constant of motion, becomes the classical magnetic moment $\mu_r = m_0\gamma^2 v_{\perp}^2/2B \rightarrow \mu = m v_{\perp}^2/2B$. The additional, purely relativistic term on the right-hand-side of equation (\ref{eq:gcastatic1}) is of the order $v^2/c^2$ compared to the other terms and hence, negligible in the Newtonian limit $v^2 \ll c^2$ (\citealt{Northropbook}). The Newtonian equations of motion for the guiding centre are then retrieved as
\begin{equation}
\frac{d\mathbf{R}}{dt} =v_{\|}\mathbf{\hat{b}}+\frac{\mathbf{\hat{b}}}{B} \times \Biggl\{ -c\mathbf{E} + \frac{cm}{q}\left(v_{\|}\frac{d\mathbf{\hat{b}}}{dt}+\frac{d\mathbf{u_E}}{dt}\right)  + \frac{\mu c}{q}\nabla B \Biggr\},
\label{eq:gcanewton1}
\end{equation}
\begin{equation}
\frac{d \left(m v_{\|}\right)}{dt} = m \mathbf{u_E}\cdot \frac{d\mathbf{\hat{b}}}{dt} +qE_{\|} -\mu\mathbf{\hat{b}}\cdot\nabla B.
\label{eq:gcanewton2}
\end{equation}
Comparing the relativistic guiding centre equations (\ref{eq:gcastatic1}) to their Newtonian limit (\ref{eq:gcanewton1}) shows how the separate drift terms are modified for relativistic particles of mass $m_0\gamma$. The first term on the right-hand-side of (\ref{eq:gcastatic1}) is the motion parallel to $\mathbf{\hat{b}}$, unmodified because the factors of $\gamma$ cancel. The second term (the first term in the cross product) is the $\mathbf{E}\times\mathbf{B}$ drift, which is also unmodified. The third term, combines the curvature drift (resulting from the static part of the inertial drift) $v_{\|}d\mathbf{\hat{b}}/dt = v_{\|}^{2}\left(\mathbf{\hat{b}}\cdot\nabla\right)\mathbf{\hat{b}}+v_{\|}\left(\mathbf{u_E}\cdot\nabla\right)\mathbf{\hat{b}}$ and the polarisation drift $d\mathbf{u_E}/dt = v_{\|}\left(\mathbf{\hat{b}}\cdot\nabla\right)\mathbf{u_E} + \left(\mathbf{u_E}\cdot \nabla\right)\mathbf{u_E}$, where non-static fields are neglected. For these drifts the gyration period increases by a factor $\gamma$. Because the mass of the gyrating particle is $\gamma$ times larger, the gyroradius is as well. Then, the magnitude of the magnetic field, in the frame of reference moving at $\mathbf{u_E}$, is $B^* = B(1-E^{2}_{\perp}/B^2)^{1/2} = B\sqrt{1-u_{E}^2/c^2}$ up to first order, explaining the factor $1/\sqrt{1-u_{E}^2/c^2}$ appearing in all terms including the magnetic field magnitude. The $\nabla B$ drift, the fourth term, is also a factor $\gamma$ larger than in equation (\ref{eq:gcanewton1}) (a factor $\gamma^2$ is hidden in $\mu_r = m_0 \gamma^{*2} v^{*2}_{\perp}/2B^*$). The $\nabla B$ drift results from differences in the radius of curvature on opposite sides of the gyro-orbit, hence the larger the gyroradius, the larger the effect of the drift velocity. The last term on the right-hand-side of equation (\ref{eq:gcastatic1}) is an additional, purely relativistic drift in the direction $\mathbf{\hat{b}} \times \mathbf{u_E}$, which is the direction of $E_{\perp}$ (\citealt{Northropbook}). 

By comparing the equation of motion for the parallel momentum (\ref{eq:gcastatic2}) to its Newtonian limit (\ref{eq:gcanewton2}) it can be seen how the guiding center acceleration is modified for relativistic particles. The first term is the acceleration of a particle of mass $m_0\gamma$ due to a change of direction of $\mathbf{B}$. The parallel electric acceleration, the second term on the right-hand-side of (\ref{eq:gcastatic2}), is unmodified relativistically since no mass is involved in that term. The third term in the parallel acceleration, the mirror deceleration is $\gamma$ times larger in the relativistic equation (\ref{eq:gcastatic2}). Again because the mass, and hence the gyroradius are larger by a factor $\gamma$.
  
Depending on the relative strength of the perpendicular electric field, the guiding centre equations for slowly varying fields (\ref{eq:gcastatic1})-(\ref{eq:gcastatic3}) can be simplified further. If the perpendicular electric field is small compared to the other terms, the purely relativistic drift vanishes and the reference frame moving with $\mathbf{u_E}$ becomes irrelevant. This is the set of equations used in recent work by~\cite{Rosdahl}, \cite{Gordovskyy} and \cite{Pinto} to analyse test particle acceleration in the solar corona. We will quantify the importance and relevance of all separate drift terms and the parallel electric field in Section~\ref{sec:particles}.

The GCA calculations have been performed using a fourth order Runge-Kutta scheme with adaptive time stepping. The typical timestep for the evolution of the guiding centre variables (the particle position in three-space, its relativistic momentum and magnetic moment) is much smaller than the MHD timestep in the cases we explore. Therefore, the MHD fields are considered static on particle timescales, and it is possible to evolve particles dynamically in an MHD snapshot. The MHD data is dimensionless and is scaled to CGS units before being used in the test particle calculations.

The particle timestep $\delta t$ is determined based on its parallel acceleration $a = d v_{\|}/dt$ and velocity $v = \sqrt{(v_{\|})^2 + (v_{\perp})^2}$ as the minimum of $\delta r / v$ and $v / a$, where $\delta r$ is the grid step restricted by the particles CFL condition of $0.8$ making sure a particle cannot cross more than one cell of the MHD grid. Then an Euler integration step is taken with this timestep, to predict the particles trajectory. Based on the results of this Euler integration a definitive timestep for the Runge-Kutta integration is chosen as the minimum of the Euler timestep used and the new timesteps determined by the new acceleration, velocity and position as $(v / a)_{Euler}$ and $(\delta r / v)_{Euler}$.

The electric and magnetic fields and their spatial derivatives, for the right-hand-side of equations~(\ref{eq:gcastatic1})-(\ref{eq:gcastatic3}), are taken from the resistive MHD simulations by linear interpolation for each particle position, within four-dimensional cells $(x,y,z,t)$ from the adjacent grid points. The interpolations in space and time are performed between the CFL limited MHD steps. However, in this work we are advancing particles in static MHD snapshots without advancing the MHD fields. The separate drift terms in the right-hand-side of equation~(\ref{eq:gcastatic1}) are also interpolated and their absolute values are computed at every timestep to analyse which mechanism accelerates individual particles. The gyroradius $R_L = \gamma m_0 v_{\perp}/Bq$ is also calculated at every timestep and compared to the typical cell size to validate the usage of the guiding centre approximation. The velocity $v_{\perp}$ corresponds to all terms of equation~(\ref{eq:gcastatic1}) perpendicular to the magnetic field. For a Maxwellian velocity distribution with $v = \sqrt{(v_{\|})^2 + (v_{\perp})^2}$, the thermal velocity is $v_{th} = \sqrt{(2 k_B T \rho_0/m_{p} p_0)} \sim 10^7 m/s$ for protons in a fluid of temperature about $T = 10^6 K$ with the proton rest mass $m_p = 1.6726 \cdot 10^{-24} g$ and dimensionless pressure $p_0$ and fluid density $\rho_0$ as defined in Section~\ref{sec:MHDtheory}. In typical astrophysical low-$\beta$ plasmas like the solar corona, for particles with thermal velocity and a typical magnetic field of $10^{-2} T$ the gyroradius is of the order $R_L \sim 10^{-3} m$. Typical cell sizes for the highest effective resolution of $4800^2$ are of the order $10^4 m$. So even in more extreme circumstances, if the particles are accelerated to non-thermal velocities, the gyroradius may increase up to 7 orders until the guiding centre approximation fails. The magnetic moment is initially determined as $\mu_{r}^{*} = m_0 \gamma^{*2} v^{*2}_{\perp}/2B^*$, where the velocity perpendicular to the magnetic field $v^{*}_{\perp}$, in the frame of reference moving at $\mathbf{u_E}$, corresponds to the gyration velocity $v_g$. The particles Lorentz factor is $\gamma = 1/\sqrt{1 - v^{2}/c^{2}}$.

At all four $x$ and $y$ boundaries we employ open boundaries. Particles are allowed to leave the physical domain and are then destroyed. In 3D setups the boundaries in the $z$-direction are periodic, in accordance with the MHD boundary conditions. In 2.5D the third direction is invariant for MHD quantities, meaning that there is no boundary for particles to travel in the $z$-direction in our setups. This can cause the particles to accelerate indefinitely in the parallel, resistive electric fields and in the current channels (up to the speed of light). 

The magnetic moment $\mu_{r}^{*}$ of each particle is conserved~(\ref{eq:gcastatic3}). Therefore we can distinguish between particles accelerated by a parallel electric field and particles accelerated by the drift terms in equation~(\ref{eq:gcastatic1}). A particle accelerated by a parallel electric field $E_{\|} = \eta J_{\|}$, has a relatively small pitch angle $\alpha = \arctan(v^*_{\perp}/v_{\|})$. Whereas a particle accelerated by the drift terms has a relatively large pitch angle. The perpendicular velocity $v_{\perp}^*$ in the frame of reference moving at velocity $\mathbf{u_E}$ is determined by equation~(\ref{eq:gcastatic3}). 
\section{Results in 2.5D configurations}
\label{sec:2Dresults}
The ideal MHD equilibrium, either in a force-free or a non-force-free description is subject to an instability in which the current channels repel one another. In 2.5D, where the dynamics in the z-direction are invariant, this linear instability consists of an antiparallel displacement of each of the current channels along the y-direction (the two channels are initially located left and right from $x=0$). After this displacement the channels undergo a rotation and a twisting motion. Based on an energy principle one can show that analytically, the combination of displacement and rotation drives instability (\citealt{Richard}). The growth rate of this instability is typically quantified by a linear growth phase in the bulk kinetic energy (\citealt{Keppens}). During the linear and the subsequent nonlinear phase of the instability, magnetic field lines can reconnect with the background magnetic field. Reconnection of magnetic field lines causes nearly-singular current sheets and secondary islands to form in which particles can get accelerated. 
\subsection{General features and convergence}
To analyse the growth rate of the tilt instability, we quantify the bulk kinetic energy in the displacing current channels. We use tracers to identify the current channels after they have started their displacement and to quantify energy measures. The growth rate indicates how fast reconnection occurs in 2.5D configurations and it is used to decide at what time and in which setup the test particles are evolved.

\begin{figure}
	\includegraphics[width=\columnwidth]{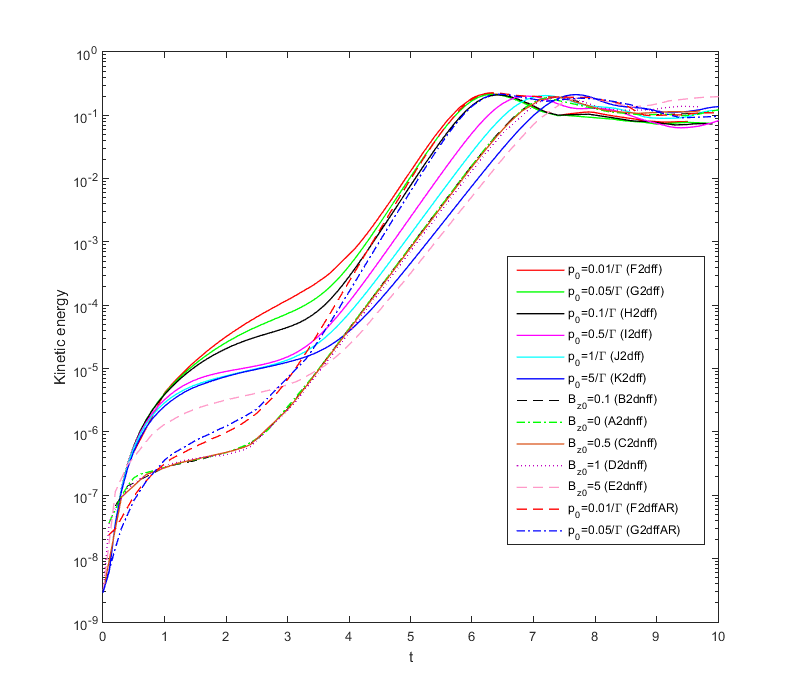}
    \caption{Kinetic energy density evolution for all 2.5D runs with effective resolution $2400^2$ (except F2dffAR2), at all times integrated over a single current channel as identified by an advected tracer, distinguished by line style and colour.}
    \label{fig:kineticenergy}
\end{figure}

\begin{figure}
	\includegraphics[width=\columnwidth]{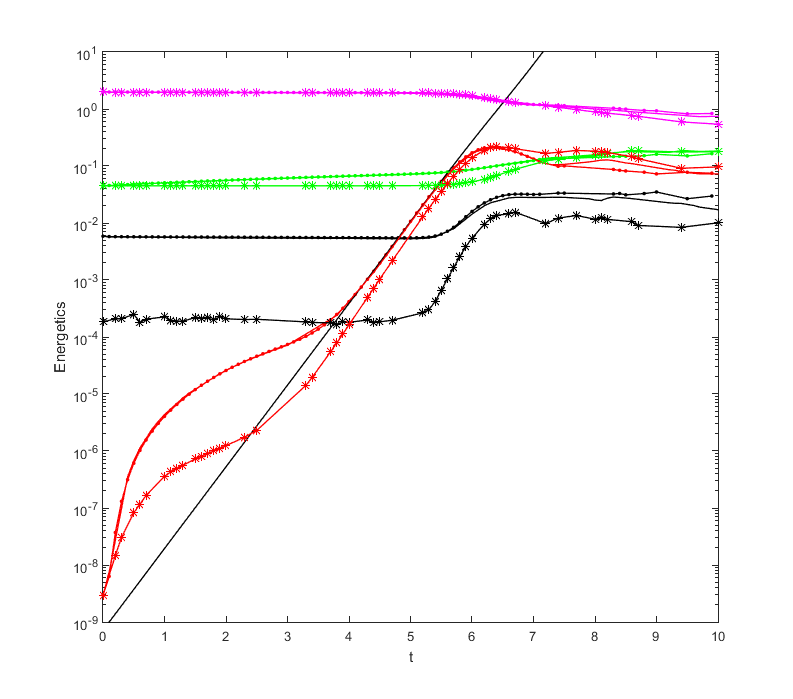}
    \caption{The evolution of kinetic (red), magnetic (magenta) and internal energy (green) density, at all times integrated over a single current channel as identified by an advected tracer, for 2.5D force-free case with $p_0 = 0.05/\Gamma$ (G2dff, GG2dff and G2dffAR). Each curve is shown at two resolutions ($2400^2$ with a solid line and $4800^2$ with dots as indicators) and for a run with anomalous resistivity applied ($2400^2$, with asterisks as indicators). The  Ohmic heating effect is quantified by the black curve. The straight thin black line is a fit to the linear growth in kinetic energy.}
    \label{fig:energeticsp005}
\end{figure}

Fig.~\ref{fig:kineticenergy} shows the evolution of a mean value as in~(\ref{eq:mean}) for a kinetic energy density $f=0.5\rho u^2$ for all equilibrium configurations. We compare the growth rates of the tilt instability in the different setups, quantified by a linear fitting routine. Half the slope of the linear fit of the kinetic energy in logscale sets the growth rate $\gamma_{tilt}$ (mentioned in Table~\ref{tab:example_table} for all runs with effective resolution $2400 \times 2400$). For force-free cases the growth rate increases with decreasing plasma-$\beta$, from cases F2dff to K2dff, namely, $\gamma^{F}_{tilt} = 1.6578$, $\gamma^{G}_{tilt} = 1.6510$, $\gamma^{H}_{tilt} = 1.6492$, $\gamma^{I}_{tilt} = 1.5793$, $\gamma^{J}_{tilt} = 1.4968$ and $\gamma^{K}_{tilt} = 1.4018$. This is in accordance with the (low-resolution) results of~\cite{Richard}. Whereas for the non-force-free cases we find the opposite trend, decreasing growth rate with decreasing plasma-$\beta$, with all values agreeing with the ones reported in~\cite{Keppens}. We have verified that this growth rate estimate does not change when using the other current channel or when using the entire domain to calculate the mean kinetic energy. Compared to the non-force-free setup from~\cite{Keppens}, the perfect linear growth phase in kinetic energy is generally shorter, spanning the time interval $t \in [4.5, 5.5]$ for the force-free configurations, compared to $t \in [3,7]$ for the non-force-free configurations. It also starts later, ends earlier and yields a faster evolution, confirming the early findings of~\cite{Richard} for a force-free equilibrium with low plasma-$\beta$. The force-free setup gives us the opportunity to reach a plasma-$\beta$ which is three times smaller than in the non-force-free cases explored by~\cite{Keppens}. 
Fig.~\ref{fig:energeticsp005} shows the evolution of (\ref{eq:mean}) for a kinetic energy density $f=0.5\rho u^2$, magnetic energy density $0.5 B^2$, internal energy density $p/(\Gamma-1)$ and an Ohmic heating term $\eta J^2$ (which is really an energy density rate change). Two curves are shown for each quantity, one with effective resolution $4800\times4800$ and one with effective resolution $2400\times2400$, for the force-free case with initial pressure $p_0 = 0.05/\Gamma$, to demonstrate convergence. A third curve is shown for each quantity, for the same case with effective resolution $2400\times2400$, with an anomalous resistivity model applied. Judging on these global energetic indicators, the results are nicely converged, with minor differences only in the far nonlinear regime of the evolution for the cases with uniform resistivity and only differing in resolution. The case shown has magnetic energy dominating internal energy (due to $\beta < 1$ conditions). The magnetic energy also clearly dominates the kinetic energy. The internal energy grows in the interval $t \in [0,6]$, which can be explained by the low $\beta$ conditions. Due to low pressure, relative to the magnetic pressure, the work done by the strong magnetic field compresses the plasma. The rise in internal energy is more significant (and visible on a log scale) in the low $\beta$ regime due to the low initial pressure, while the magnetic field configuration is the same in all force-free cases. The Ohmic heating also shows a transition around $t \approx 5$, lagging the growth phase of the kinetic energy.
\begin{figure}
	\includegraphics[width=\columnwidth]{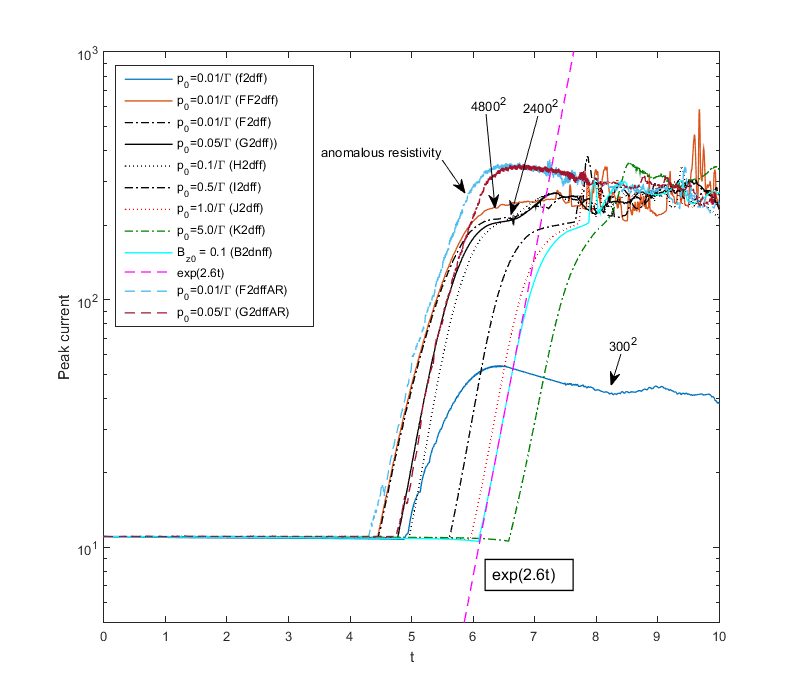}
    \caption{Peak current evolution for all force-free 2.5D runs, distinguished by line style and colour. Non-force-free run B2dnff ($B_{z0}=0.1$) is shown as a comparison. Different resolutions are shown for force-free case with $p_0=0.01/\Gamma$. Runs with anomalous resistivity are shown for force-free cases with $p_0=0.01/\Gamma$ and $p_0=0.05/\Gamma$. Exponential growth at $\exp(2.6t)$ is indicated to guide the eye.}
    \label{fig:peakcurrent}
\end{figure}
To explore how well the MHD evolution converges, we analyse the evolution in time of the maximum current $\log(max(J_z))$ that is reached in all 2.5D simulations, for different resolutions. This peak current density also indicates the start of the linear growth phase of the tilt instability. Based on these results we can determine which configuration is the best candidate for fast reconnection and hence in which we will find the most efficient particle acceleration. In all cases (force-free and non-force-free) a linear growth phase of the logarithm of the peak current density is established, in accordance with results for non-force-free configurations in~\cite{Keppens}. The linear growth for $\log(max(J_z))$ is accurately resolved with our block-AMR strategy in combination with our third-order spatio-temporal treatment, showing up to secondary tearing-type instabilities. In Fig.~\ref{fig:peakcurrent}, the evolution of the peak current $\log(max(J_z))$ is shown in a log-linear scale for all force-free 2.5D cases, the non-force-free case with $B_{z0}=0.1$ is shown for comparison. For the force-free case with $p_0=0.01/\Gamma$ we show all three resolutions (f2dff, F2dff, FF2dff) to show a local measure of convergence. In the island-dominated phase, beyond $t\approx 6$ for the non-force-free case with the fastest kinetic energy density evolution (F2dff), there are still noticeable differences for the highest resolutions. The low resolution run (f2dff) reaches much lower peak current values than the two high resolution runs. Comparing runs at identical resolutions of $2400^2$, we can detect a trend of the current evolution for different $\beta$. For the force-free equilibrium there is a systematic delay in the onset of the singular current development when raising $\beta$ (which corresponds to raising the initial pressure $p_0$ in the force-free cases). This is opposite of what~\cite{Keppens} found for non-force-free equilibriums (delay when lowering $\beta$), but agrees with the findings of~\cite{Richard} for their low-resolution force-free runs. For comparison, $\log(max(J_z))$ for the non-force-free cases with fastest evolution is plotted as well in Fig.~\ref{fig:peakcurrent}. All cases, both force-free and non-force-free, roughly have the same slope for the onset of this singularity, indicated by $\log(max(J_z)) \sim 2.6t$ (plotted by the dashed, magenta line). The systematic delay observed is consistent with the tilt mode growth rates, which are quantified by the linear growth phase of the kinetic energy in Fig.~\ref{fig:kineticenergy}.
\subsection{Effects of anomalous resistivity}
Using an anomalous resistivity model, rather than uniform resistivity, has an effect on the energetics. In Fig.~\ref{fig:kineticenergy} we also show the evolution of the kinetic energy for cases F2dffAR and G2dffAR with anomalous resistivity rather than uniform resistivity. Both cases have the exact same settings as F2dff and G2dff, except for the resistivity only being nonzero in regions with current density $j > j_c=12$ with $j_c$ chosen to be larger than the peak current at equilibrium. The onset of the tilt instability in these cases starts at earlier time compared to F2dff and G2dff and lasts for a longer period. The slope of the linear growth phase of the tilt instability is steeper than for their respective uniform resistivity equivalents, $\gamma^{FAR}_{tilt} = 1.8206$ and $\gamma^{GAR}_{tilt} = 1.8012$, and for a case with the anomalous resistivity threshold $j_c = 500 > max(J_z (t=0))$ and hence, just numerical resistivity, $\gamma^{FAR2}_{tilt} = 1.6404$. The peak value of the kinetic energy that is reached is the same as in the cases with uniform resistivity, such that the tilt instability has a longer time to grow, with a larger growth rate, to develop strong current sheets.

For a critical current density $j_c = 12$, larger than the largest value of the current in equilibrium, the effects on kinetic energy and Ohmic heating are clearly visible in Fig.~\ref{fig:energeticsp005} for case G2dffAR. The onset of the linear growth of the kinetic energy (red line with asterisks as a marker) starts earlier and lasts longer. This means that the instability has a longer time to grow and that larger current density can be reached during the linear growth. The integrated ohmic heating starts at a lower value, because the lack of resistivity in regions with a current density $j < j_c$. Because the linear growth phase lasts longer, the ohmic heating can build up for a longer time and it reaches almost the same value in the nonlinear phase as the case with uniform resistivity. The magnetic energy and the internal energy show less to no effect due to anomalous resistivity.

Given the ideal character of the tilt instability, there is hardly any effect of (anomalous) resistivity on the peak current during the phase before the linear tilt growth. The evolution of the peak current (Fig.~\ref{fig:peakcurrent}) for cases with anomalous resistivity is similar to the cases with uniform resistivity. However, during the linear growth phase, the growing gradient of the magnetic field causes the current to grow in specific areas. In these areas resistivity is switched on, but in the surroundings, resistivity is still zero. This means that the current cannot diffuse into the surroundings and it will build up for a longer time. In the setups with anomalous resistivity therefore, a higher peak current is reached and the linear growth phase lasts longer. Due to this effect, there are also visual differences in the total current density. In the models with anomalous resistivity, secondary islands and narrow current sheets grow earlier in the simulation and do not diffuse into the surroundings. As long as the threshold for anomalous resistivity to be nonzero is larger than the peak current value at equilibrium ($j_c > 11$), there are no differences in the peak currents reached during the nonlinear phase, nor are there any visual differences in the topology of the current distribution. However, there is no parallel electric field component $E_{\|} = \eta \mathbf{J} \cdot \mathbf{\hat{b}}$ anymore in the regions with a low current density. This has an expected effect on particle acceleration in the ambient and inside the current channels, but not at reconnection sites with a large peak current density. The behaviour for different plasma-$\beta$ (at equilibrium) does not change. The same peak current is reached for all force-free cases with anomalous resistivity and in the nonlinear phase there are minor differences just as for the cases for uniform resistivity.

\begin{figure*}
  \centering
    \subfloat[F2dff ($\bar{\beta}(t=0) = 0.04$), $t=6$]{\includegraphics[width=0.667\columnwidth, trim= 10cm 2cm 10cm 2cm, clip=true]{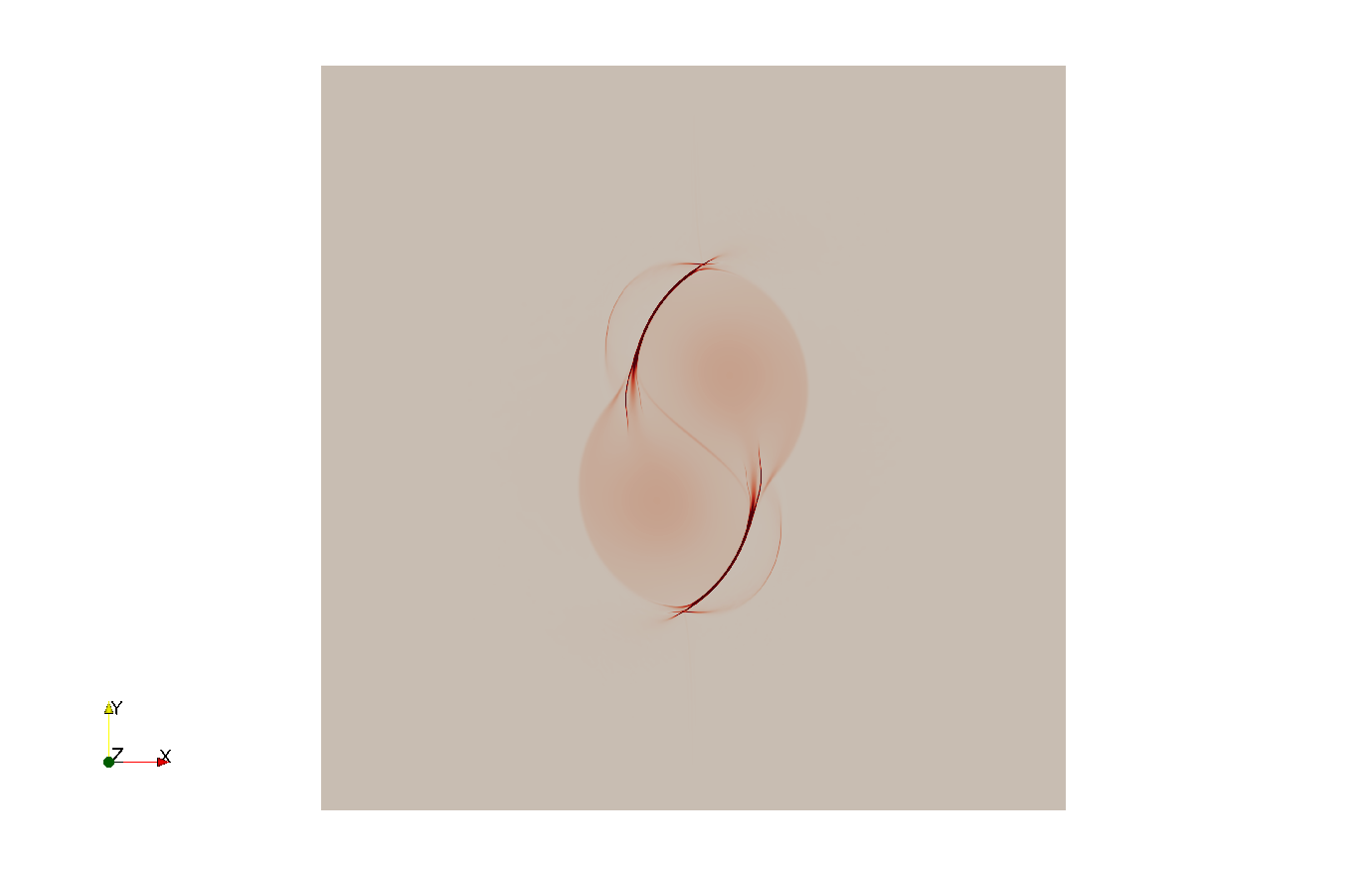}}
		\subfloat[F2dffAR ($\bar{\beta}(t=0) = 0.04$), $t=6$]{\includegraphics[width=0.667\columnwidth, trim= 10.5cm 2.4cm 11cm 0.3cm, clip=true]{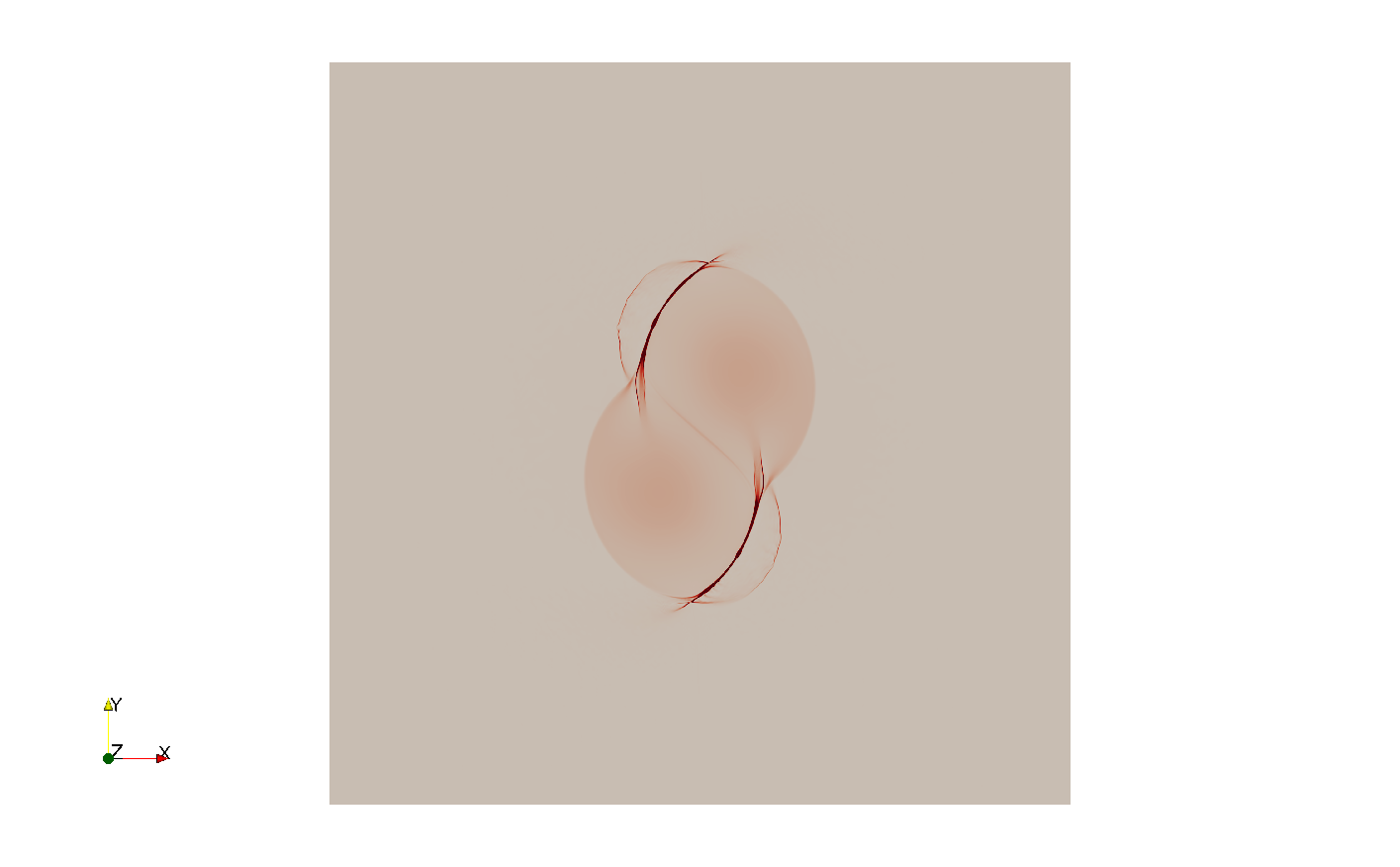}}\subfloat[B2dnff ($\bar{\beta}(t=0) = 5.8$), $t=6$]{\includegraphics[width=0.667\columnwidth, trim= 10cm 2cm 10cm 2cm, clip=true]{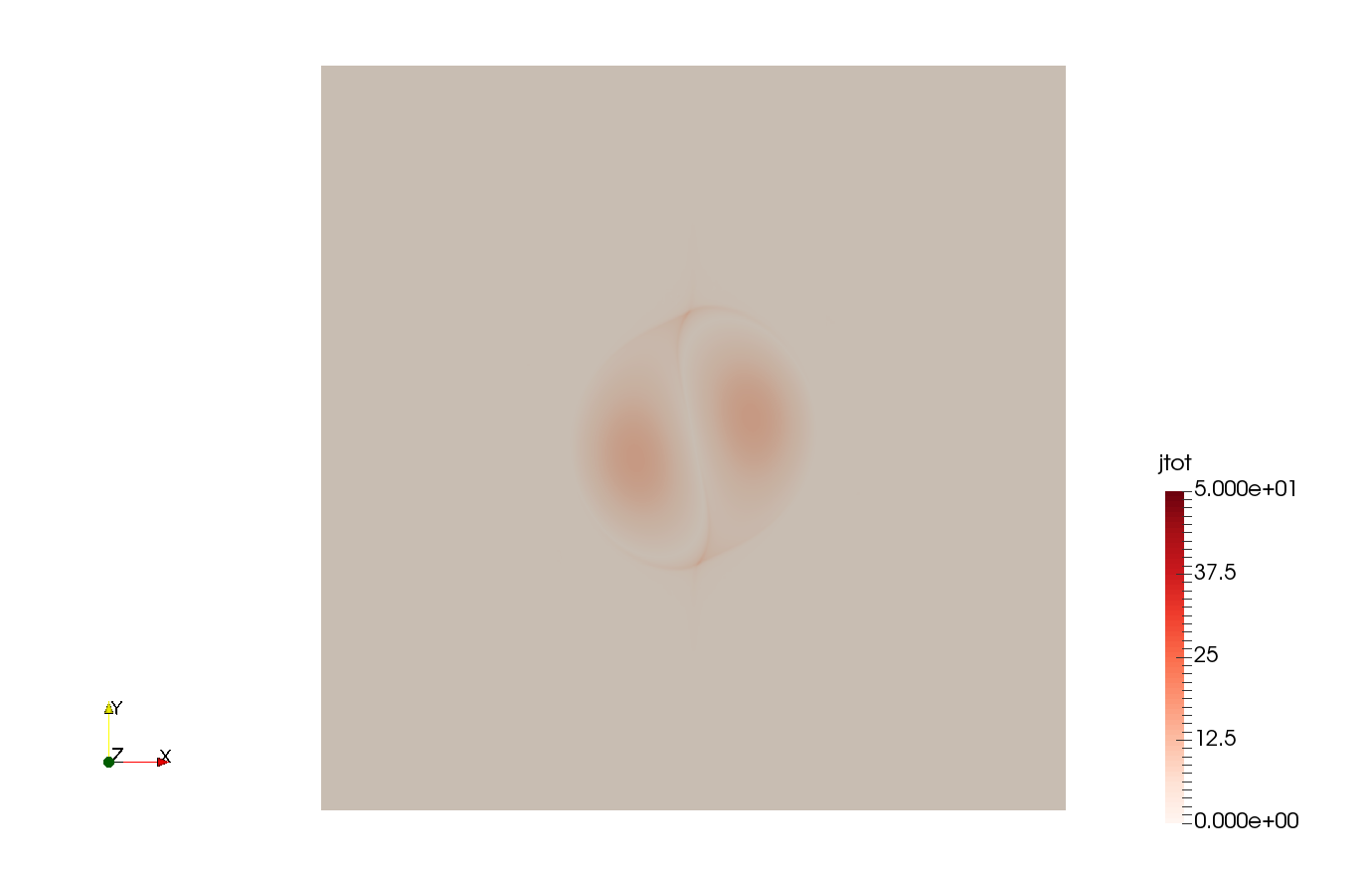}}
		
  \subfloat[F2dff ($\bar{\beta}(t=0) = 0.04$), $t=8$]{\includegraphics[width=0.667\columnwidth, trim= 10cm 2cm 10cm 2cm, clip=true]{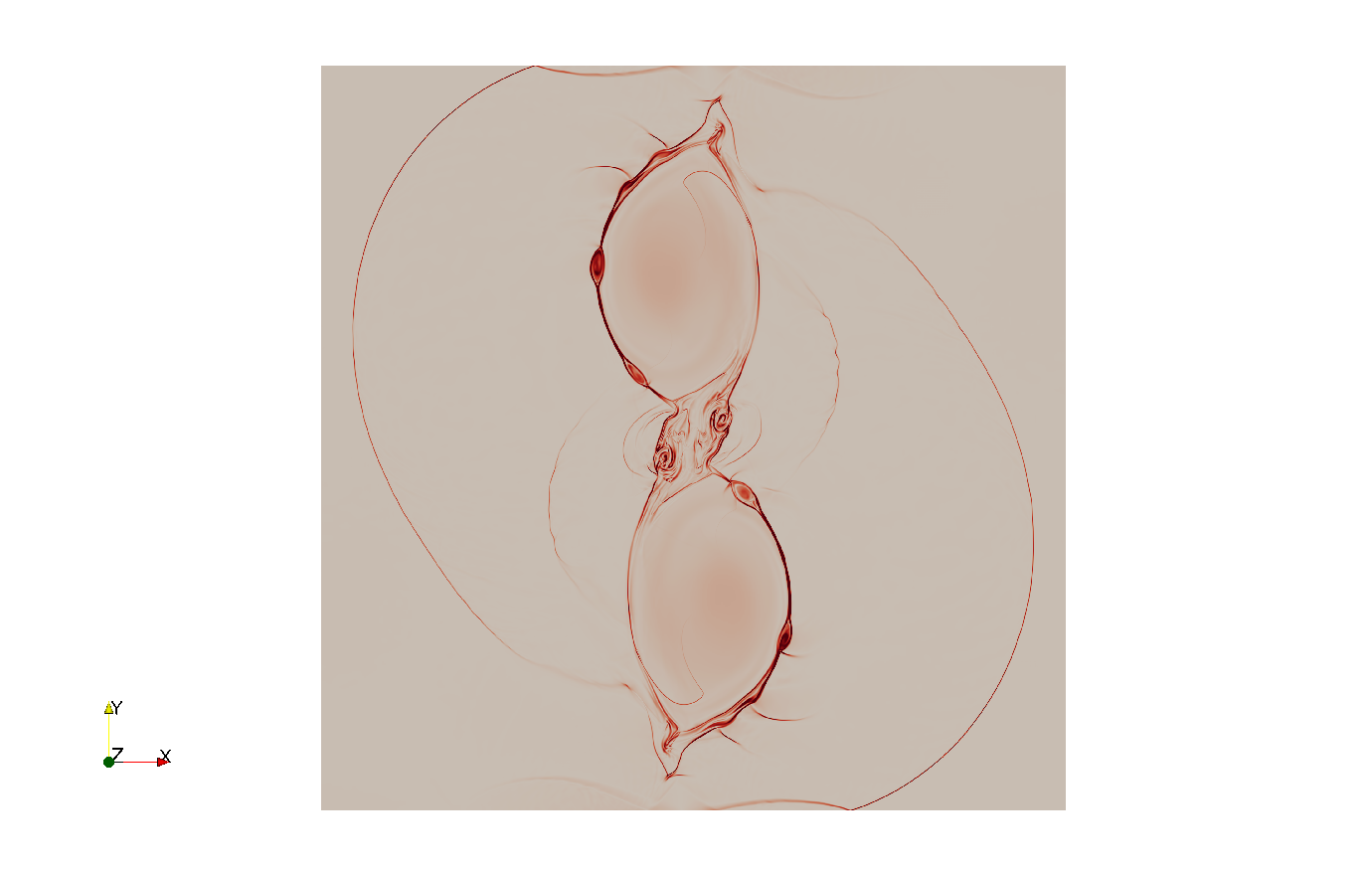}}\subfloat[F2dffAR ($\bar{\beta}(t=0) = 0.04)$, $t=8$]{\includegraphics[width=0.667\columnwidth, trim= 10.5cm 2.4cm 11cm 0.3cm, clip=true]{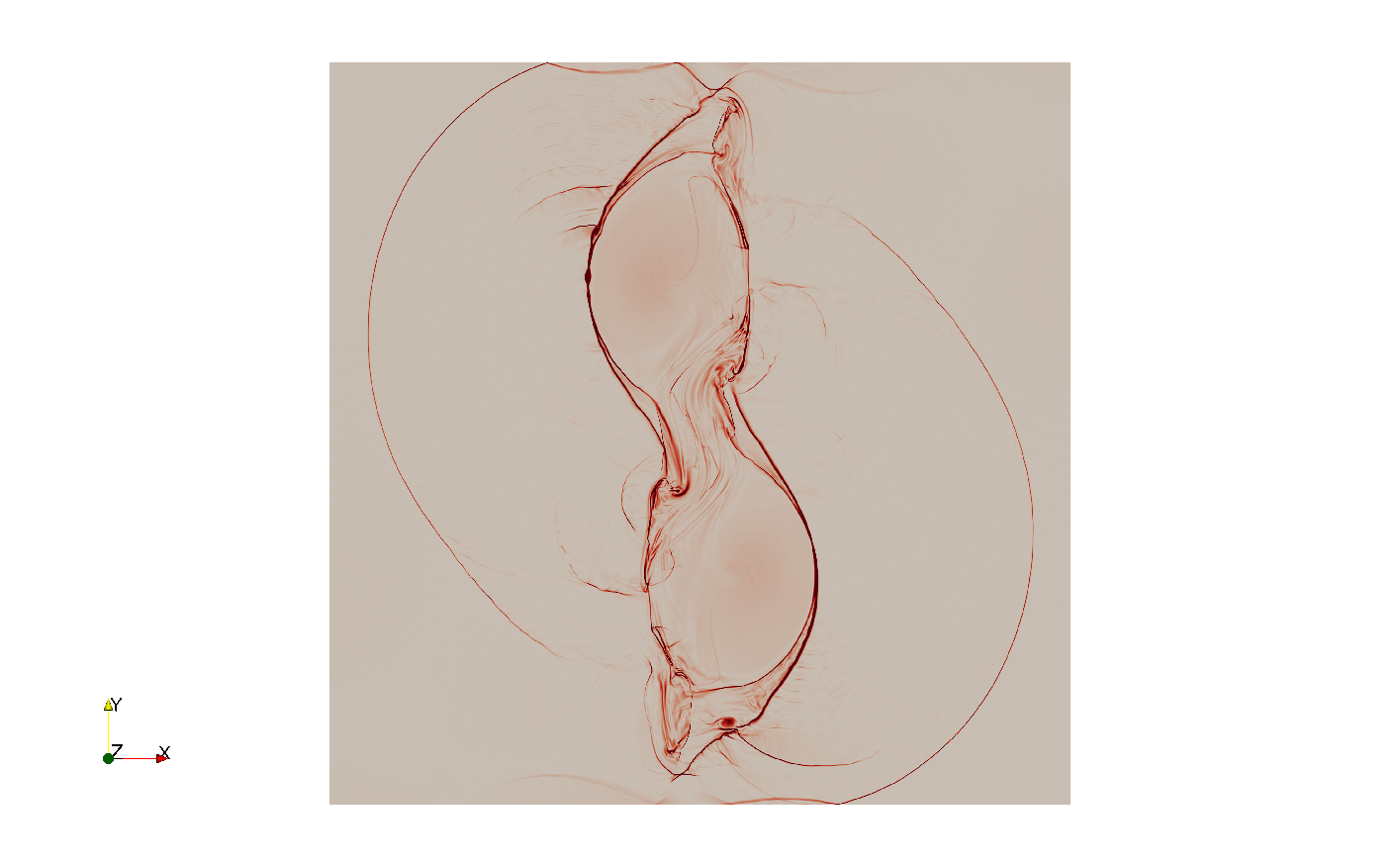}}\subfloat[B2dnff ($\bar{\beta}(t=0) = 5.8$), $t=8$]{\includegraphics[width=0.667\columnwidth, trim= 10cm 2cm 10cm 2cm, clip=true]{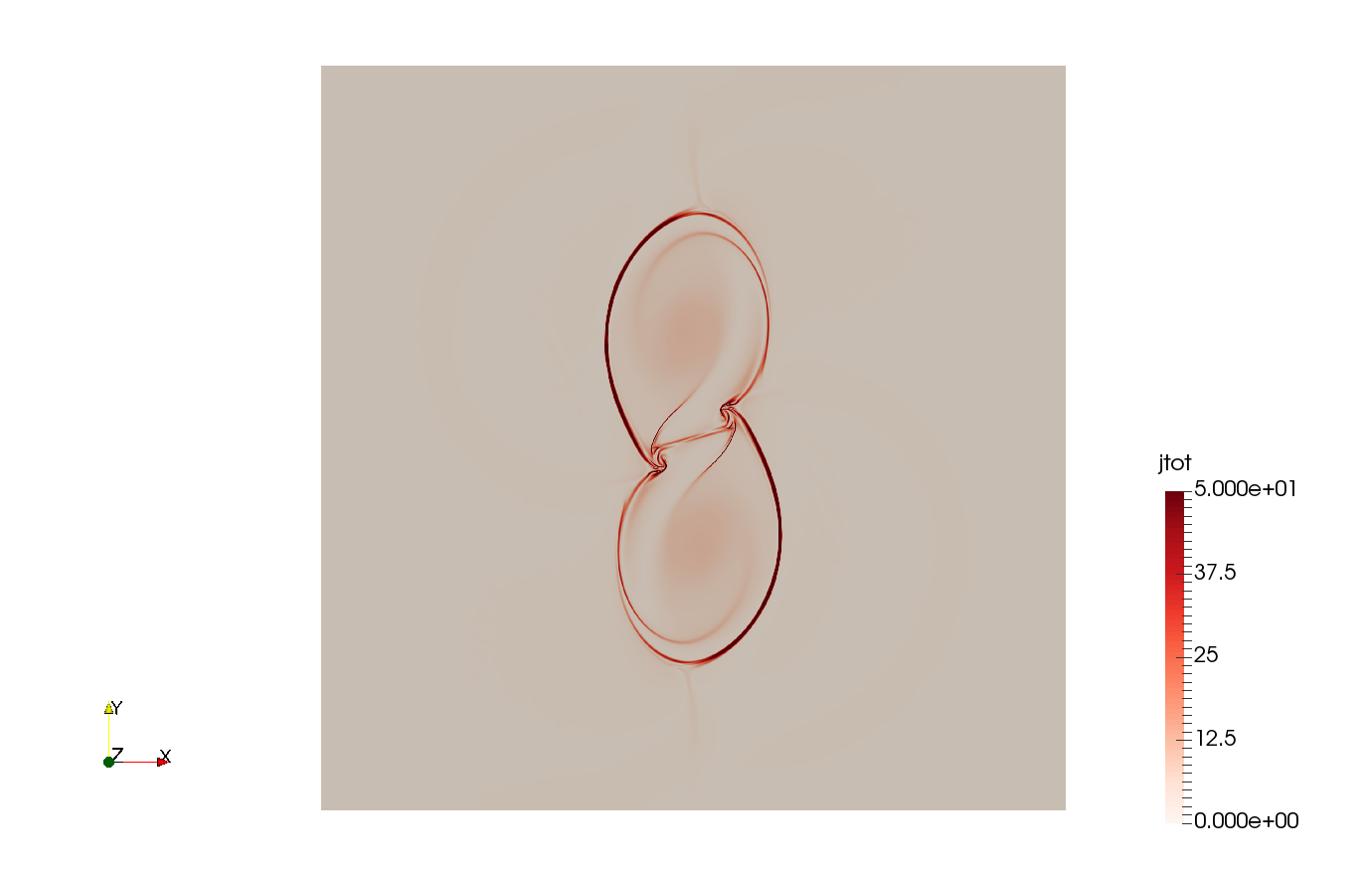}}
\caption{Evolution of the total current magnitude $\lvert\mathbf{J}\rvert$ for the three 2.5D cases with fastest reconnection, force-free cases F2dff with uniform resistivity (left), F2dffAR with anomalous resistivity (middle) and non-force-free case B2dnff with uniform resistivity (right). The rows correspond to times $t=6$ (top) and $t=8$ (bottom), at a time before the instability starts and after the fully nonlinear regime has been reached. A linear colour scale is saturated to show values between $[0-50]$, showing all of the structure.
}
\label{fig:jtot}
\end{figure*}

Fig.~\ref{fig:jtot} shows the total current magnitude $J$ for the two fastest force-free cases (F2dff with uniform resistivity and F2dffAR with anomalous resistivity applied) next to each other at times $t=6$ (top) and $t=8$ (bottom) and again, the 2.5D non-force-free case with the fastest peak current evolution (B2dnff) as comparison. A linear colour scale (with values between 0-50) is used for all frames, such that all structure is visible. Near-singular current sheets are formed at the boundaries of the rotating islands where antiparallel field lines can reconnect, causing heating of the plasma. The non-force-free case (B2dnff reaches the disruption of the current sheets by tearing-type chaotic reconnection beyond $t=6$, whereas the force-free cases shown, with a plasma $\beta < 1$, show the start of this turbulent stage already at $t=6$. The chaotic formation of secondary islands on the disrupting current sheets can also be seen in Fig.~\ref{fig:jtot} at $t=8$ for cases F2dff and F2dffAR. Cases F2dff and F2dffAR only differ in the resistivity model applied. Case F2dffAR shows a similar evolution of the current up to $t=6$, but hereafter the current diffuses in a different manner due to the anomalous resistivity model.
\subsection{Reconnection}
\begin{figure}
\centering
\includegraphics[width=0.75\columnwidth, trim= 10cm 2cm 10cm 2cm, clip=true]{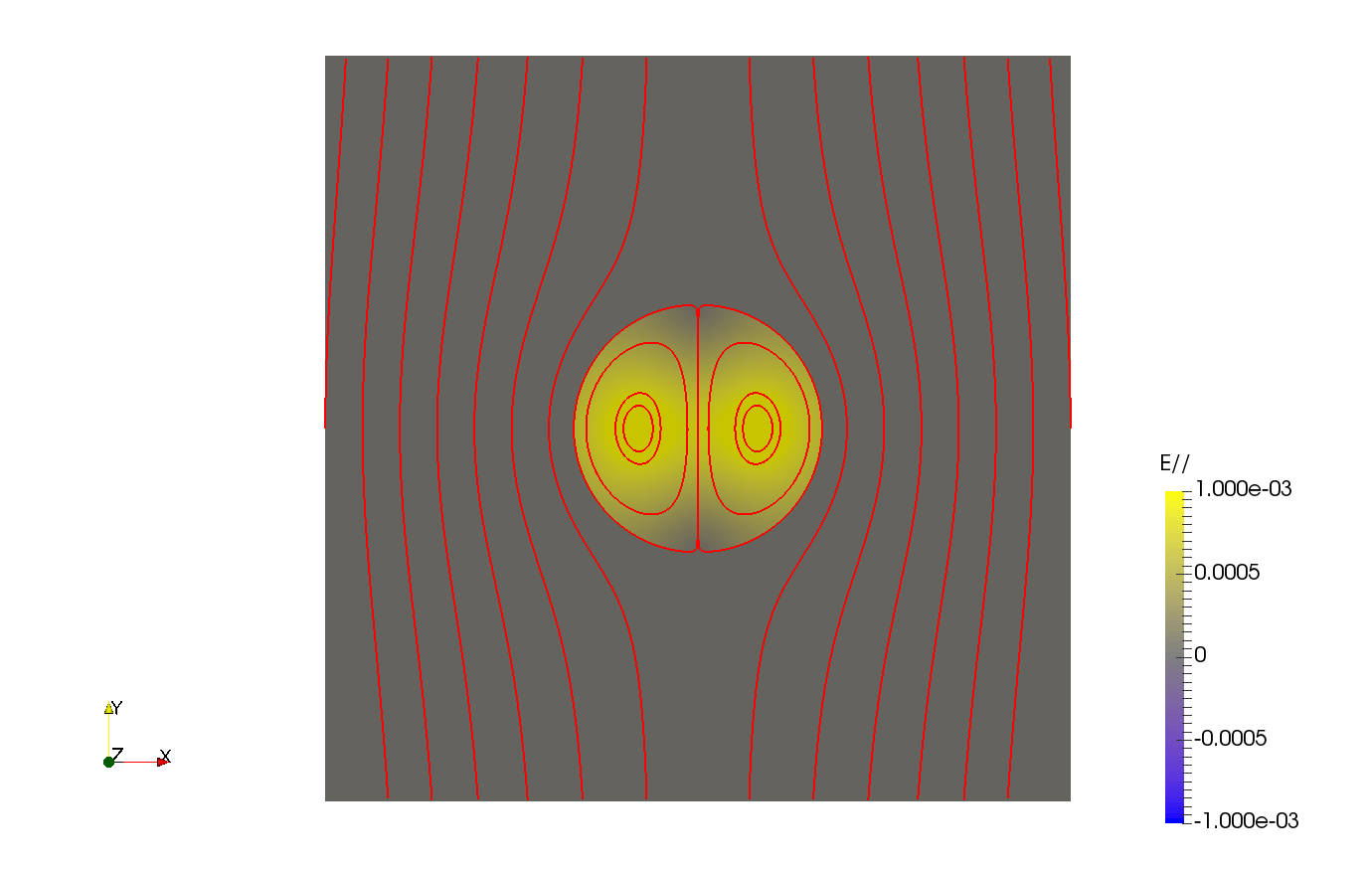}
\caption{For all force-free cases, we show the parallel electric field $E_{||} = \mathbf{E}\cdot\mathbf{B}/B$ at equilibrium ($t=0$) and selected field lines (in red) indicating the magnetic field structure. A linear colour is saturated to show values between $[-0.001, 0.001]$.}
\label{fig:Epart0}
\end{figure}

\begin{figure*}
\centering
\subfloat{\includegraphics[width=0.8\columnwidth, trim= 10cm 2cm 10cm 2cm, clip=true]{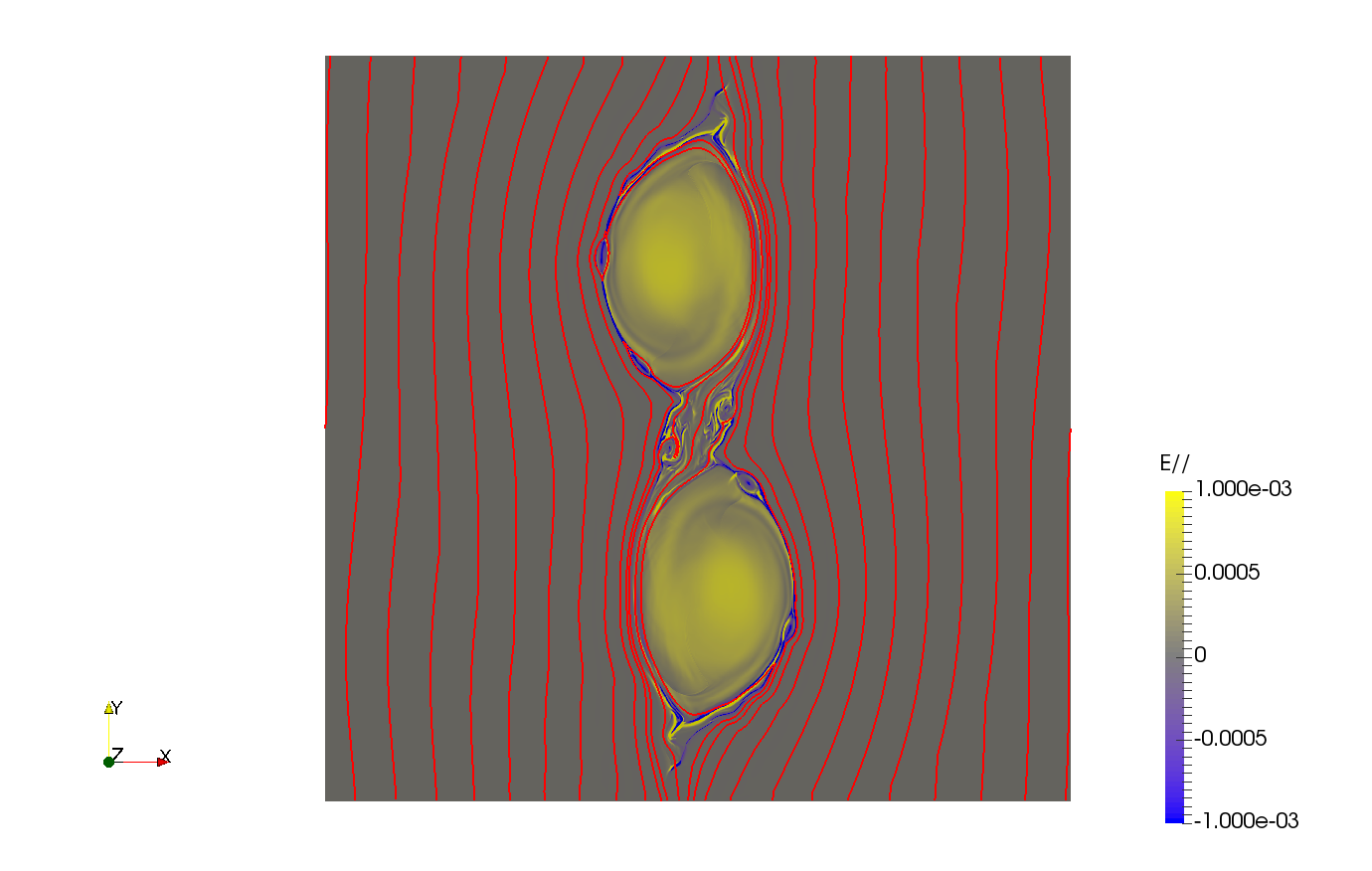}}
\subfloat{\includegraphics[width=0.84\columnwidth, trim= 10cm 1.85cm 10cm 2.3cm, clip=true]{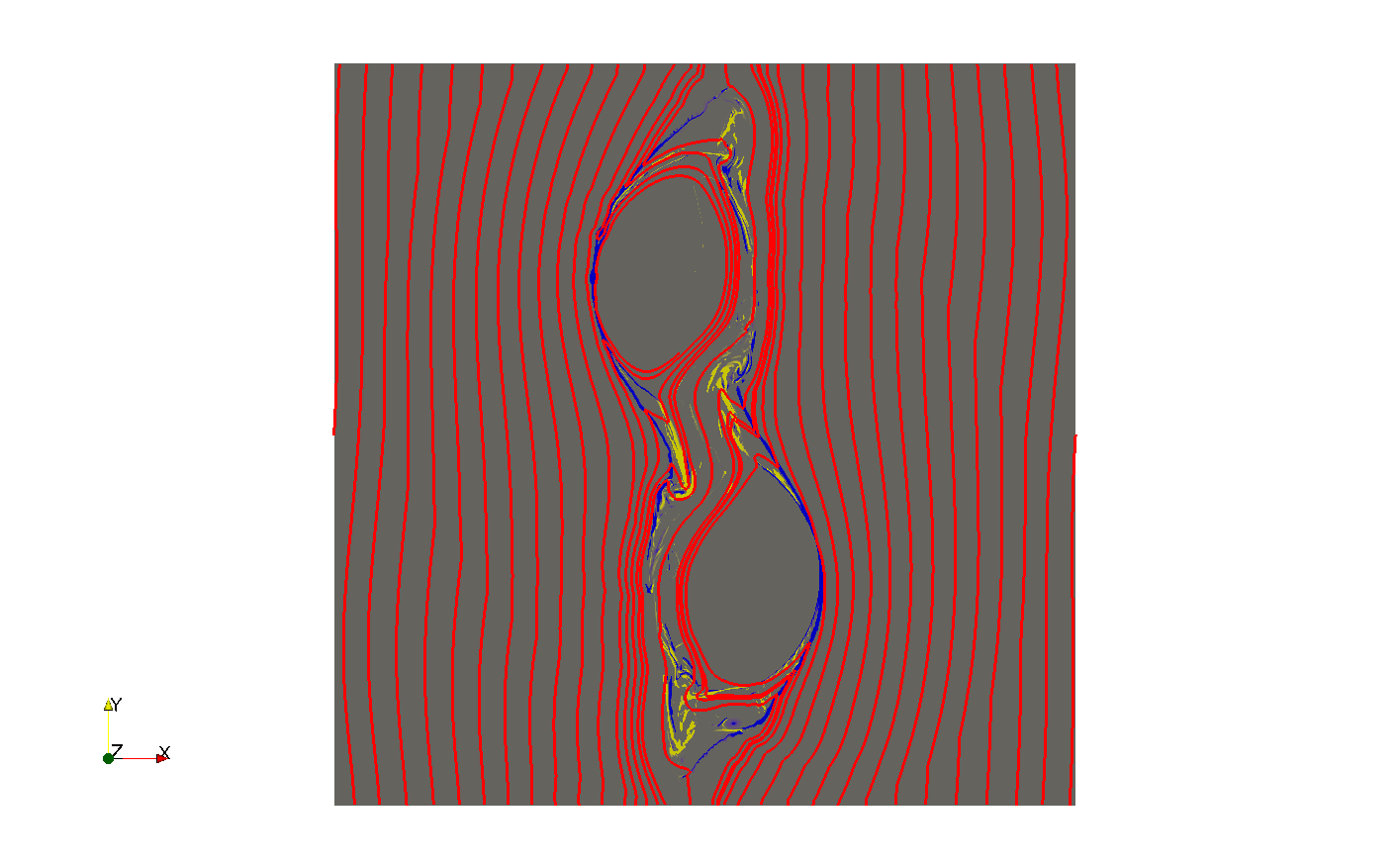}}
\caption{For force-free cases F2dff (left panel) and F2dffAR (right panel) (with $\bar{\beta}(t=0) = 0.04$ and $p_0=0.01/\Gamma$), we show the parallel electric field $E_{||} = \mathbf{E}\cdot\mathbf{B}/B$ at $t=8$ and selected field lines (in red) indicating the magnetic field structure. A linear colour is saturated to show values between $[-0.001, 0.001]$.}
\label{fig:Epart8}
\end{figure*}

\begin{figure*}
\centering
\subfloat{\includegraphics[width=0.8\columnwidth, trim= 10cm 2cm 10cm 2cm, clip=true]{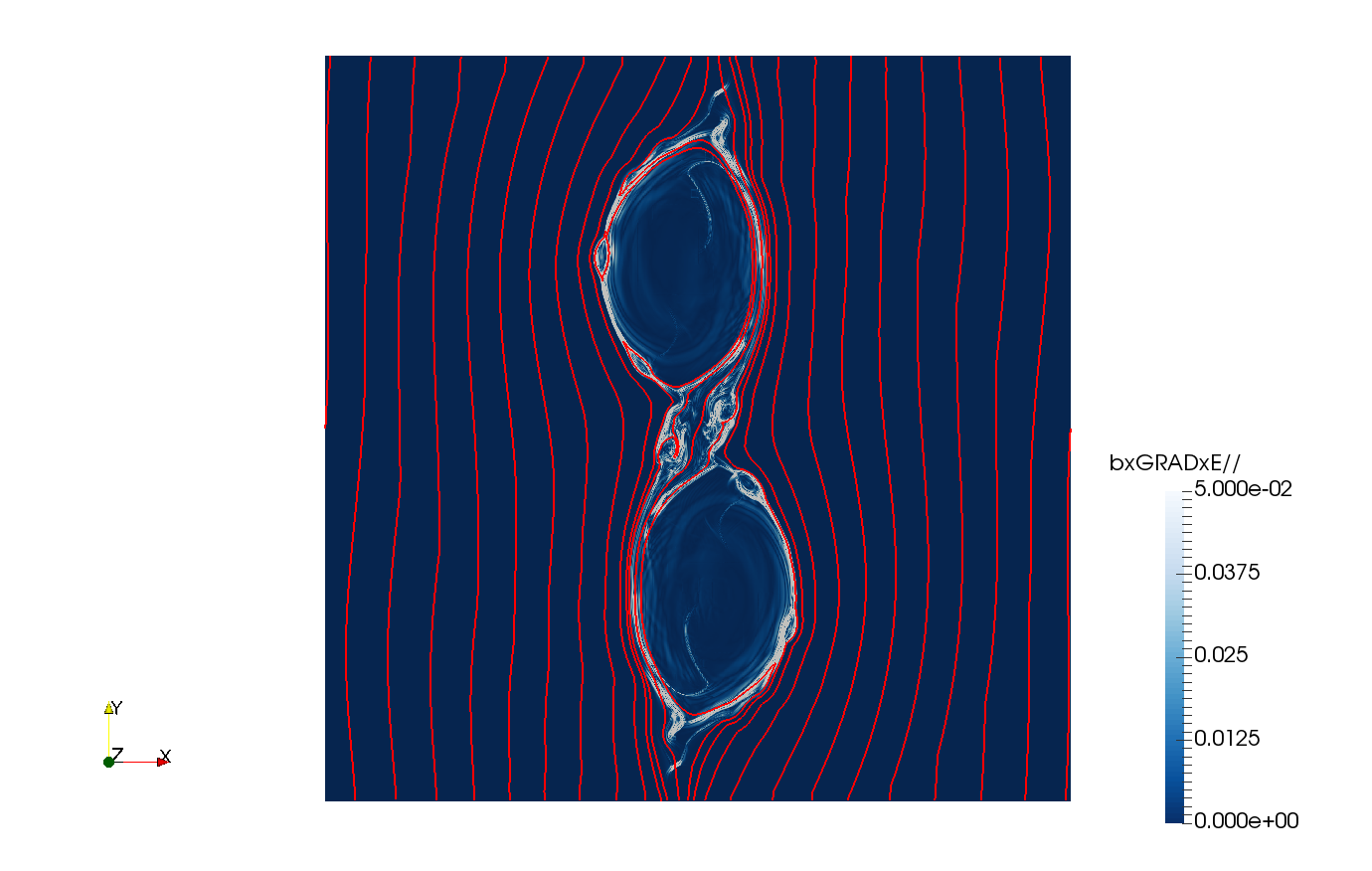}}
\subfloat{\includegraphics[width=0.84\columnwidth, trim= 10cm 1.85cm 10cm 2.3cm, clip=true]{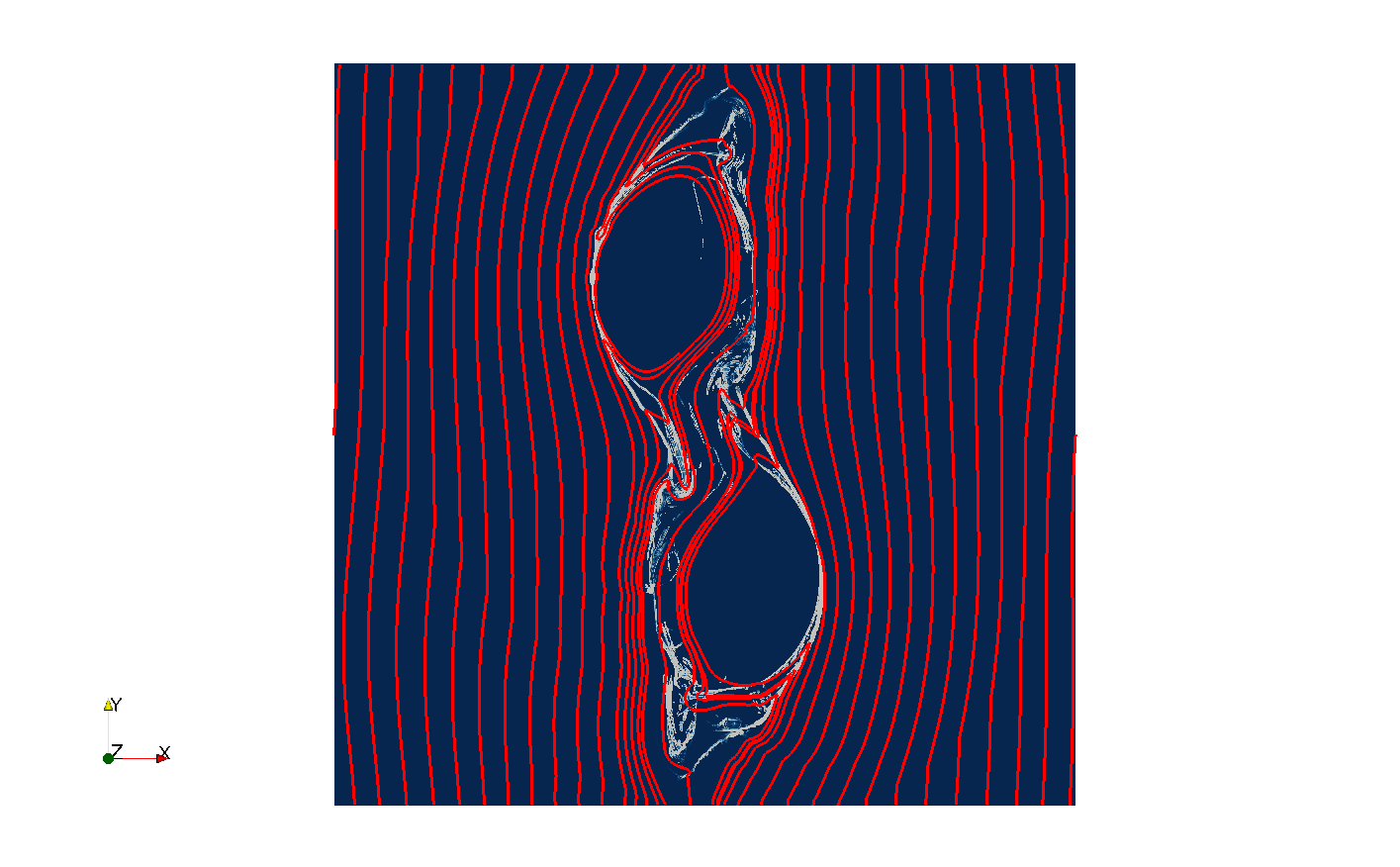}}
\caption{For force-free cases F2dff (left panel) and F2dffAR (right panel) (with $\bar{\beta}(t=0) = 0.04$ and $p_0=0.01/\Gamma$), we 
show the topological measure of field line breakage, normalised by the magnetic field magnitude $B$: $\lvert\mathbf{\hat{b}} \times \left(\nabla \times E_{||}\mathbf{\hat{b}}\right)\rvert$ at $t=8$. Selected magnetic field lines (in red) are shown, indicating the magnetic field structure. A linear colour is saturated to show values between $[0, 0.05]$.}
\label{fig:rect8}
\end{figure*}

Besides the different equilibrium setups, the dynamics are affected by differences in initial conditions (pressure for force-free setups and perpendicular magnetic field component for non-force-free setups), hence different plasma-$\beta$. This leads to differences in where secondary islands appear and how they deform. And hence, also where reconnection occurs exactly. To analyse where particles will get accelerated most efficiently, we locate reconnection sites based on several indicators. Reconnection regions in this essentially 2D system can be identified by a non-zero electric field, parallel to the magnetic field, $E_{||}$, indicated by a parallel current density, i.e., $E_{||}=\mathbf{E}\cdot \mathbf{B}/B = \eta \mathbf{J}\cdot\mathbf{B}/B \neq 0$~(\citealt{Priest}). A region of non-zero parallel electric field is strongly suggesting reconnection. However, mathematically, a more stringent, formal criterion for reconnection can be written as~(\citealt{Biskamp}; \citealt{Lapenta})
\begin{equation}
\lvert\mathbf{\hat{b}} \times \left(\nabla \times E_{||}\mathbf{\hat{b}}\right)\rvert \neq 0
\label{eq:reconnection}
\end{equation}
normalised by $B = \lvert\mathbf{B}\rvert$, the magnitude of the magnetic field. This is a direct measure of the violation of conservation of frozen-in magnetic field lines, indicating reconnection. This measures the rotation of the magnetic field lines, induced by the parallel electric field. The most direct evidence is obtained by finding field lines that started in the direction from top to bottom in $(y)$, but have reconnected and changed topology.  We show the parallel electric field including selected magnetic field lines at $t=0$ in Fig.~\ref{fig:Epart0} to compare the topology at equilibrium and in the nonlinear phase. Selected magnetic field lines are plotted on top of the reconnection indicators $E_{||} \neq 0$ and equation~(\ref{eq:reconnection}) in the left panel of Fig.~\ref{fig:Epart8} and Fig.~\ref{fig:rect8} respectively, for force-free case F2dff with uniform resistivity and in the right panels of Fig.~\ref{fig:Epart8} and Fig.~\ref{fig:rect8} for F2dffAR with anomalous resistivity. For case F2dff, there is only positive parallel electric field inside the current channels initially and no reconnecting field lines, nor $\lvert\mathbf{\hat{b}} \times \left(\nabla \times E_{||}\mathbf{\hat{b}}\right)\rvert \neq 0$ at $t=0$. For F2dffAR there is no parallel electric field in the equilibrium phase due to absent resistivity.
At $t=8$, negative parallel electric field has developed at the boundaries of the tilted current channels and in between for both cases. This confirms that likely reconnection sites develop in the strong current layers and in between the current channels. Anti-parallel field lines break and reconnect in these regions and topological changes occur, including the formation of secondary islands. The global topological rearrangements establish field lines reconnecting between $x>0$ (right of the initial current channels) and $x<0$ (left of the initial current channels) regions. 
There is one major difference between cases with uniform resistivity and with anomalous resistivity, expected to have a major effect on particle acceleration. For case F2dff, in Fig.~\ref{fig:Epart8}, there is a strong parallel electric field inside the current channels whereas for F2dffAR, in the right panel of Fig.~\ref{fig:Epart8}, there is no parallel electric field inside the current channels. 
The location of strongest violation of field line topological connectivity as indicated by equation~(\ref{eq:reconnection}), in Fig.~\ref{fig:rect8}, coincides with the reconnection regions identified by the non-zero parallel electric field. However, one should be careful judging the occurrence of reconnection based on the presence of parallel electric field, as is clear from the equilibrium F2dff shown in Fig.~\ref{fig:Epart0} and from the nonzero parallel electric field inside the current channels in Fig.~\ref{fig:Epart8}. In the reconnection regions as indicated by equation~(\ref{eq:reconnection}), there is however a strong parallel electric field in both cases F2dff and F2dffAR. These differences due to resistivity are expected to have a large effect on particle acceleration due to parallel electric field in the regions where no reconnection occurs. Similar observations hold for all force-free and non-force-free cases, with differences in the exact topology of the magnetic field and the time the reconnection develops.
\section{Results in 3D configurations}
\label{sec:3Dresults}
In 3D scenarios the $B_z$ component of the magnetic field is expected to have a stronger effect on stability of the current channels. In 2.5D configurations the role of $B_z$ is minimal since the translational invariance prevents potential (de)stabilisation due to field line bending (\citealt{Keppens}). In 3D configurations field lines may bend with respect to the $z$-direction. A second additional effect is the fact that each current channel may be liable to an ideal kink instability. From the Kruskal-Shafranov limit we know that for $K_{cr} \equiv \lvert\bar{J_z}\rvert/\bar{B_z} < 2 a / R_0 = 4 \pi a/L \approx 1$ the equilibrium is stable, where we use a plasma column radius $a \approx 0.5$ and length $L=6$. We quantify the ratio $K_{cr} \equiv \lvert\bar{J_z}\rvert/\bar{B_z}$, from the initial condition for all cases (mentioned in Table~\ref{tab:example_table} for all 3D runs) and we find that all force-free cases are liable to a kink instability since $K_{cr}= 3.83$ (The $B_{z0}$ component is the same for all force-free cases, resulting in the same Kruskal-Shafranov limit). For the non-force-free setup, only the lowest plasma-$\beta$ case, E2dnff, is stable, as was confirmed numerically by~\cite{Keppens}. For non-force-free cases B2dnff to E2dnff, we find, respectively, 43.5, 8.7, 4.35 and 0.87 for the Kruskal-Shafranov limit. Therefore we expect that all force-free 3D cases will show kink unstable behaviour, since they have an insufficient vertical magnetic field $B_{z}$ to stabilise kink deformations. This allows us to explore the low plasma-$\beta$ regime without the limitation of strong vertical magnetic fields stabilising the equilibrium. We will analyse what the result is of the kink instability in a force-free setup on magnetic reconnection.
\subsection{3D effects of the kink instability on energy conversion and reconnection}
\begin{figure}
	\includegraphics[width=\columnwidth]{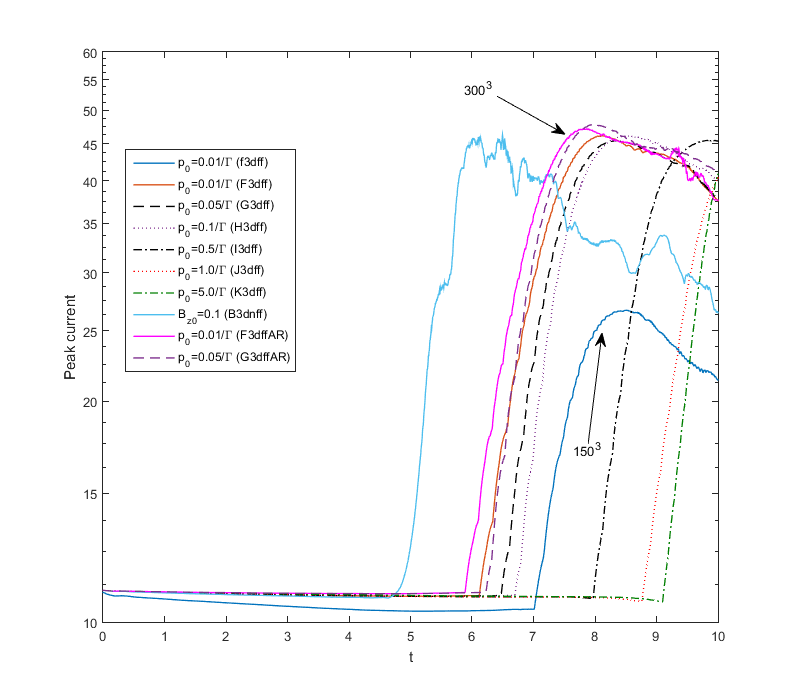}
    \caption{Peak current evolution for all force-free 3D runs, distinguished by line style and colour. Non-force-free run B3dnff ($B_{z0}=0.1$) is shown as a comparison. Different resolutions are shown for the force-free case with $p_0=0.01/\Gamma$ (F3dff).}
    \label{fig:peakcurrent_3D}
\end{figure}
In Fig.~\ref{fig:peakcurrent_3D} we show the evolution of the peak current $\log(max(J_z))$ for all force-free 3D runs. Non-force-free case B3dnff (with $B_{z0} = 0.1$ and the largest, finite, Kruskal-Shafranov limit $K_{cr} = 43.5$, realising the fastest evolution of the peak current for non-force-free setups), is also included for comparison. In all 3D force-free cases, the onset of the instability is delayed (compared to the 2.5D cases in Fig.~\ref{fig:peakcurrent}) by the magnetic tension due to the $B_z$ component. And this effect is stronger, for higher pressure and hence plasma-$\beta$. However, unlike in the non-force-free cases, where the setup with lowest plasma-$\beta$ (E3dnff, with $B_{z0}=5.0$) completely suppresses any instability development~(\citealt{Keppens}), all force-free cases develop a combination of tilt and kink instabilities. If the initial velocity perturbation has no $z$ dependence, the evolution of the 3D cases behaves identically to the 2.5D cases and the tilt instability develops at earlier time. Another important feature to notice in Fig.~\ref{fig:peakcurrent_3D} is the difference for two similar setups with varying resolution. For the force-free cases F3dff and f3dff, both with $p_0 = 0.01/\Gamma$ but with effective resolutions of $300^3$ and $150^3$ respectively, the peak current evolution for lower resolution is delayed. The saturation levels attained in the far nonlinear regime are higher for the higher-resolution runs. Based on the higher-resolution runs in 2.5D this was to be expected and~\cite{Keppens} shows that at even higher resolutions, higher saturation levels are reached. However, visual data inspection and more global convergence measures confirm that at a resolution of $300^3$ sufficient detail is captured.

\begin{figure*}
  \centering
      \subfloat{\includegraphics[width=0.8\columnwidth, clip=true]{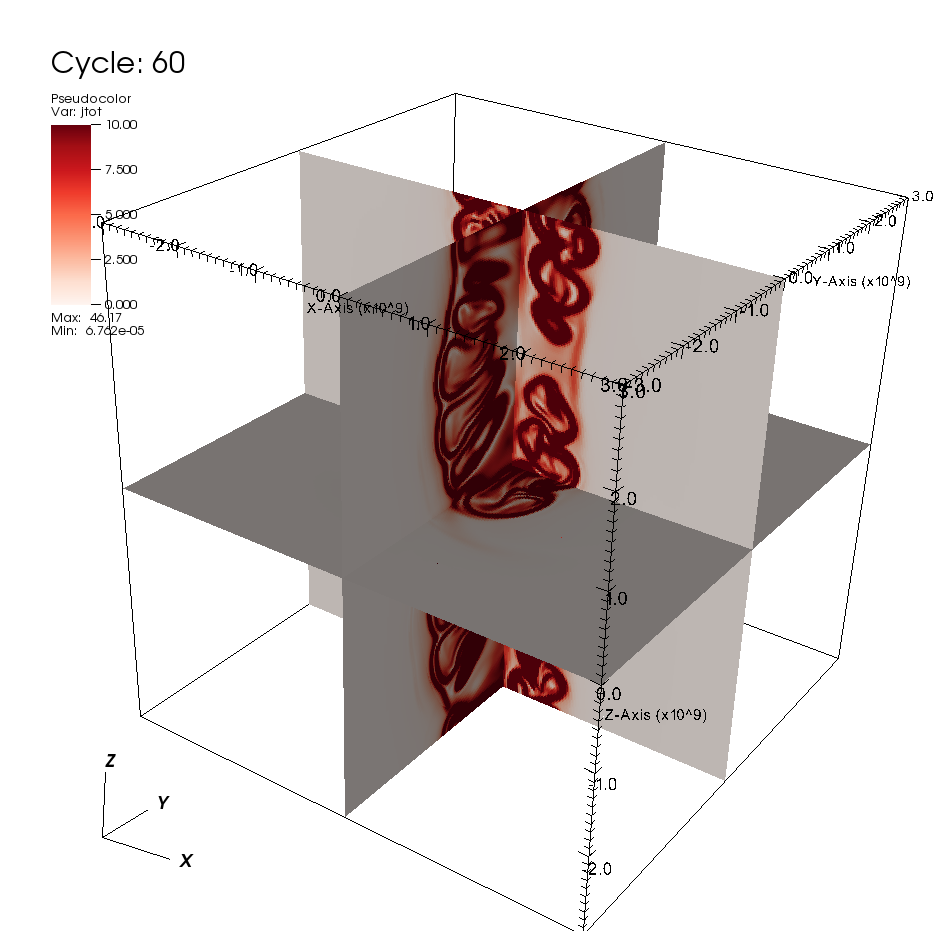}}
		\subfloat{\includegraphics[width=0.8\columnwidth, clip=true]{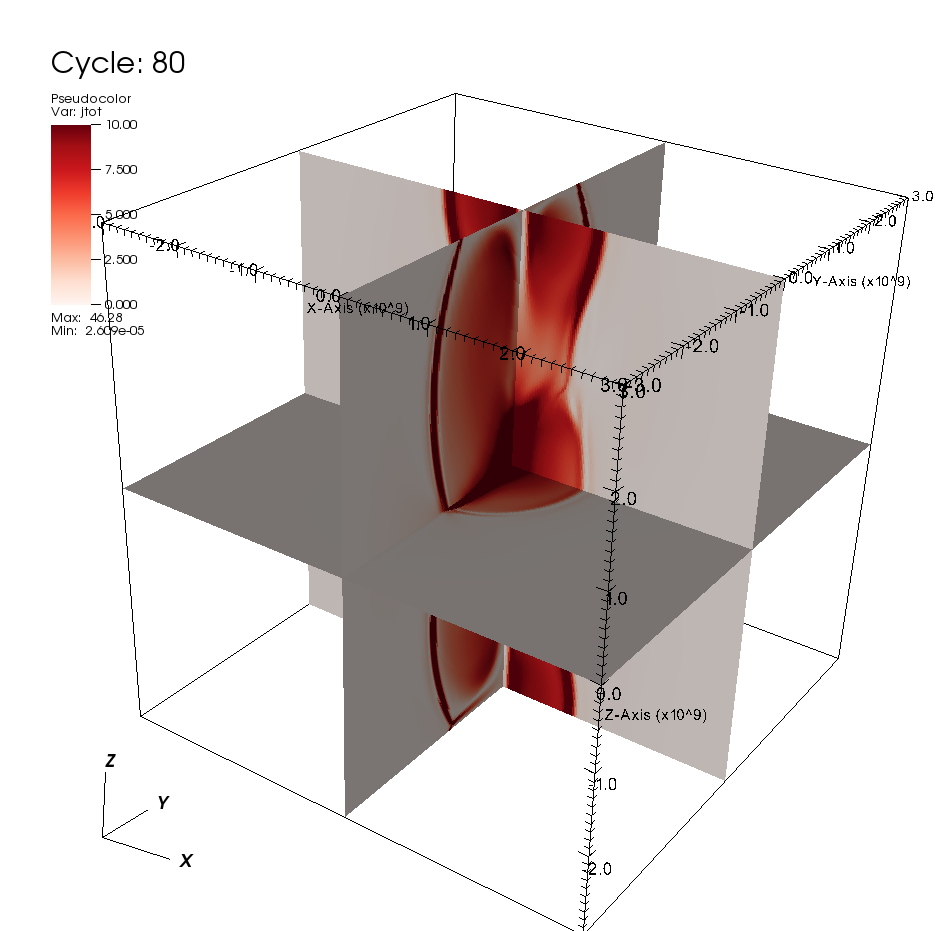}}
\caption{For the fastest non-force-free case B3dnff (left panel) and the fastest force-free case F3dff (right panel) we show a 3D view of the total current density with slices cut through the three axes, at $t=6$ (left) and $t=8$ (right). The instability in the non-force-free case develops faster and the helical structure is clearly visible. The colour scale is saturated at $j=10$. The box size is $6L\times 6L \times 6L$ with $L = 10^9 cm$.}
\label{fig:3Djtot_p001}
\end{figure*}

\begin{figure*}
  \centering
      \subfloat{\includegraphics[width=0.8\columnwidth, clip=true]{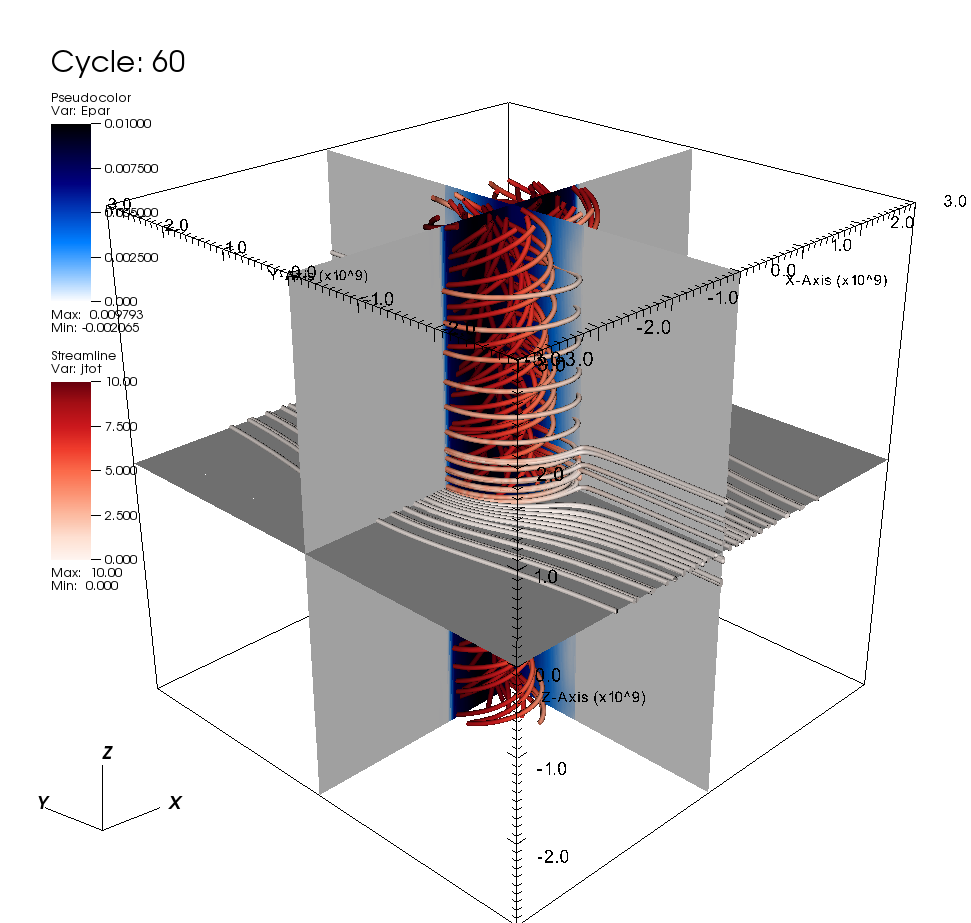}}
		\subfloat{\includegraphics[width=0.8\columnwidth, clip=true]{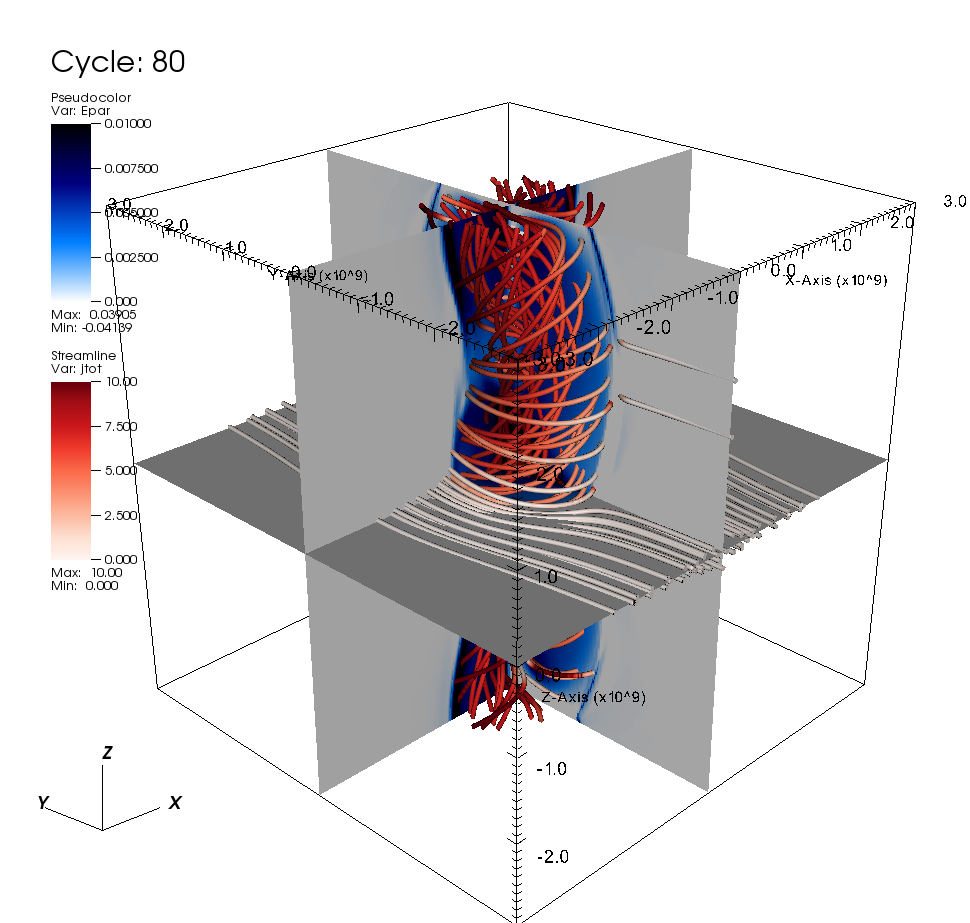}}
\caption{For the fastest force-free case F3dff we show a 3D view of the parallel, resistive electric field $E_{||}$ with slices cut through the three axes, at $t=6$ (at the onset of the instability, in the left panel) and $t=8$ (at the end of the linear growth phase of the instability, in the right panel). The colour scale is saturated at $0.001$. Selected field lines are shown, coloured by their total current density value, saturated at $j=10$. The box size is $6L\times 6L \times 6L$ with $L = 10^9 cm$.}
\label{fig:Epar3D}
\end{figure*}

\begin{figure}
  \centering
      \subfloat{\includegraphics[width=0.5\columnwidth, clip=true]{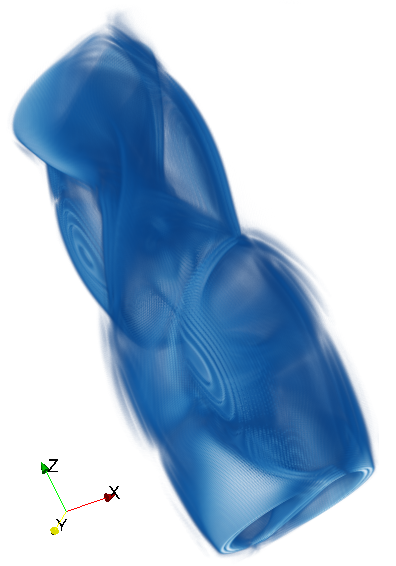}}
		\subfloat{\includegraphics[width=0.5\columnwidth, clip=true]{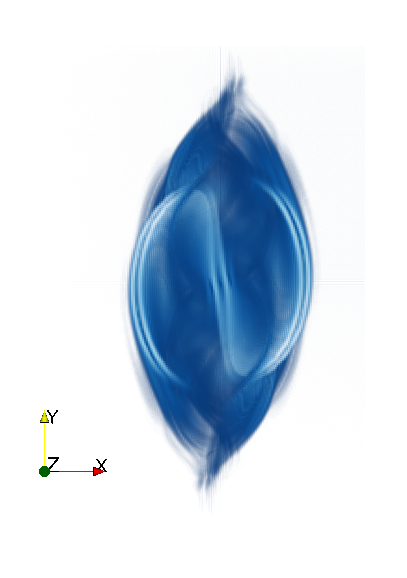}}
\caption{Same case as shown in Fig.~\ref{fig:Epar3D}, at $t=9$. Shown is a volume rendering of the change of topology of the magnetic field according to equation~(\ref{eq:reconnection}), satured at 0.05, in a side view (left) and a top view (right).}
\label{fig:Rec3D}
\end{figure}

In all 3D force-free cases peak current enhancement develops due to the tilt instability. An additional kink deformation and accompanying fine-structure develops in both current channels. In true 3D renderings of the total current density, the force-free cases are compared to the non-force-free evolution as described in~\citealt{Keppens}. In Fig.~\ref{fig:3Djtot_p001}, we show three-slices of the total current density at the end of the linear growth phase, for the fastest non-force-free case B3dnff (at $t=6$, in the left panel) and the fastest force-free case F3dff (at $t=8$, in the right panel), respectively. A very different pattern is observed for both cases. The kinking of the current channels is visible in both cases, although the helical structure is not as clearly visible for the force-free case in the right panel of Fig.~\ref{fig:3Djtot_p001} as for the non-force-free case in the left panel. The current channels repel each other and are displaced in the $y$-direction. The observed patterns in the $x,y$-plane are in agreement with the results from 2.5D simulations in Fig.~\ref{fig:jtot}. The higher axial field component in the force-free case, and hence, the smaller Kruskal-Shafranov limit suppresses some, but not all of the secondary mushroom-like features as are seen in the non-force-free case. In the force-free case, there is no $z$-component of the magnetic field in the ambient, outside the current channels. The diffusion of the current in the surroundings of the current channels is therefore completely different from the non-force-free cases. Both cases show development of thin sheet-like structures, indicated by a strong current.
The development of strong currents in these areas can also be seen in Fig.~\ref{fig:Epar3D}, for the same force-free case F3dff. Here we show the parallel, resistive electric field on the same slices as in the right panel of Fig.~\ref{fig:3Djtot_p001}, with selected magnetic field lines coloured by total current density, at $t=6$ at the onset of the instability (left panel) and at $t=8$ at the end of the linear growth phase of the instability (right panel).
Based on 2.5D results, the presence of parallel, resistive electric field is not the most accurate measure of reconnection occurring. Therefore we show a volume rendering of the change of topology of the magnetic field as measured by equation~(\ref{eq:reconnection}) at $t=9$ in Fig.~\ref{fig:Rec3D}. From the top view on the right it is clear that the whole area of the current channels, including the boundaries and the area in between the channels is a reconnection site. From the side view on the left, the kinking of the current channels is visible and it shows that the reconnection region stretches all the way along the current channel in the $z$-direction.
\section{Test particle acceleration in 2.5D configurations}
\label{sec:particles}
We evolve 200.000 test particles in 2.5D MHD snapshots with effective resolution of $2400\times2400$, obtained with 4 AMR levels. To analyse particle dynamics we solve the guiding centre equations of motion~(\ref{eq:gcastatic1}-\ref{eq:gcastatic3}) with the new GCA module in MPI-AMRVAC to get the particles position, velocity and magnetic moment. The particles are evolved in the force-free 2.5D setups with fastest reconnection (as indicated by the growth rate of the tilt instability and the reconnection indicator in Fig.~\ref{fig:rect8} for case F2dff (left panel) with uniform resistivity and for case F2dffAR (right panel) with anomalous resistivity. These two cases showed the fastest reconnection, for the lowest plasma-$\beta$ in a force-free equilibrium, being most in accordance with the conditions in the solar corona. Particles are initiated in static MHD snapshots at three different times. Shortly after the start of the linear growth of the tilt instability at $t=5$ (corresponding to $5 t_S \approx 417$ seconds), at the moment the instability reaches the nonlinear regime at $t=8$ ($8 t_S \approx 680$ seconds) and its peak current and far in the saturated, nonlinear regime at $t=9$ ($9 t_S \approx 765$ seconds). The particles are evolved for the typical time $t = t_S = L/c_S \approx 85.26$ seconds, with $c_S$ the sound speed outside the current channels in the MHD snapshots. The MHD quantities are static on the particles timescales and are not evolved. We analyse 200.000 electrons and protons to see differences caused by the mass difference and the opposite charge of the particles.

Particles are initialised randomly over the domain in the $x-y$ plane, with a fraction of $0.99$ of the ensemble uniformly distributed in space in a rectangular block, encapsulating the two (displaced) current channels and the areas with the largest current density, $x \in [-1,1]$, $y \in [-3,3]$. The other fraction of $0.01$ of the ensemble is uniformly distributed over the full domain $x \in [-3,3]$, $y \in [-3,3]$, including the surrounding background. This way, most of the particles are uniformly distributed in the region around the current channels and for a resolution of $2400^2$ and $200.000$ particles we have an average density of more than one particle per cell. Initially these particles represent a spatially uniform, thermal plasma. The remaining particles are in the ambient, where no current sheets are formed, representing the thermal background plasma. Simulations with a uniform initial distribution over the whole domain (including the ambient) have been conducted and the particles in the ambient follow the field lines (see e.g. Fig.~\ref{fig:Epart8}) and leave the domain on a timescale much shorter than $1t_S$. Therefore we neglect these particles and from Figures~\ref{fig:dist_ekin_t5}-\ref{fig:dist_ekin_t8} we can conclude that the particles in the area $x \in [-1,1]$, $y \in [-3,3]$ remain representative for the total plasmas, including the thermal part of the distribution, even at $t=9$. Typical simulation box lengths are of the order $\mathcal{O}(10^7 m)$, meaning that with an effective grid resolution of $2400^2$, the orbiting motion takes place on approximately $10^{-3}$ of a grid cell, for protons and for electrons even a factor 1836 smaller. The magnetic field as seen by a particle in orbit, will be constant across the gyro-orbit, if the magnetic field is obtained with the aforementioned resolution. The particles orbit will only change due to changes in the direction of the magnetic field and the guiding centre approximation is an accurate description of the particle dynamics. The major advantage of this approach is that the time step taken in the simulations can be several orders of magnitude larger in comparison with solving the full equation of motion. Initially, particles have a Maxwellian velocity distribution for $v = \sqrt{(v_{\|})^2 + (v_{\perp})^2}$. This corresponds to the thermal velocity $v_{th} = \sqrt{(2 k_B T \rho_0/m_{p} p_0)}$ of protons in a fluid of temperature about $T = 10^6 K$ with the proton rest mass $m_p = 1.6726 \cdot 10^{-24} g$ and dimensionless pressure $p_0$ and fluid density $\rho_0$ as defined in Section~\ref{sec:MHDtheory}. The particles have a uniform pitch angle distribution with $\alpha \in [-\pi/2, \pi/2]$. The initial gyroradius of all particles is several orders smaller than the smallest cell size. After one time $t_S$, there are particles which are accelerated significantly, such that their gyroradius becomes large, although still small compared to a grid cell of $25000 m$. The maximum gyroradius reached in the simulations carried out is $17 \cdot 10^{-2} m$ for electrons and $8.5 \cdot 10^2m$ for protons, both after $1 t_S$. Mainly for protons it is therefore necessary to be cautious about obtained results at late times since gyration effects may become important and cannot be neglected.
\subsection{Particle energy}
We will quantify the particle dynamics by looking at the energy distribution of the whole ensemble of particles, by analysing the relative importance of the separate drifts in particular regions of the domain and by following the evolution of the kinetic energy of individual particles, identified to show interesting behaviour.

\begin{figure*}
  \centering
    \subfloat[]{\includegraphics[width=\columnwidth, clip=true]{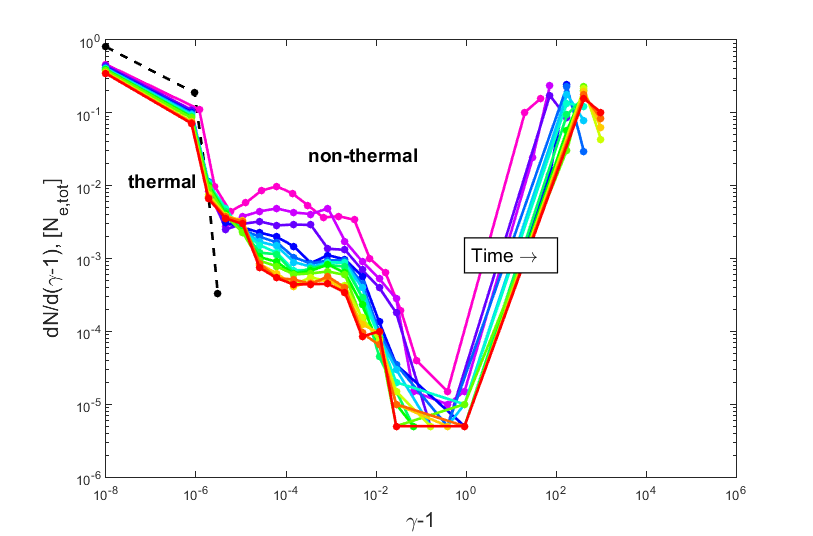}}
		\subfloat[]{\includegraphics[width=\columnwidth, clip=true]{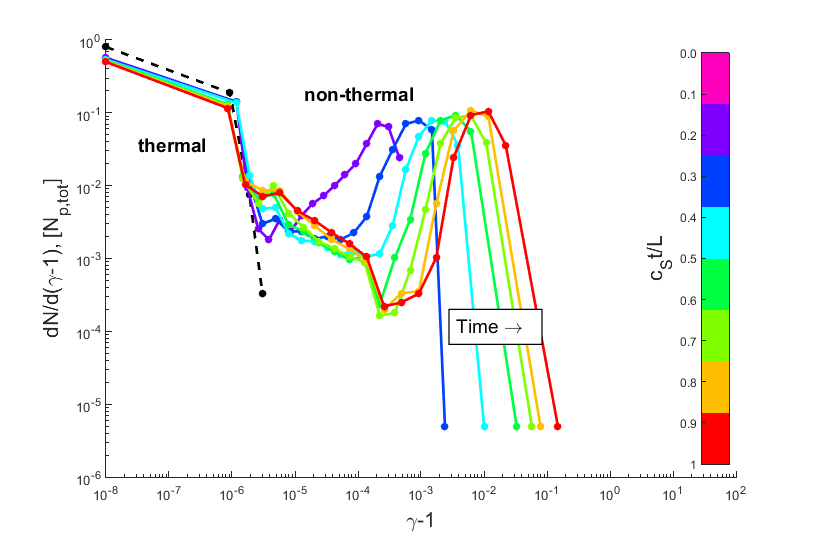}}
\caption{Kinetic energy distribution counted by particle number at $t=5$ plotted up to $t=6$ for 200.000 electrons (left panel) and protons (right panel) for equilibrium F2Dff. Time is measured in units of $L/c_S$, see the colour bar at the right. The initial Maxwellian is depicted with a dashed, black line. For the electrons in the left panel two decades per tick are used for the x-axis.}
\label{fig:dist_ekin_t5}
\end{figure*}

\begin{figure*}
 \subfloat[]{\includegraphics[width=\columnwidth, clip=true]{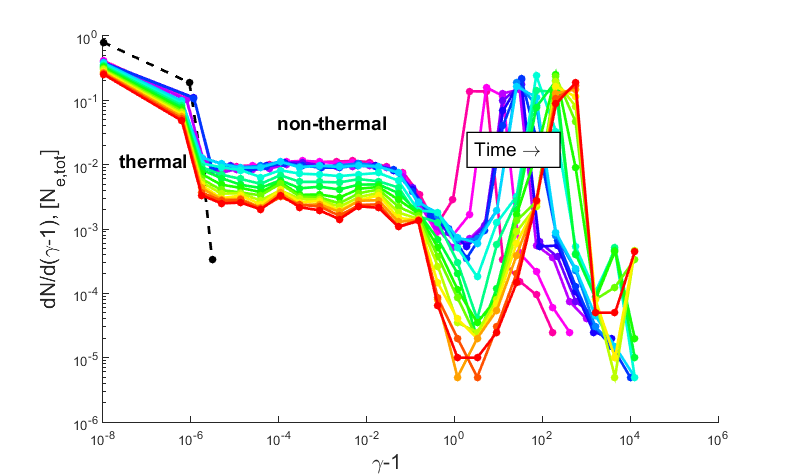}}
		\subfloat[]{\includegraphics[width=\columnwidth, clip=true]{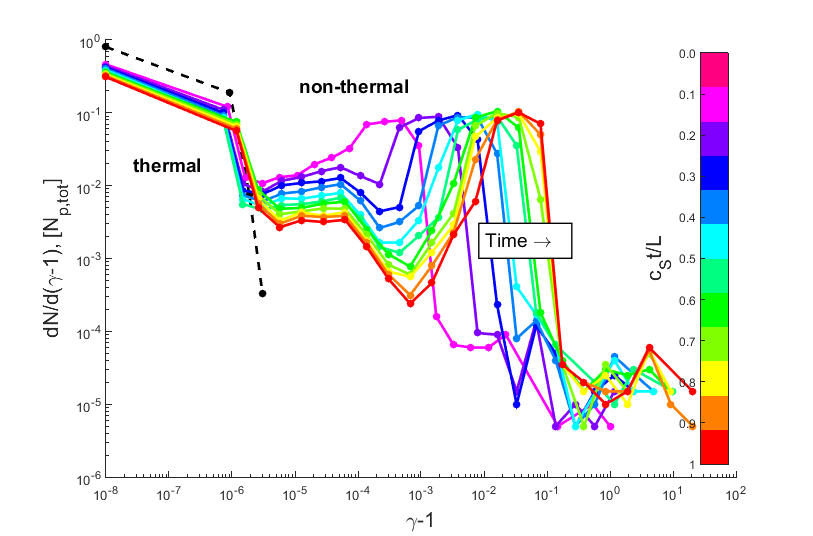}}
\caption{Kinetic energy distribution counted by particle number at $t=8$ plotted up to $t=9$ for 200.000 electrons (left panel) and protons (right panel) for equilibrium F2Dff. Time is measured in units of $L/c_S$, see the colour bar at the right. The initial Maxwellian is depicted with a dashed, black line. For the electrons in the left panel two decades per tick are used for the x-axis.}
\label{fig:dist_ekin_t8}
\end{figure*}

The kinetic energy of a particle is $E_{k} = (\gamma - 1) m_0 c^2 \approx m_0 c^2 (\frac{1}{2}\frac{v^2}{c^2} + \frac{3}{8} \frac{v^4}{c^4})$. In Figures \ref{fig:dist_ekin_t5}-\ref{fig:dist_ekin_t8} the kinetic energy (divided by $m_0c$) distribution counted by number of particles is shown for both electrons (in the left panels) and protons (in the right panels), normalised by the total number of particles, at $t=5$ before the linear growth phase of the instability and at $t=8$ far in the strongly nonlinear regime of case F2dff. The initial Maxwellian is depicted as a black dashed line, showing the initial thermal distribution. The total number of bins is set to the square root of total number of particles, on a logarithmic scale. The distribution is plotted in every panel at representative time intervals up to $1 t_S$, indicated in the colour bar on the right. All particles, both initiated in the ambient and in the regions of the (displaced) current channels are initially thermal. In the current channels and in the reconnection zones, non-thermal distributions form. Outside the current channels and the reconnection zones, from now on referred to as the ambient, the particles remain thermal. The electron dynamics are much faster and a high energy tail already develops withing $0.1 t_S$, whereas the proton dynamics are slower, in accordance with the mass difference. The spectra are initially purely thermal, but all quickly develop a middle part at medium energy and a high energy tail.

\begin{figure*}
 \subfloat[]{\includegraphics[width=\columnwidth, clip=true]{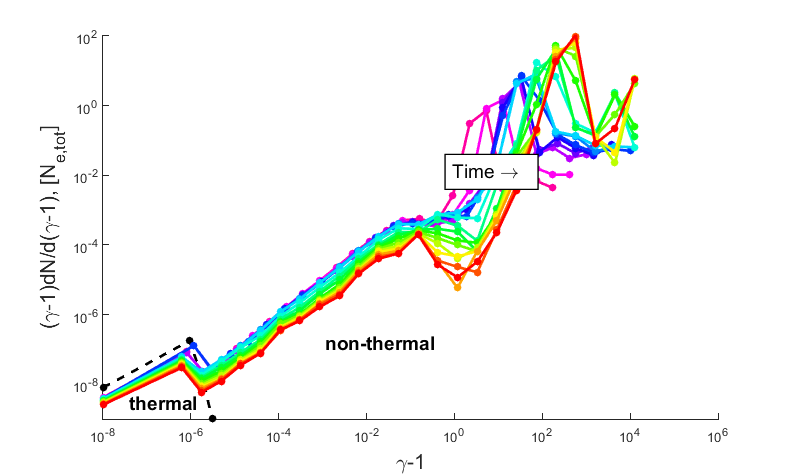}}
		\subfloat[]{\includegraphics[width=\columnwidth, clip=true]{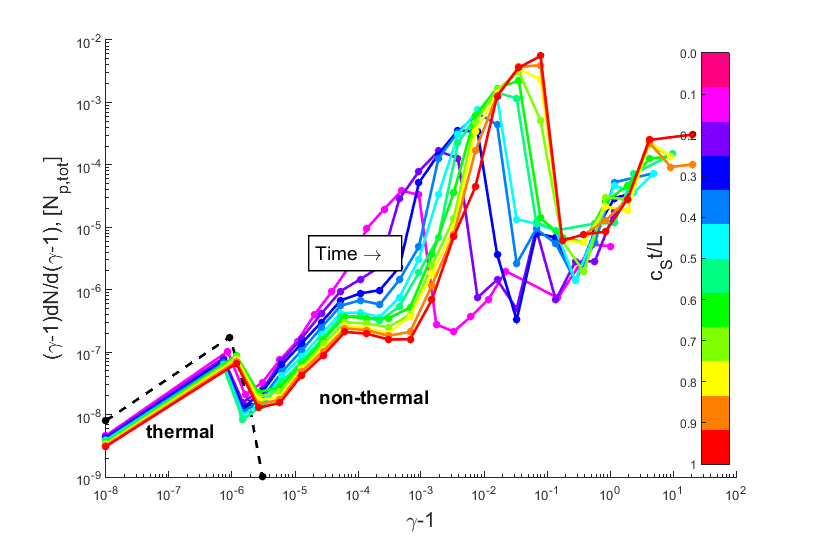}}
		\caption{Kinetic energy distribution counted by particle energy at $t=8$ plotted up to $t=9$ for 200.000 electrons (left panel) and protons (right panel) for equilibrium F2Dff. Time is measured in units of $L/c_S$, see the colour bar at the right. The initial Maxwellian is depicted with a dashed, black line. For the electrons in the left panel two decades per tick are used for the x-axis.}
\label{fig:dist_ekin_number_t8}
\end{figure*}

At $t=5$, the high energy tail developing is mainly due to acceleration parallel to the magnetic field, for $\gamma-1 > 1$. Because of uniform resistivity, there is a parallel, resistive electric field in the whole domain, accelerating particles accordingly. There is a medium tail forming as well between the thermal distribution and the high energy tail, for $10^{-5} < \gamma - 1 < 10^0$. Parallel acceleration due to resistive electric field is the dominant mechanism compared to the drift terms in equation~(\ref{eq:gcastatic1}). At $t=8$ in Fig.~\ref{fig:dist_ekin_t8} for electrons and for protons, particles are still accelerated up to very high energies ($\gamma\sim 10^4$) due to parallel acceleration, but a more pronounced tail has developed at medium energies as well $\gamma \sim 1 - 10^2$ due to other acceleration mechanisms. This middle part of the spectrum reaches a Lorentz factor of maximum $1.5$ to $2$ times the thermal Lorentz factor $\gamma=1$, confirming the findings of \cite{Zhou} for acceleration due to magnetic curvature drifts and magnetic gradient drifts perpendicular to the magnetic field.
The high energy particles dominate not only in number, seeing that almost all particles pick up a parallel acceleration at later times, they also dominate in energy. This can be seen from Fig.~\ref{fig:dist_ekin_number_t8} where the kinetic energy distribution is plotted at $t=8$, now not counted by the number of particles but by the kinetic energy of the particle. The high energy tail is now even more pronounced at all times, both for protons (right panel) and electrons (left panel). The plasma energy content is dominated by the particles at high energy, contrasting with the expected energy spectrum in which low energy particles dominate both by number and energy (\citealt{Rosdahl}).
The maximum energy reached in these simulations based on 2.5D MHD runs is orders of magnitude larger than expected from observations of particles acceleration in solar flares (\citealt{Rosdahl}). This effect is mainly due the acceleration in the invariant direction and due to the high resistivity $\eta=10^{-4}$ (more realistic, lower values would severely limit the computation time and no longer minimise the numerical dissipation for MHD evolutions), and the subsequent electric field accelerating particles in the direction parallel to the magnetic field. Therefore the maximum energies reached by the particles are  unrealistic in 2.5D simulations. Neglecting the second high energy tail caused by parallel acceleration in the current channels, electrons reach maximum Lorentz factors of $\gamma \sim 2$ and protons of $\gamma \sim 1.01$, in the reconnection zones in both cases, and these values are more realistic for the solar corona. 
\subsection{Effects of anomalous resistivity on particle energy}
Evolving test particles in static MHD snapshots in a 2.5D configuration shows that the dominant processes are particle acceleration in either of the two current channels or acceleration by the parallel, resistive electric field in the reconnection regions. Both processes cause a large acceleration in the direction parallel to the magnetic field, which is dominant over all other acceleration and drift processes. This means that either the particles leave the domain through one of the open $x,y$ boundaries, or they are accelerated indefinitely in the current channels, in the translationally invariant $z$-direction. To counteract this artifact, we also evolve particles in the setups with anomalous resistivity. The particles are then expected not to accelerate due to resistive electric field in the ambient, where resistivity is absent. However, they still will be in the reconnection areas around the current channels and inside the current channels.

Particles, mainly trapped inside the current channels but also in the ambient, reach extremely high energies due to indefinite acceleration in the translationally invariant $z$-direction due to the 2.5D nature of the simulations. This numerical artifact is counteracted by employing the anomalous resistivity model described by equation~(\ref{eq:anomalousres}). In this case (F2dffAR) there is no resistivity, and hence no parallel electric field $E_{||} = \eta \mathbf{J} \cdot \mathbf{\hat{b}}$, in regions with a low current (low meaning equilibrium values). Particles are now still accelerated indefinitely in the invariant $z$-direction, but not anymore due to resistive electric fields in the ambient. In general protons and electrons behave similarly, with the main difference that protons are much slower due to the difference in mass. It takes a longer time for protons to form a smooth high energy tail and the maximum Lorentz factor reached is two orders of magnitude smaller than for electrons in all cases.

\begin{figure*}
  \centering
   
 		\subfloat[]{\includegraphics[width=\columnwidth, clip=true]{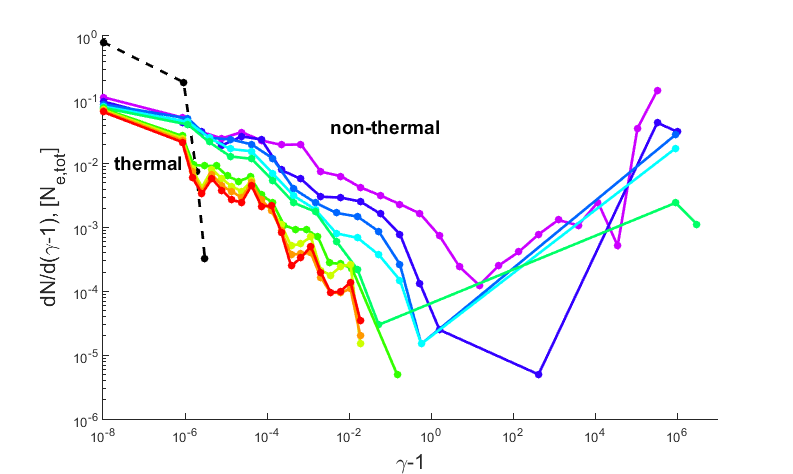}}
 		\subfloat[]{\includegraphics[width=\columnwidth, clip=true]{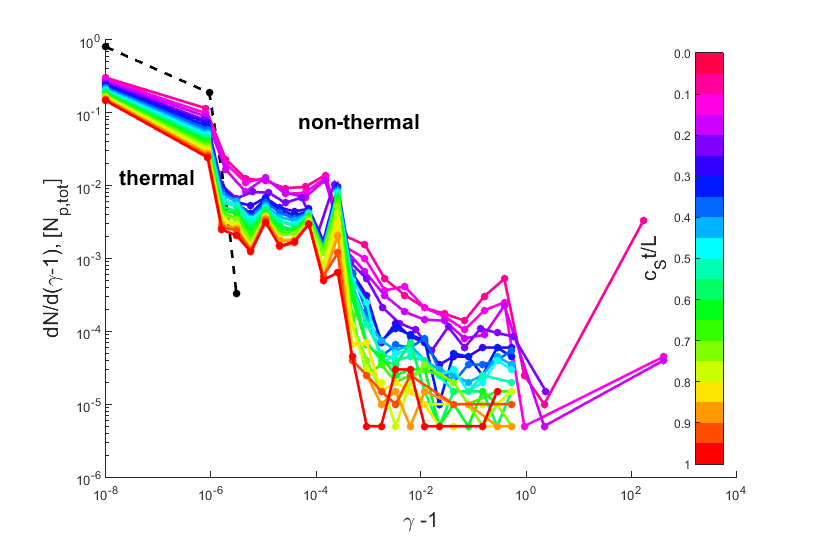}}		
	\caption{Kinetic energy distribution counted by particle number at $t=8$ plotted up to $t=9$ for 200.000 electrons (left panel) and protons (right panel) for equilibrium F2DffAR. Time is measured in units of $L/c_S$, see the colour bar at the right. The initial Maxwellian is depicted with a dashed, black line. Two decades per tick are used for the x-axis.}
\label{fig:dist_AR_ekin}
\end{figure*}

\begin{figure*}
  \centering
    		\subfloat[]{\includegraphics[width=\columnwidth, clip=true]{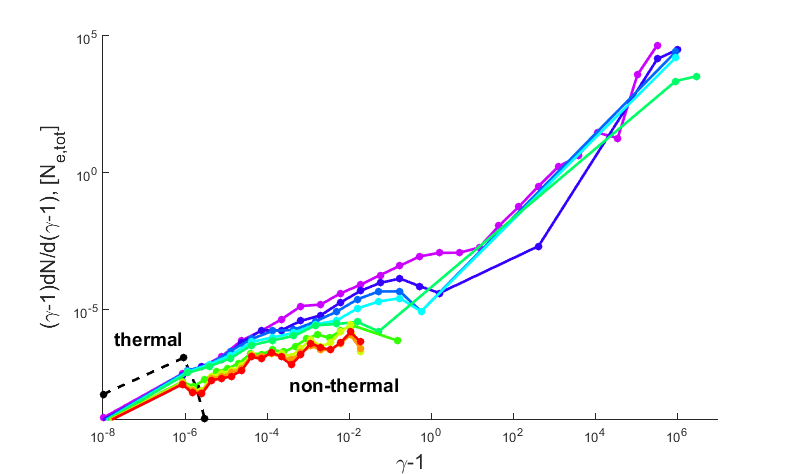}}
		\subfloat[]{\includegraphics[width=\columnwidth, clip=true]{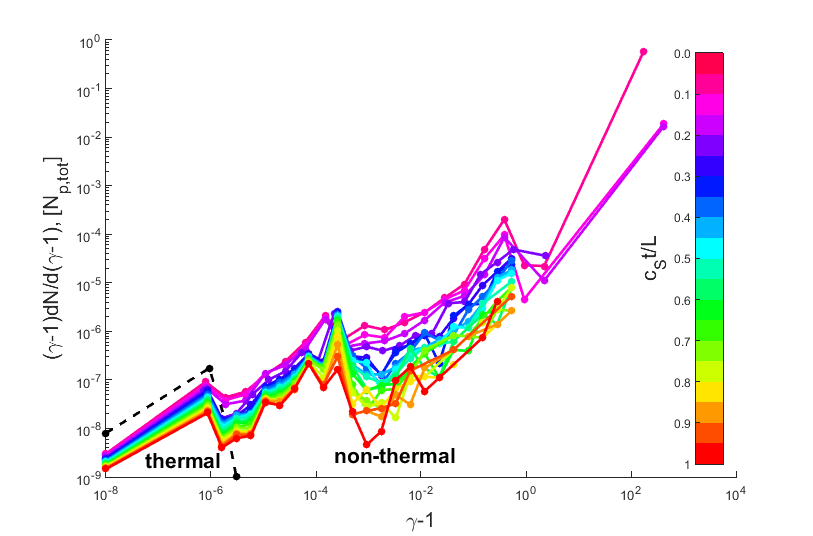}}
	\caption{Kinetic energy distribution counted by particle energy at $t=8$ plotted up to $t=9$ for 200.000 electrons (left panel) and protons (right panel) for equilibrium F2DffAR. Time is measured in units of $L/c_S$, see the colour bar at the right. The initial Maxwellian is depicted with a dashed, black line. Two decades per tick are used for the x-axis.}
\label{fig:dist_AR_ekin_number}
\end{figure*}

The expected effect of anomalous resistivity on particle acceleration is that less particles are accelerated by resistive electric field, since it is not present uniformly anymore. This avoids strong particle acceleration in the ambient, meaning that particles can still accelerate up to high energies, but high energy particles should not dominate in number anymore. This effect can be observed from the kinetic energy distributions. In Fig.~\ref{fig:dist_AR_ekin} the number distribution is plotted and both for electrons and protons there is still a second high energy tail due to parallel acceleration, but it is less dominant than in the case with uniform resistivity (Fig.~\ref{fig:dist_ekin_t8}). 
However, in Fig.~\ref{fig:dist_AR_ekin_number}, for the same case, the kinetic energy distribution is plotted and the high energy particles are even more dominant in energy. Due to the anomalous resistivity present only in regions with strong current, large gradients in the electromagnetic fields arise. A larger peak current is reached, already concluded from the MHD simulations, and particles are accelerated up to even higher kinetic energy ($\gamma \sim 10^6$ for electrons and $\gamma \sim 10^2$ for protons). We conclude that with uniform resistivity applied, high energy particles are dominant in the energy distribution both in number and counted by energy per particle. Whereas with anomalous resistivity applied, high energy particles are mainly dominant in the energy distribution counted by energy but less in number.
\subsection{Pitch angle}
\begin{figure}
	\includegraphics[width=0.9\columnwidth,trim= 10cm 2cm 4cm 2cm, clip=true]{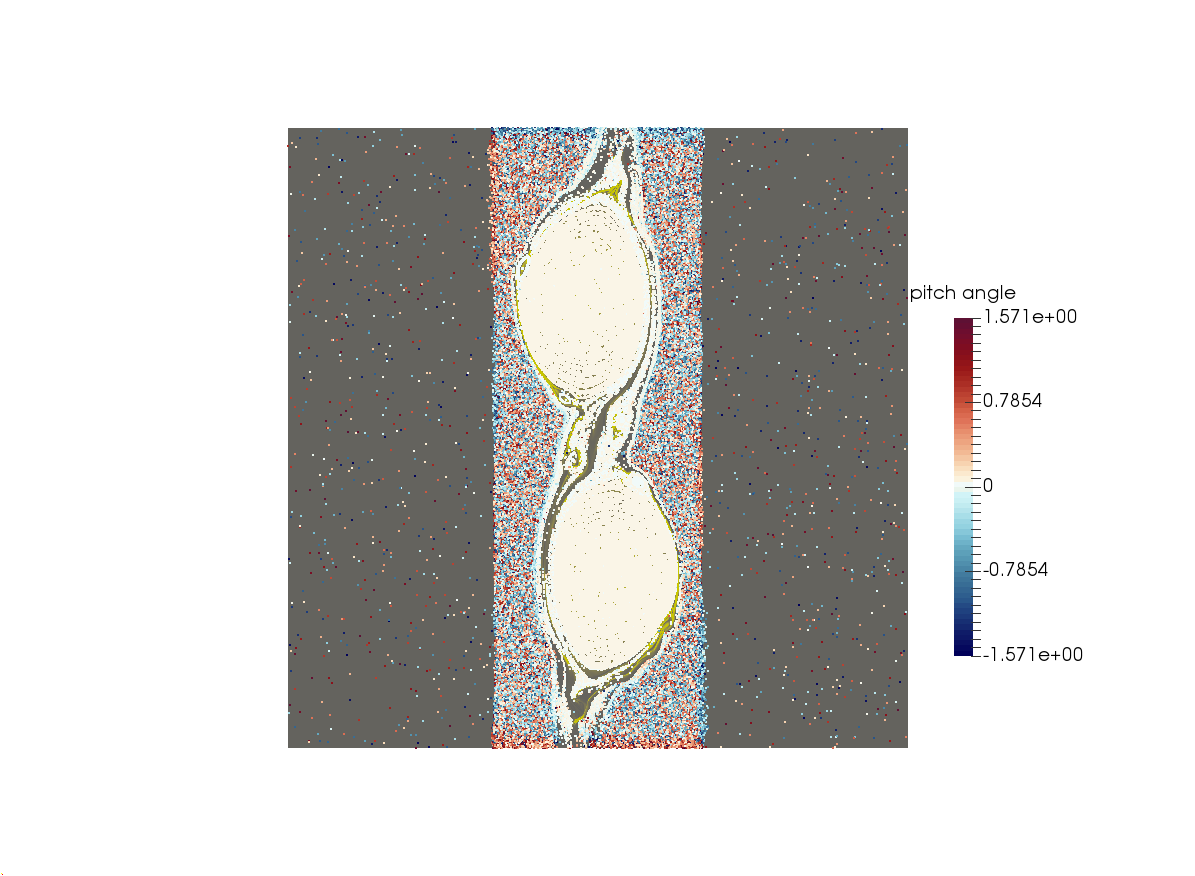}
    \caption{Spatial distribution of pitch angle $\alpha$ at $8t_S + 0.05t_S$ for electrons in case F2dff. Particles are visualised as dots plotted on top of the magnitude of the parallel electric field as obtained from MHD, where the yellow coloured parts indicate a nonzero parallel electric field and hence reconnection occurring, and grey parts indicate absent parallel electric field (see Fig.~\ref{fig:Epart8} for the resistive electric field without particles plotted on top). For the parallel electric field a linear colour is saturated to show values between $[0, 0.001]$. Particles in and around the current channels have a pitch angle close to zero (mainly coloured white) and thus a dominant parallel velocity.}
    \label{fig:alpha_noar}
\end{figure}

\begin{figure}
	\includegraphics[width=0.9\columnwidth,trim= 10cm 2cm 4cm 2cm, clip=true]{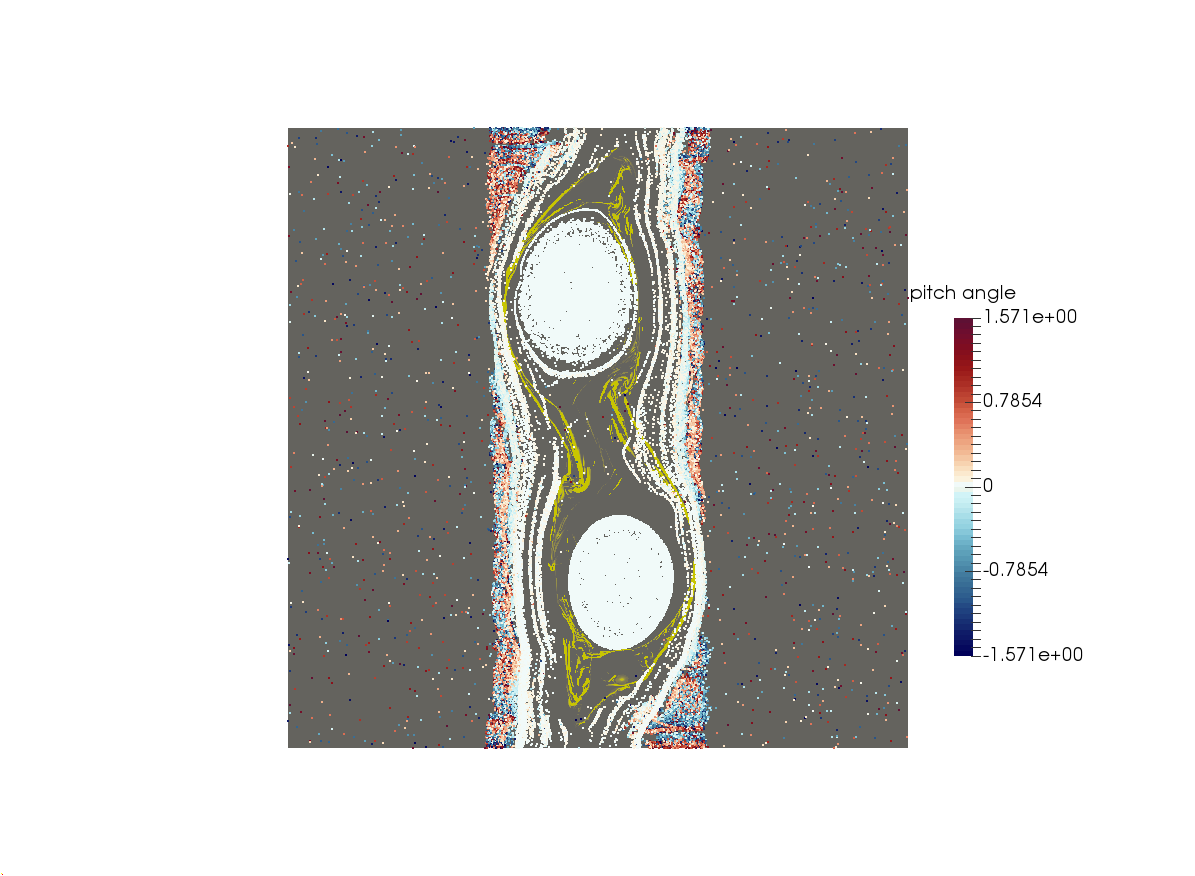}
    \caption{Spatial distribution of pitch angle $\alpha$ for protons at $8t_S + 0.05t_S$ in case F2dffAR. Particles are visualised as dots plotted on top of the magnitude of the parallel electric field as obtained from MHD, where the yellow coloured parts indicate a nonzero parallel electric field and hence reconnection occurring, and grey parts indicate absent parallel electric field (see Fig.~\ref{fig:Epart8} for the resistive electric field without particles plotted on top). For the parallel electric field a linear colour is saturated to show values between $[0, 0.001]$. Particles in the current channels have a pitch angle close to zero (mainly coloured white) and thus a dominant parallel velocity.}
    \label{fig:alpha_ar}
\end{figure}

The dominance of parallel acceleration is also depicted by the spatial pitch angle distribution $\alpha = \arctan(v^*_{\perp}/v_{\|})$, in Fig.~\ref{fig:alpha_noar} for electrons in setup F2dff and in Fig.~\ref{fig:alpha_ar} for protons in setup F2dffAR, plotted on top of the absolute value of the resistive electric field as obtained from MHD, as an indicator of reconnection regions (see Fig.~\ref{fig:Epart8} for the respective resistive electric field without particles plotted on top). In the regions with strong resistive electric field (marked by yellow background colour in Fig.~\ref{fig:alpha_ar}) the pitch angle is small, as expected, however, also in the current channels (with absent resistive electric field, indicated by the grey background colour) the pitch angle is close to zero (indicated by the particles coloured white) already shortly after the particles are initialised. The dominance of the resistive electric field acceleration makes high energy electrons move antiparallel to the magnetic field and high energy protons parallel to the magnetic field. In a setup with anomalous resistivity, this is still the case. The anomalous resistivity solves the issue of parallel acceleration in regions without a high current, but not in the current channels. The particles are even accelerated up to higher energies due to the higher peak current reached in the case with anomalous resistivity (see Fig.~\ref{fig:peakcurrent}). Therefore at $8t_S + 0.05t_S$, the particles have travelled further into the current channels (in the direction perpendicular to the plane shown) and built up a higher energy. Particles are dragged into the current channels from regions where resistivity, and hence parallel electric field is absent. This is visualised by the `gaps' in Figures~\ref{fig:alpha_ar} and \ref{fig:drifts_ar} for case F2dffAR, with anomalous resistivity, where no particles are present anymore in the areas without resistive electric field around the current channels, but energetic particles (with $\alpha \approx 0$ and large parallel velocity) are residing in the current channels.
\subsection{Particle drifts}
\begin{figure*}
  \centering
    \subfloat{\includegraphics[width=0.66667\columnwidth,trim= 10cm 2cm 4cm 2cm, clip=true]{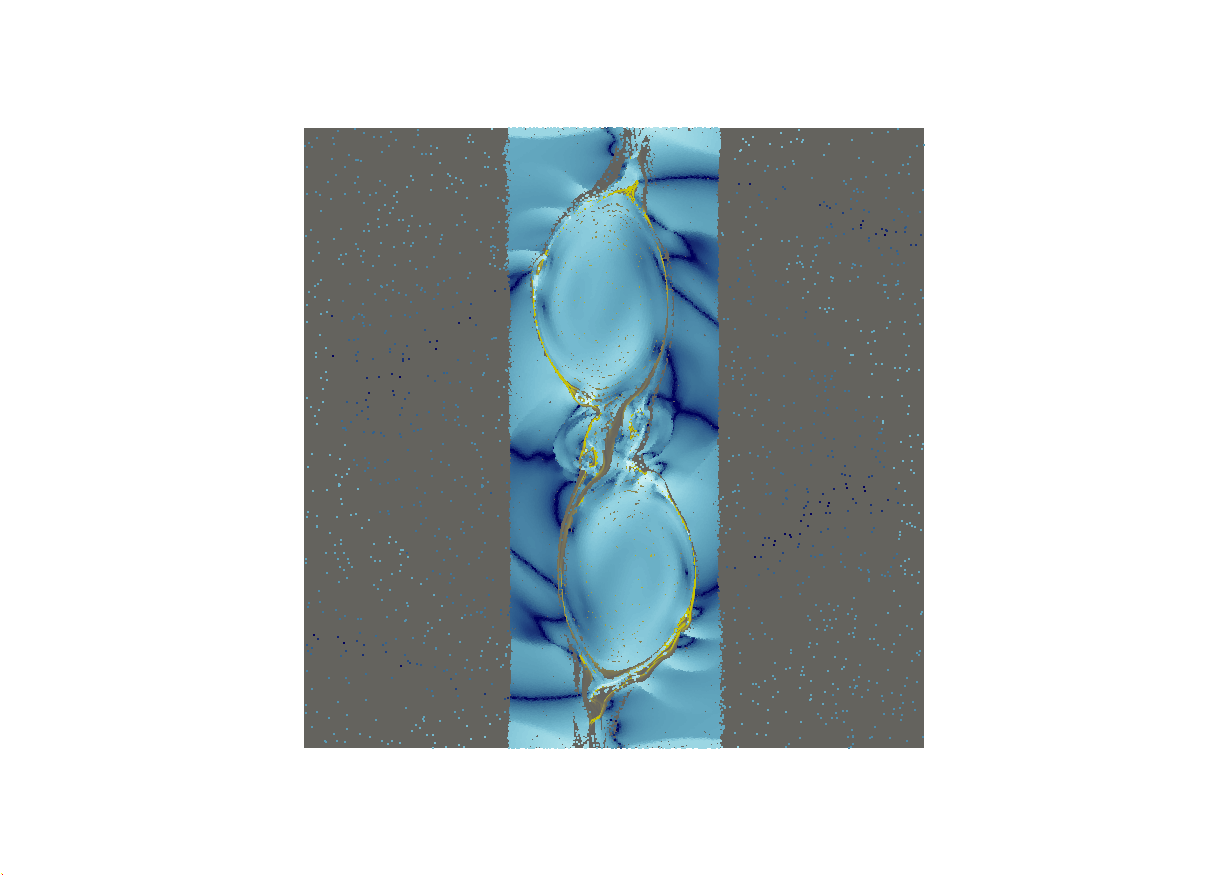}}\subfloat{\includegraphics[width=0.66667\columnwidth,trim= 10cm 2cm 4cm 2cm, clip=true]{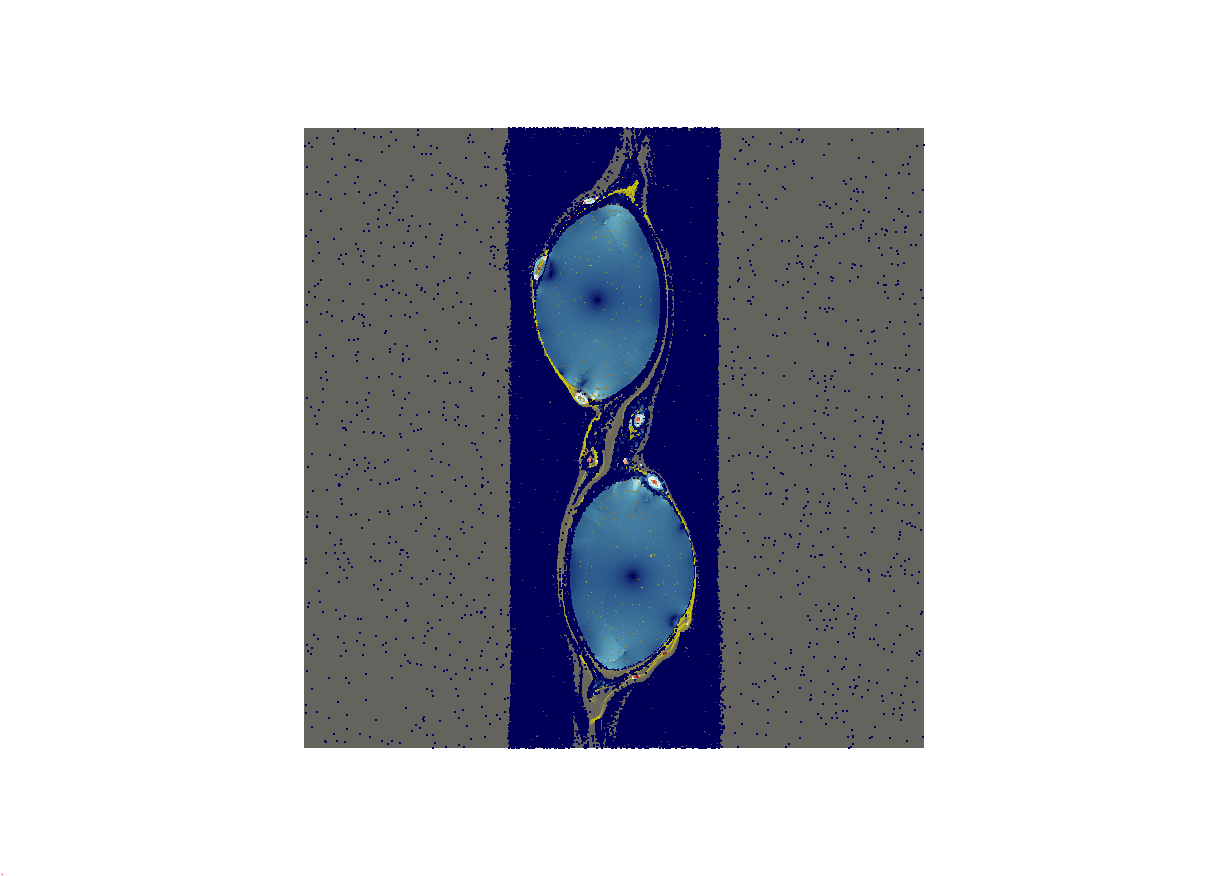}}\subfloat{\includegraphics[width=0.66667\columnwidth,trim= 10cm 2cm 4cm 2cm, clip=true]{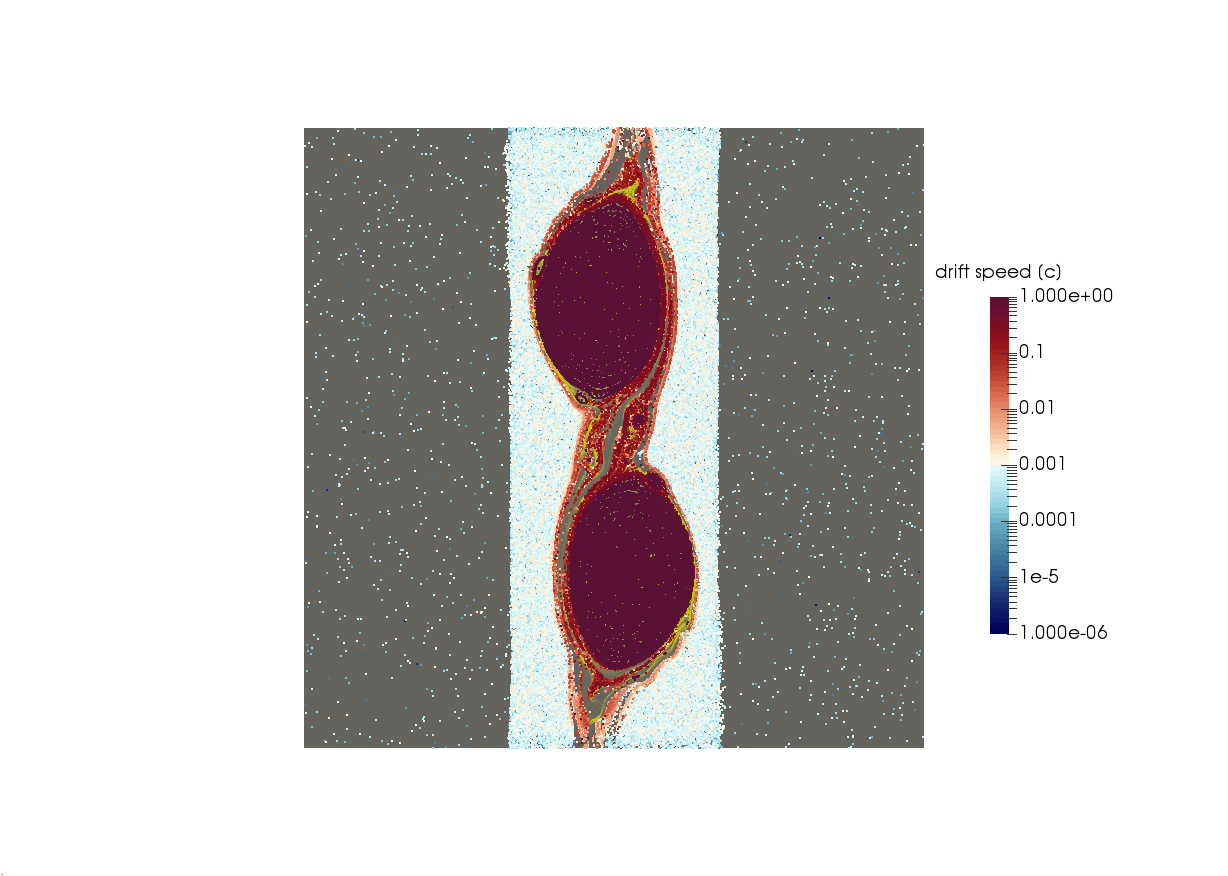}}
\caption{Magnitude of the relativistic $\mathbf{E}\times\mathbf{B}$ drift ($-\mathbf{\hat{b}}/B \times c\mathbf{E}$, the second term on the right-hand-side in equation (\ref{eq:gcastatic1})), the relativistic curvature $\mathbf{B}$ drift $\frac{\mathbf{\hat{b}}}{B\left(1-E_{\perp}^{2}/B^2\right)} \times (cm_0\gamma/q)\left(v_{\|}^{2}\left(\mathbf{\hat{b}}\cdot\nabla\right)\mathbf{\hat{b}}\right)$ (the third term in equation (\ref{eq:gcastatic1})) and the parallel velocity $v_{\|}\mathbf{\hat{b}}$ (the first term in equation (\ref{eq:gcastatic1})) from left to right at $8t_S + 0.05t_S$ in case F2dff. The drifts are normalised to the speed of light and plotted to the same scale. Particles visualised as dots are plotted on top of the magnitude of the parallel electric field as obtained from MHD, where the yellow coloured parts indicate a nonzero parallel electric field and hence reconnection occurring, and grey parts indicate absent parallel electric field (see Fig.~\ref{fig:Epart8} for the resistive electric field without particles plotted on top). For the parallel electric field a linear colour is saturated to show values between $[0, 0.001]$. The parallel velocity is clearly dominant compared to the drift velocities.}
\label{fig:drifts_noar}
\end{figure*}

\begin{figure*}
  \centering
    \subfloat{\includegraphics[width=0.66667\columnwidth,trim= 10cm 2cm 4cm 2cm, clip=true]{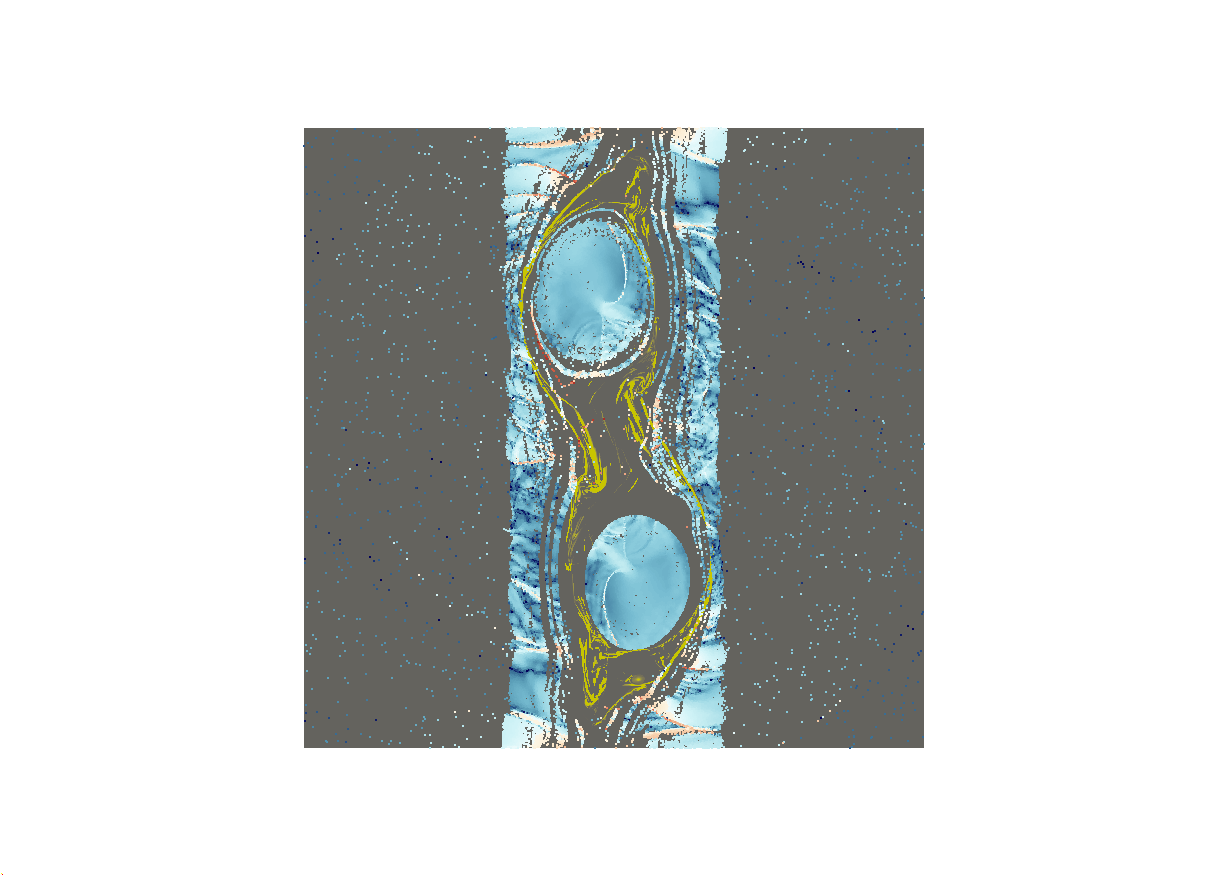}}\subfloat{\includegraphics[width=0.66667\columnwidth,trim= 10cm 2cm 4cm 2cm, clip=true]{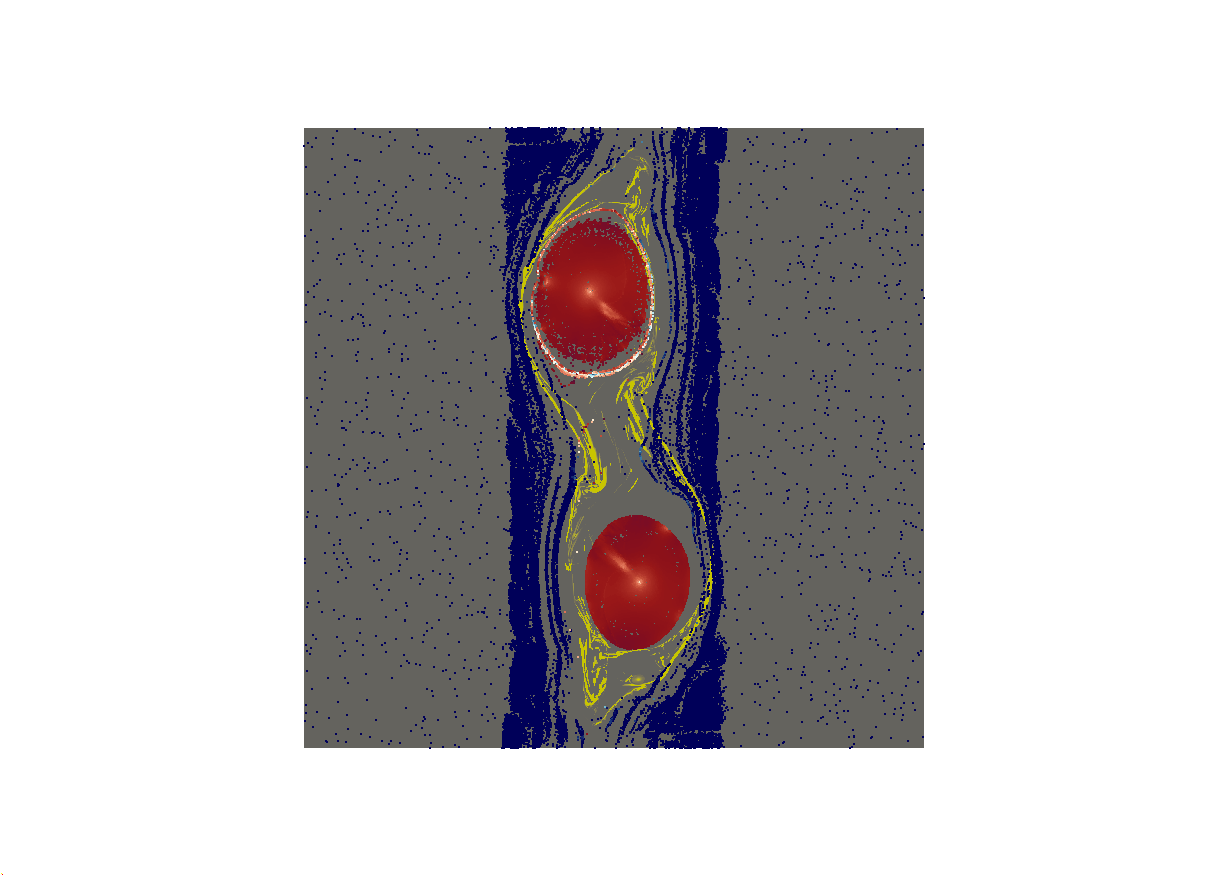}}\subfloat{\includegraphics[width=0.66667\columnwidth,trim= 10cm 2cm 3cm 2cm, clip=true]{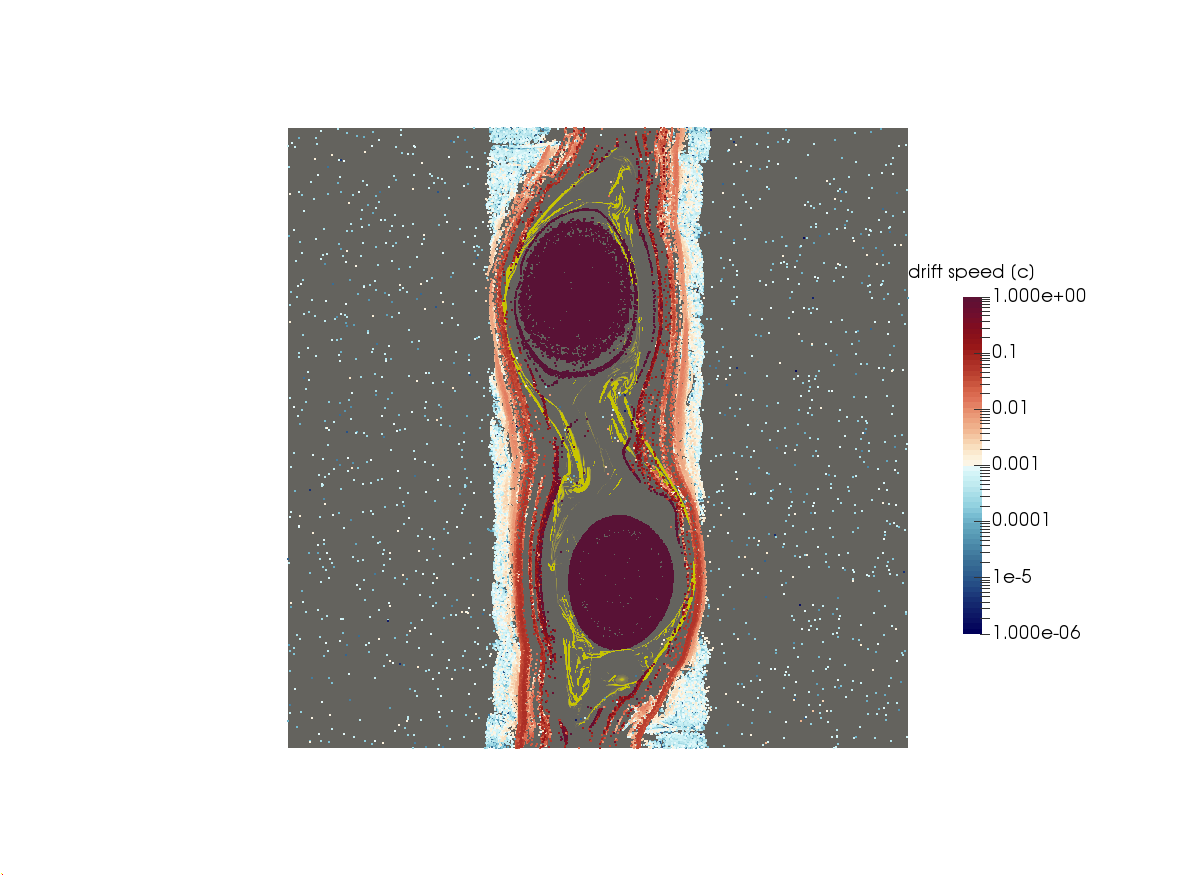}}
\caption{Magnitude of the relativistic $\mathbf{E}\times\mathbf{B}$ drift ($-\mathbf{\hat{b}}/B \times c\mathbf{E}$, the second term on the right-hand-side in equation (\ref{eq:gcastatic1})), the relativistic curvature $\mathbf{B}$ drift $\frac{\mathbf{\hat{b}}}{B\left(1-E_{\perp}^{2}/B^2\right)} \times (cm_0\gamma/q)\left(v_{\|}^{2}\left(\mathbf{\hat{b}}\cdot\nabla\right)\mathbf{\hat{b}}\right)$ (the third term in equation (\ref{eq:gcastatic1})) and the parallel velocity $v_{\|}\mathbf{\hat{b}}$ (the first term in equation (\ref{eq:gcastatic1})) from left to right at $8t_S + 0.05t_S$ in case F2dffAR. The drifts are normalised to the speed of light and plotted to the same scale. Particles visualised as dots are plotted on top of the magnitude of the parallel electric field as obtained from MHD, where the yellow coloured parts indicate a nonzero parallel electric field and hence reconnection occurring, and grey parts indicate absent parallel electric field (see the right panel of Fig.~\ref{fig:Epart8} for the resistive electric field without particles plotted on top). For the parallel electric field a linear colour is saturated to show values between $[0, 0.001]$. Compared to Fig.~\ref{fig:drifts_noar} the curvature drift inside the current channels has gained importance.}
\label{fig:drifts_ar}
\end{figure*}

To analyse the relative importance of the drift terms in the evolution equation of the guiding centre position, we plot the magnitudes of the three most dominant vectorial terms in equation~(\ref{eq:gcastatic1}) in Figures \ref{fig:drifts_noar} and \ref{fig:drifts_ar} for 200.000 electrons in a snapshot with uniform resistivity (F2dff) and with anomalous resistivity (F2dffAR) respectively at $t=8+0.05$, again measured in units of the speed of sound $t_S$. The particles are coloured by magnitude of the, from left to right, $\mathbf{E} \times \mathbf{B}$ drift ($-\mathbf{\hat{b}}/B \times c\mathbf{E}$, the second term on the right-hand-side of equation (\ref{eq:gcastatic1})), the curvature drift ($\left(\mathbf{\hat{b}}/\left(B\left(1-E_{\perp}^{2}/B^2\right)\right)\right) \times (cm_0\gamma/q)\left(v_{\|}^{2}\left(\mathbf{\hat{b}}\cdot\nabla\right)\mathbf{\hat{b}}\right)$, the third term on the right-hand-side of (\ref{eq:gcastatic1})) and the parallel velocity ($v_{\|}\mathbf{\hat{b}}$, the first term on the right-hand-side of (\ref{eq:gcastatic1})), normalised by the speed of light $c$, plotted on top of the absolute value parallel, resistive electric field obtained from the MHD snapshot at $t=8$, as an indicator of reconnection regions (see Fig.~\ref{fig:Epart8} for the respective resistive electric field without particles plotted on top). 

In regions with strong resistive electric field, particles are accelerated strongly, mainly in the direction parallel to the magnetic field, corresponding to the $v_{\|}\mathbf{\hat{b}}$ term in equation~(\ref{eq:gcastatic1}). However, the relativistic $\mathbf{E} \times \mathbf{B}$ drift (the second term on the right hand side of equation~(\ref{eq:gcastatic1})) and the relativistic curvature drift $\mathbf{B} \times \nabla \mathbf{B}$ (the third term on the right hand side of equation~(\ref{eq:gcastatic1})) have a non-negligible contribution to particle acceleration. The other drift terms in equation~(\ref{eq:gcastatic1}) are at least five orders of magnitude smaller. The relative importance of the curvature drift on the total particle velocity is due to the proportionality to $v_{\|}^2$ (see equations (\ref{eq:gcastatic2}) and (\ref{eq:gcanewton2})). \cite{Zhou2} finds similar results, where the curvature drift is less dominant in case the resistive electric field acceleration is neglected and enhanced when parallel electric field is taken into account. 
The spatial distribution of the particles does not change much compared to the initial uniform distribution. Fast particles are trapped inside the current channels, both in the case with uniform and in the case with anomalous resistivity. The particles in the ambient (indicated by the grey background, with absent resistive electric field) remain thermal. In case F2dffAR the particles are pushed away from the regions without resistivity around the current channels due to the higher magnetic field gradients building up. The magnetic islands trap particles even more efficiently in this case, enabling them to reach higher parallel velocities in the $z$-direction, than in cases with uniform resistivity. There is a strong correlation between the energy of the particles and the displacement in the $z$-direction (the direction depends on the current channel and on the charge of the particle). This is another indication that the particle acceleration is dominated by resistive electric field, confirming the results of \cite{Zhou2} for electron acceleration in 2.5D reconnection in the solar corona.
\subsection{Particle trajectories}
To gather more insight in the details of individual particle behaviour, few typical examples of the most energetic particles moving through reconnection regions and inside the current channels are identified. We show the trajectories for electrons in case F2dff with uniform resistivity in Fig.~\ref{fig:drifts_ar} and for protons in case F2dffAR with anomalous resistivity in Fig.~\ref{fig:drifts_noar}, plotted on top of the magnitude of the parallel electric field as obtained from MHD. Again, the yellow coloured parts indicate a nonzero parallel electric field and hence reconnection occurring, and grey parts indicate absent parallel electric field (see Fig.~\ref{fig:Epart8} for the respective resistive electric field without particles plotted on top). The most efficient acceleration happens to particles trapped inside the current channels, moving parallel to the magnetic field, reaching Lorentz factors up to $\gamma \sim 10^2$ within $0.01 t_S$ to $0.1 t_S$ both for protons and electrons. Since the electric field is static in the snapshots taken, the particles Lorentz factor cannot grow faster than $\gamma \sim qEct$ with $E$ the constant amplitude of the electric field. Particles reaching medium energies (Lorentz factor of the order of $\gamma \sim 2$) are observed to undergo repeated acceleration and deacceleration passing through reconnecting fields in the current sheets, in accordance with the 3D results of \cite{Rosdahl}. 
\begin{figure}
	\includegraphics[width=\columnwidth]{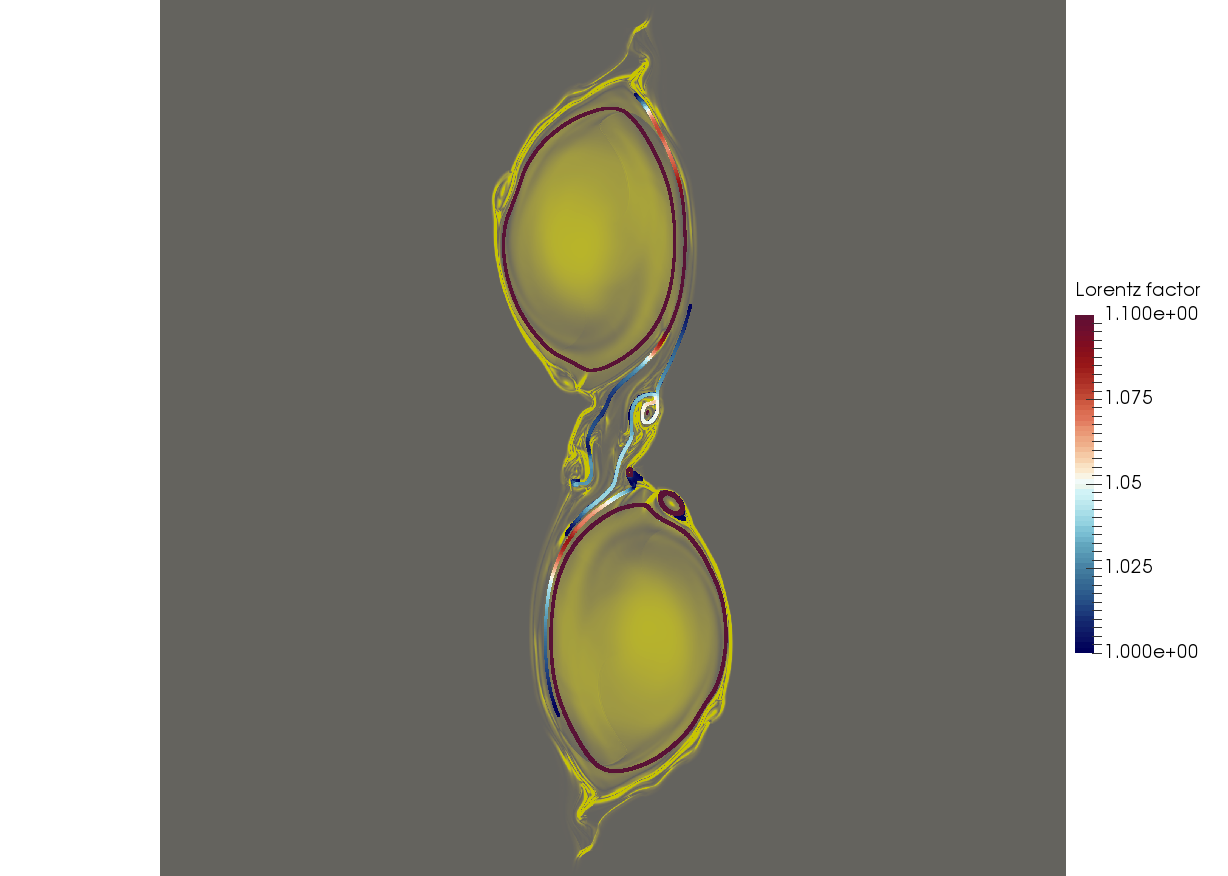}
    \caption{Selected electron trajectories, coloured by their Lorentz factor $\gamma$, in a snapshot with uniform resistivity from $8t_S$ to $8t_S + 0.1t_S$. Again plotted on top of the magnitude of the parallel electric field as obtained from MHD, where the yellow coloured parts indicate a nonzero parallel electric field and hence reconnection occurring, and grey parts indicate absent parallel electric field (see Fig.~\ref{fig:Epart8} for the resistive electric field without particles plotted on top). For the parallel electric field a linear colour is saturated to show values between $[0, 0.001]$. The same particles are used in Fig.~\ref{fig:gammavst_electrons} to quantify $\gamma(t)$. These particles belong to the non-thermal kinetic energy population.}
    \label{fig:individualelectrons}
\end{figure}
\begin{figure}
	\includegraphics[width=\columnwidth]{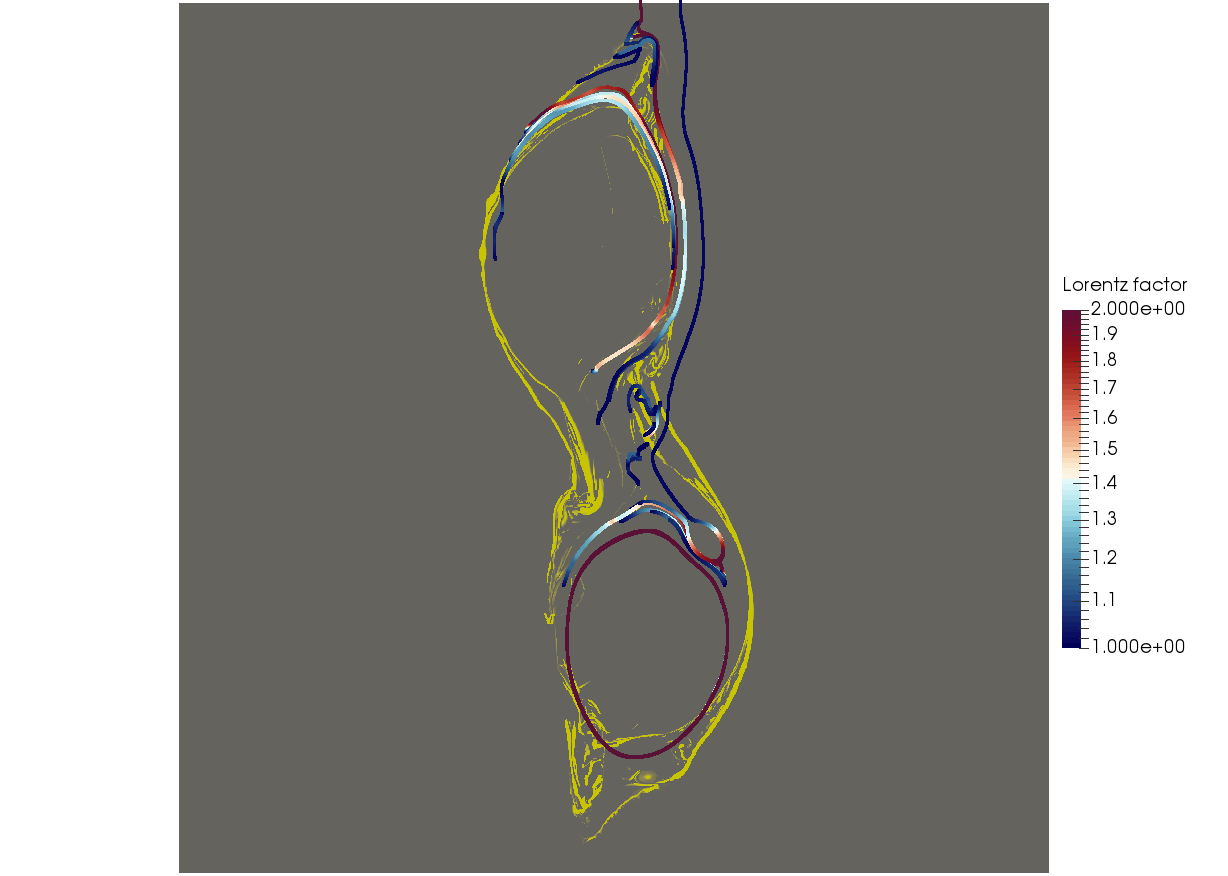}
    \caption{Selected proton trajectories, coloured by their Lorentz factor $\gamma$, in a snapshot with anomalous resistivity from $8t_S$ to $8t_S + t_S$. Again plotted on top of the magnitude of the parallel electric field as obtained from MHD, where the yellow coloured parts indicate a nonzero parallel electric field and hence reconnection occurring, and grey parts indicate absent parallel electric field (see the right panel of Fig.~\ref{fig:Epart8} for the resistive electric field without particles plotted on top). For the parallel electric field a linear colour is saturated to show values between $[0, 0.001]$. The same particles are used in Fig.~\ref{fig:gammavst_protons} to quantify $\gamma(t)$. These particles belong to the non-thermal kinetic energy population.}
    \label{fig:individualprotons}
\end{figure}
For several of the selected particles the temporal evolution of $\gamma$ is depicted in Fig.~\ref{fig:gammavst_electrons} for electrons at $t=8t_S$ to $t=8t_S+0.1t_S$ in case F2dff and in Fig.~\ref{fig:gammavst_protons} for protons at $t=8 t_S$ to $t=8t_S+0.01t_S$ in case F2dffAR. Dashed lines growing linearly in time are plotted to guide the eye. The repetitive acceleration and deacceleration is stronger in cases with anomalous resistivity, due to the evolution of regions with zero resistivity and hence no resistive electric field to accelerate the particles, and regions with resistivity causing strong gradients in the fields and hence a strong resistive electric field to accelerate the particles again. The particles reaching medium $\gamma$ are located in the current sheets and reconnection zones, whereas the particles reaching higher $\gamma$ are trapped inside the current channels or in the secondary islands, confirming the picture obtained from the energy distributions. 
\begin{figure}
	\includegraphics[width=\columnwidth]{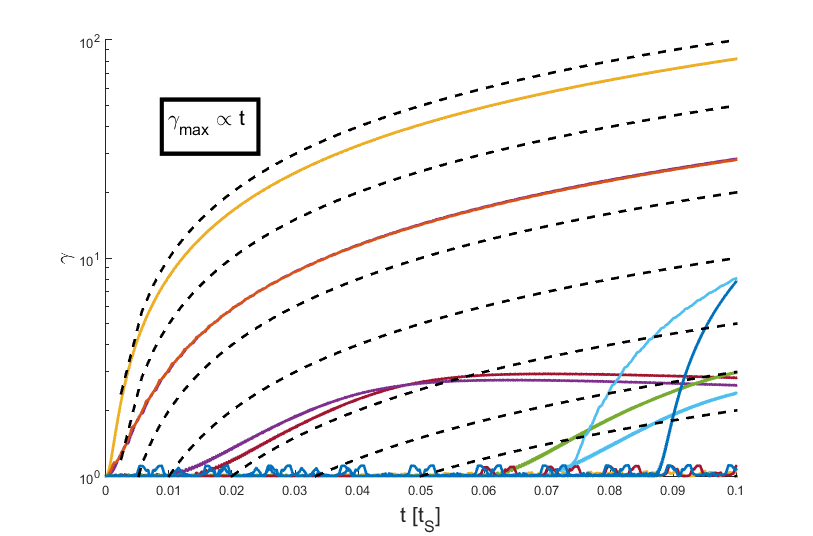}
    \caption{$\gamma \propto t$ for selected electrons in a snapshot with uniform resistivity from $8t_S$ to $8t_S + 0.1t_S$. These particles belong to the non-thermal kinetic energy population. Note the difference for particles mainly affected by parallel acceleration (quickly growing $\gamma(t)$), versus those that repeatedly visit reconnection regions (fluctuating $\gamma(t)$). All particles located in the reconnection zones (secondary islands and thin current sheets) are also shown in Fig.~\ref{fig:individualelectrons} as well as two particles located inside the two current channels.}
    \label{fig:gammavst_electrons}
\end{figure}
\begin{figure}
	\includegraphics[width=\columnwidth]{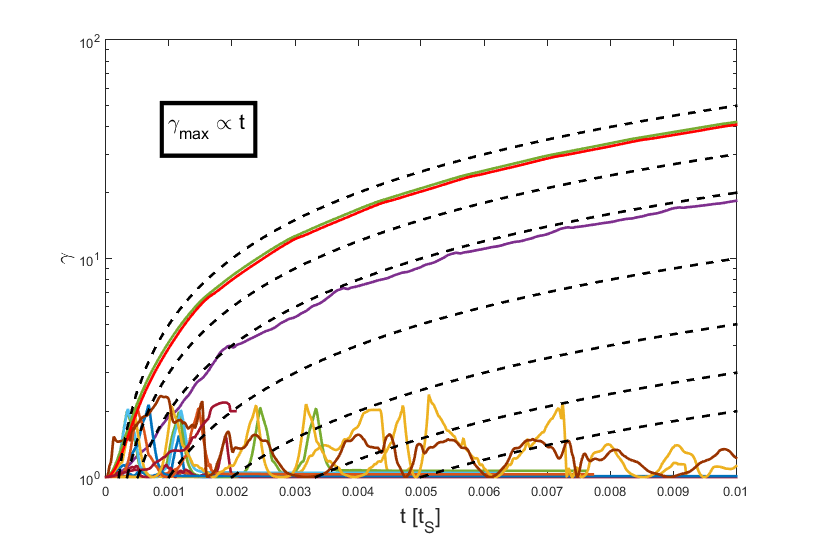}
    \caption{$\gamma \propto t$ for selected protons in a snapshot with anomalous resistivity from $8t_S$ to $8t_S + 0.01t_S$. These particles belong to the non-thermal kinetic energy population. Note the difference for particles mainly affected by parallel acceleration (quickly growing $\gamma(t)$), versus those that repeatedly visit reconnection regions (fluctuating $\gamma(t)$). All particles located in the reconnection zones (secondary islands and thin current sheets) are also shown in Fig.~\ref{fig:individualprotons} as well as one particle located inside the bottom current channel.}
    \label{fig:gammavst_protons}
\end{figure}
\section{Conclusions}
The first part of this work treats the numerical analysis of reconnection induced by a tilt instability in 2.5D setups and the interaction of a tilt and a kink instability in 3D setups. The resistive MHD equations have been solved in plasmas with very high to very low plasma-$\beta$ for force-free configurations and the results were compared to results for non-force-free equilibrium configurations in \cite{Keppens}. The goal was to select the setup with the fastest reconnection occuring in conditions most realistic for a stellar corona (i.e. low plasma-$\beta$ and force-free conditions). In all cases the instability grows linearly and causes reconnection. The onset of the instability and the peak currents reached depend on the initial setup strongly, but the general behaviour is similar in all cases. For a force-free setup, the instability grows faster and starts earlier for lower plasma-$\beta$, confirming the low resolution results of \cite{Richard}. For a non-force-free setup, with a constant guide field in the $z$-direction and a nonzero pressure gradient, the trend is opposite. The tilt instability occurs earlier for higher plasma-$\beta$, confirming results of \cite{Keppens}. Our high resolution results allow us to follow secondary island formation in the regions with strong currents. The effect of a kink instability, in 3D configurations, can enhance or delay the tilt instability depending on the strength of the guide field $B_z$ in non-force-free cases. In force-free setups the kink instability occurs in all cases but the magnetic tension in the $z$-direction delays the tilt instability compared to the 2.5D with similar plasma-$\beta$.

In all cases magnetic reconnection is prone to efficiently accelerate particles. We chose a force-free setup with the lowest plasma-$\beta$ considered to meet the conditions in the solar corona most realistically. We ran this case both with uniform resistivity and anomalous resistivity only present in regions with a strong current (sufficiently high such that there is no resistivity in the equilibrium phase), to show the effect of resistivity on reconnection. In both cases nearly-singular current sheets and secondary islands form in the high resolution 2.5D runs, but the results are different in the non-linear regime. In the case with anomalous resistivity a higher peak current is reached due to magnetic gradients that can build at places where there is a transition from nonzero resistivity to zero resistivity. 

We have analysed the behaviour and dynamics of 200.000 test particles with a relativistic guiding centre approach in static MHD snapshots from a tilt instability, causing magnetic reconnection. In all cases a Lorentz factor $\gamma \gg 1$ is reached by particles accelerated in the reconnection regions and the current channels. Even excluding the high-energy tail due to the indefinite acceleration in the current channels, particles accelerating in the reconnection zones reach $\gamma ~ \mathcal{O}(10)$, indicating that relativistic modifications are of major importance for guiding centre dynamics in the low plasma-$\beta$ conditions applied. However, the purely relativistic drift, the last term in equation (\ref{eq:gcastatic1}) $\left(\mathbf{\hat{b}}/\left(B\left(1-E_{\perp}^{2}/B^2\right)\right)\right) \times v_{\|}E_{\|}\mathbf{u_E}/c$ is several orders smaller than the curvature drift, the $\mathbf{E}\times\mathbf{B}$ drift and the parallel velocity and therefore negligible in our settings. This is in accordance with the results of \cite{Rosdahl}, \cite{Gordovskyy} and \cite{Pinto} for solar corona conditions.
The grid resolutions used in the MHD simulations are still a lot larger than the gyroradius of the test particles, such that the guiding centre approach is valid. However, there are particles reaching a gyroradius much larger than expected under the physical conditions applied. This is mainly due to the high Lorentz factors reached and by limiting the fast acceleration due to resistive electric fields, this issue will also be solved. However, in follow-up work the full equation of motion (\ref{eq:lorentztens}) should be solved to compare results and to confirm the validity of the guiding centre approximation. This also gives the opportunity to analyse test particle dynamics in relativistic plasma environments like magnetospheres of compact objects, pulsar wind nebulae and active galactic nuclei, where typical gyroradii are comparable in size to the scale of spatial variations of electromagnetic fields.
As is well known, from 2.5D MHD simulations (\citealt{Rosdahl}, \citealt{Zhou2}), test particles accelerate in the direction of the magnetic field, due to the presence of a parallel, strong, resistive electric field. We found that both electrons and protons are accelerated efficiently within MHD timescales in reconnection zones outside the current channels, starting from a Maxwellian distribution function developing a medium energy tail with $1 < \gamma < 10^2$. On top of that, particles can accelerate indefinitely in the translationally invariant direction, producing a high energy tail with energies up to $\gamma = 10^6$. The medium tail develops quickly after the onset of reconnection, whereas the high energy tail due to resistive electric fields develops instantly even in (near) equilibrium snapshots. Both acceleration in reconnection zones and in the infinitely long current channels is dominated by resistive electric fields compared to all other means of acceleration, confirming the results of \cite{Zhou2}. The energies reached are highly sensitive to the resistivity parameter chosen for the MHD simulations. To moderate the dominant resistive electric field acceleration, we applied anomalous resistivity with magnitude $\eta=0.0001$, such that resistive electric fields are only present in regions with a current larger than the equilibrium value. The distribution function now changed in the sense that there are still high energy particles dominating in energy, but not dominating in number anymore. There are less regions with resistive electric fields and hence less particles accelerated. However, the peak current reached in the MHD simulations with anomalous resistivity is higher, causing the maximum Lorentz factor for both protons and electrons to be even higher in setups with anomalous resistivity. Particle acceleration is sensitive to the spatial distribution of the resistive electric fields, meaning that particles accelerate strongly parallel to the magnetic field everywhere with non-zero resistivity. In previous studies with PIC methods (e.g. \cite{SironiPorth} for relativistic plasmas and \cite{Li}) for low-$\beta$ plasmas) it was found that magnetic curvature was the dominant acceleration mechanism. In all our simulations the contribution of resistive field acceleration is at least one order of magnitude higher than all other means of acceleration. Magnetic curvature acceleration, conform the third term in equation (\ref{eq:gcastatic1}) $\left(\mathbf{\hat{b}}/\left(B\left(1-E_{\perp}^{2}/B^2\right)\right)\right) \times (cm_0\gamma/q)\left(v_{\|}^{2}\left(\mathbf{\hat{b}}\cdot\nabla\right)\mathbf{\hat{b}}\right)$, is enhanced by the acceleration caused by resistive electric fields parallel to the magnetic fields due to the proportionality to the parallel velocity of particles squared and is therefore the second most important acceleration mechanism. This is again in accordance with the findings of \cite{Zhou2}. 
Analysing results of 2.5D test particle simulations in static snapshots helps us to tackle the issue of the indefinite acceleration in the $z$-direction and consequent hard energy spectra for future 3D setups with periodic boundary conditions and 2.5D setups in which the electric and magnetic fields are not constant (i.e. particles evolved simultaneously alongside the MHD evolution and not in static snapshots). A strategy based on anomalous resistivity alone did not yet avoid the artificial build-up of an extreme high energy tail. The effect of particles reaching too high energies due to resistive electric fields can be moderated in several ways. A potential solution is to separate the resistivity appearing in the MHD equations and in the particle equations of motion, e.g. \cite{Zhou2}, where an anomalous resistivity with magnitude $\eta=0.003$ is chosen for MHD evolutions and an anomalous resistivity with magnitude $\eta=1.0 \times 10^{-7}$ is chosen for test particle simulations. Including 3D effects would solve the issue of an invariant $z$-direction, however, periodic boundary conditions would have the same effect on the indefinite acceleration (limited by the speed of light) of particles. A solution would be to apply a thermal bath for particles leaving the simulation box in the periodic direction. A particle would then enter at the opposite periodic boundary with a random thermal velocity. This would solve the extreme high energy tail in the distribution function and the issue of particles moving far away from the initial location in the $x,y$-plane in 2.5D simulations. This solution is considered for follow-up work in which particles are evolved in fully 3D MHD setups and for the full MHD simulation period, rather than in static snapshots to analyse the effect of the kink instability and the effect of dynamic electric and magnetic fields. In these simulations the temporal evolution of the current sheet is taken into account and particles are expected to be expelled from the current channels on this timescale due to the kink instability occurring in 3D. We also aim to explore more realistic models of the solar corona, including Hall MHD effects and curved flux ropes. Another solution would be to decouple acceleration from resistive electric fields from all other types of acceleration (magnetic gradients, magnetic curvature, perpendicular electric fields et cetera). This has been done by \cite{Zhou} and \cite{Zhou2}. This nicely shows the separate effects, but it does not solve the high energies obtained due to resistive electric field acceleration. The effect of collisions moderates acceleration as well, this has been done by \cite{Gordovskyy} in the context of coronal flux loops and has to be considered for future work in less idealised setups.
In the test particle approach, interaction between particles and feedback of the particles on the fields is neglected. Even with collisions included the feedback of highly accelerated particles which are not moderated by collisions would severely influence the electromagnetic fields, which in turn affect the acceleration of particles again. The combination of strong resistive electric fields due to a high resistivity, and the absence of back reaction on the macroscopic fields causes too many particles to reach too high energy. Another way is to adopt a fully kinetic approach, however in the full domain considered in our setup that will be computationally expensive and less flexible. A way to improve this is to evolve particles in particularly interesting regions, like reconnection zones, with a particle-in-cell approach and the thermal plasma with an MHD approach (e.g. \citealt{Daldorff}, \citealt{Markidis2}, \citealt{Markidis3}, \citealt{Markidis}, \citealt{Vaidya}). 
\section*{Acknowledgements}
This research was supported by projects GOA/2015-014 (2014-2018 KU Leuven) and the Interuniversity Attraction Poles Programme by the Belgian Science Policy Office (IAP P7/08 CHARM)
The computational resources and services used in this work were provided by the VSC (Flemish Supercomputer Center), funded by the Research Foundation Flanders (FWO) and the Flemish Government - department EWI.
BR likes to thank Lorenzo Sironi, Fabio Bacchini, Norbert Magyar, Jannis Teunissen, Kirit Makwana, Matthieu leroy and Dimitris Millas for fruitful discussions and comments.




\bibliographystyle{apalike}
\bibliography{mylib} 










\label{lastpage}
\end{document}